\documentstyle[12pt,epsfig]{article}
\newlength{\dinwidth}
\newlength{\dinmargin}
\newcommand{\ba}{\begin{array}}
\newcommand{\ea}{\end{array}}
\newcommand{\beq}{\begin{equation}}
\newcommand{\eeq}{\end{equation}}
\newcommand{\bea}{\begin{eqnarray}}
\newcommand{\eea}{\end{eqnarray}}

\def\S{{\bf S}}

\def\bce{\begin{center}}
\def\ece{\end{center}}

\def\nonu{\nonumber}

\def\pa{\partial}
\def\al{\alpha}
\def\be{\beta}
\def\ga{\gamma}
\def\Ga{\Gamma}
\def\de{\delta}

\def\la{\lambda}
\def\La{\Lambda}

\def\si{\sigma}

\def\S{{\bf S}}

\begin{document}
\thispagestyle{empty}
\addtocounter{page}{-1}
\begin{flushright}
{\tt hep-th/0109010}\\
revised(April, 2002) \\
\end{flushright}
\vspace*{1.3cm}
\centerline{\Large \bf Domain Wall}
\vskip0.3cm
\centerline{\Large \bf from Gauged  $d=4, {\cal N}=8$ Supergravity:
Part I}
\vspace*{1.5cm}
\centerline{\bf Changhyun Ahn {\rm and} Kyungsung Woo }
\vspace*{1.0cm}
\centerline{\it Department of Physics,
Kyungpook National University, Taegu 702-701 Korea}
\vskip0.3cm
\vspace*{0.8cm}
\centerline{\tt ahn@knu.ac.kr, \qquad a0418008@rose0.knu.ac.kr}
\vskip2cm
\centerline{\bf abstract}
\vspace*{0.5cm}

By studying already known extrema of non-semi-simple Inonu-Wigner
contraction $CSO(p, q)^{+}$ and
non-compact $SO(p, q)^{+}$($p+q=8$) gauged
${\cal N}=8$ supergravity in 4-dimensions developed by Hull sometime ago,
one expects
there exists nontrivial flow in the 3-dimensional
boundary field theory. We find that these
gaugings provide first-order domain-wall solutions from direct extremization
of energy-density.

We also consider the most general $CSO(p, q, r)^{+}$ with $p+q+r=8$
gauging of ${\cal N}=8$ supergravity by two successive
$SL(8, {\bf R})$ transformations of the de Wit-Nicolai theory, that is,
compact $SO(8)$
gauged supergravity. The theory found earlier has local
$SU(8) \times CSO(p, q, r)^{+}$ gauge symmetry
as well as local ${\cal N}=8$ supersymmetry.
The gauge group $CSO(p, q, r)^{+}$ is spontaneously reduced to
its maximal compact subgroup $SO(p)^{+} \times SO(q)^{+}
\times U(1)^{+r(r-1)/2}$.
The T-tensor we obtain describes a
two-parameter family of gauged ${\cal N}=8$ supergravity from which one can
construct $A_1$ and $A_2$ tensors. The effective nontrivial
scalar potential can be written as the difference of positive
definite terms. We examine the scalar potential for critical points
at which the expectation value of the scalar field is $SO(p)^{+} \times
SO(q)^{+} \times SO(r)^{+}$ invariant.
It turns out that there is no new extra critical point.
However, we do have flow equations and domain-wall
solutions for the scalar fields are the gradient flow equations of the
superpotential that is one of the eigenvalues of $A_1$ tensor.


\baselineskip=18pt
\newpage

\section{Introduction}

One of the interesting issues in recent research
is the domain wall(DW)/quantum field theory(QFT)
correspondence initiated by
\cite{bst} between supergravity,
in the near horizon region of the corresponding
supergravity brane solution, compactified
on domain wall spacetimes that are locally isometric to
Anti-de Sitter(AdS) space
but different from it
globally
and quantum(nonconformal) field theories describing the internal
dynamics of branes and living on the boundary of
such spacetimes.
DW/QFT correspondence was
motivated by the fact that the AdS metric in horospherical coordinates
is a special case of the domain wall metric \cite{lpt,cgr}.
R-symmetry of the supersymmetric QFT on
the boundary of domain worldvolume
should match with the gauge group of the corresponding gauged supergravity.
Compact gaugings are not the only ones for extended supergravities but
there exists a rich structure of non-compact and non-semi-simple
gaugings (Note that the unitarity property is preserved since in all extrema of
scalar potential, non-compact gauge symmetry is reduced to
some residual compact subgroup).
Such a theory plays a fundamental role in the description of the
DW/QFT correspondence
as the maximally compact gauged supergravity has played in the
AdS/conformal field theory(CFT) duality \cite{maldacena,witten,gkp} that
is a correspondence between certain compact gauged supergravities
and certain
conformal field theories.
It would be interesting to identify the
appropriate non-compact and non-semi-simple gauged supergravities
corresponding to each choice of brane configuration.

One of the questions we addressed was whether the maximal supergravity theories with
non-compact
gauge groups can be obtained from higher dimensional theory.
${\cal N}=8$ gauged supergravity theories have been constructed in
4-dimensions
with gauge groups $SO(p, 8-p)$ where $p=0, 1, 2, 3,$ and $4$ or with
non-semi-simple
 contractions of these gauge groups
\cite{hullprd,hullplb1,hullphysica,hullplb2,hullcqg,hullwarner1,hullwarner2,hullwarner3}.
In 7-dimensions, ${\cal N}=4$ gauged
supergravity theories have been constructed
with gauge group $SO(p, 5-p)$ with $p=0, 1, 2$ \cite{ppvw}.
In five-dimensions there exist gauged ${\cal N}=8$ supergravity theories with
gauge groups $SO(p,6-p)$ with $p=0, 1, 2, 3$ or $SU(3,1)$ \cite{grw}.
Although odd-dimensional
gauged   supergravity theories do not appear to allow gaugings of
non-semi-simple
contractions, some researchers have attempted to resolve the difficulties in
five-dimensions \cite{freetal1,freetal2}.
It has been shown that the $SO(p,q)$ gaugings and
their non-semi-simple contractions
can be obtained from the appropriate higher dimensional supergravity
theories. The
spheres
used to compactify the $SO(p)$ gaugings are replaced by hyperboloids
for the non-compact $SO(p,q)$ gaugings and generalized cylinders for the
non-semi-simple contractions \cite{hullwarner3}.

Since embedding or consistent truncation of gauged supergravity is known
to characterize $\S^7$ compactification of eleven-dimensional
supergravity\footnote{ By
generalizing compactification vacuum ansatz to the nonlinear level,
solutions of eleven-dimensional supergravity were obtained
directly from the scalar and pseudo-scalar expectation values at
various critical points of the ${\cal N}=8$ supergravity potential \cite{dnw}.
They reproduced all known Kaluza-Klein solutions of
the eleven-dimensional supergravity: round $\S^7$ \cite{dp},
$SO(7)^-$-invariant, {\sl parallelized} $\S^7$ \cite{englert},
$SO(7)^{+}$-invariant vacuum \cite{dn}, $SU(4)^{-}$-invariant vacuum
\cite{pw}, and a supersymmetric one with $G_2$ invariance.
Among them, round $\S^7$- and $G_2$-invariant vacua are
stable, while $SO(7)^{\pm}$-invariant ones are known to be unstable
\cite{dn1}.
In \cite{ar,ar1} three dimensional conformal field theories
were classified by using AdS/CFT correspondence.
In particular, researchers \cite{cpw}
have studied the $SU(3) \times U(1)$
critical point, from the point of higher dimensional analysis,
which does not belong to the classification \cite{dnw} but
is a supersymmetric critical point of four-dimensional
gauged supergravity.  }, we also
are interested in the domain-wall solution in four-dimensional gauged
supergravity.
In \cite{ap},
a renormalization group flow from ${\cal N}=8$,
$SO(8)$ invariant UV fixed point to ${\cal N}=2$, $SU(3) \times U(1)$
invariant IR fixed point was found
by studying the de Wit-Nicolai potential which is
invariant in the $SU(3) \times U(1)$ group.
For this interpretation it was crucial
to know the form of superpotential that was encoded in the structure of
the T-tensor of the theory.
Very recently, the lift to M-theory of the solution described in \cite{ap}
was constructed \cite{cpw}(See also \cite{jlp}).
Moreover,
it was natural and illuminating
to ask whether one can construct the most general
superpotential
for critical points in four-dimensional ${\cal N}=8$
gauged supergravity: 1) $SU(3)$-invariant sectors,
2) $SO(5)$-invariant sectors
and 3) $SO(3) \times SO(3)$-invariant sector \cite{aw}. In order
to find and study
BPS domain-wall solutions by  minimization of the energy-functional, one has to
reorganize it into complete squares. Then one should expect
that the scalar potential takes the form of squares of physical quantities.
One important feature of the de Wit-Nicolai $d=4, {\cal N}=8$ supergravity
is that the scalar potential can be written as the difference of two
positive square terms.
Together with kinetic terms this implies that one may construct
the energy-functional in terms of complete squares.

The other gaugings of ${\cal N}=8$ supergravity could be obtained
in the same way the $SO(8)$ gauging. One could proceed in the same way as
the de Wit-Nicolai theory by
changing the supersymmetry transformations and adding them to
the Lagrangian. Contrary to ${\cal N}=4$ supergravity in four or
seven dimensions, as a result of the complicated nonlinear tensorial
structure, it is necessary to prove that the modified $A_1$ and $A_2$ tensors
satisfy a number of rather involved and lengthy
quantities as in \cite{dn2}, to
demonstrate the supersymmetry of the theory.
However, in
\cite{hullprd,hullplb1,hullphysica,hullplb2,hullcqg,hullwarner1,hullwarner2,hullwarner3},
an indirect and simple method which uses some results known in
the de Wit-Nicolai theory
was found to generate different gaugings
from $SO(8)$ compact-gauged supergravity theory in such a way that one obtains
the full nonlinear structure automatically and is guaranteed gauge invariance
and supersymmetry.
The first step was to construct a real, self-dual anti-symmetric
$SO(p)^{+} \times SO(q)^{+}$-invariant four-form
tensor using both the generator of $SL(8,{\bf R})$ and
$SO(8)$ $\Ga$ matrices. Next, it was necessary to describe the projectors that project
the $SO(8)$ Lie algebra onto its subalgebras in terms of a four-form
tensor in order to provide a convenient way to deal explicitly with the
$SL(8, {\bf R})$ transformation.  Then exhaustive manipulations
of the invariance  of four-form tensor were
crucial for the existence of those gaugings and
finiteness of coupling constant-dependent, covariant derivative terms
as we take the infinity limit of some real parameter.
Then we possess an explicit form for the T-tensor
in terms of the standard parametrization of the scalar coset space.

In this paper, in section 2, we analyze known vacua of
four-dimensional ${\cal N}=8$ non-compact and non-semi-simple gauged
supergravity developed by Hull
\cite{hullprd,hullplb1,hullphysica,hullplb2,hullcqg,hullwarner1,hullwarner2,hullwarner3}.
We claim no originality for most of the results presented in sections
2.1-2.3, although our derivations are hopefully illuminating.
What we will do is to find out

$\bullet$ a superpotential from given T-tensor
for known gauged supergravity theory and

$\bullet$  BPS domain-wall solutions
from an energy-functional written  in terms of complete squares. \\
In section 3, we will consider most general gaugings
$CSO(p, q, r)^{+}$ initiated by Hull
where $p+q+r=8$ by using two successive $SL(8, {\bf R})$
transformations on the compact gauged supergravity.
What we will do is to find

$\bullet$ a T-tensor in $SU(8)$-basis,

$\bullet$  a superpotential
and a scalar potential  from our findings of T-tensor and

$\bullet$  domain-wall solutions.\\
In section 4, we present our main results.
In appendices, we present some details which are necessary for the
calculations in sections 2 and 3.

\section{Domain Wall from $SO(p)^{+} \times SO(q)^{+}$
Sectors  of ${\cal N}=8$ Supergravity }

Let us consider an ungauged supergravity theory with ${\cal N}$ local Majorana
supersymmetries, $4 \leq {\cal N} \leq 8$ given by Cremmer-Julia theory
\cite{cremmer} who constructed it by dimensionally reducing 11-dimensional
supergravity.
Recall that since a Majorana spinor in four-dimensions has
four real components, the total number of supercharges
for the maximal ${\cal N}=8$ theory becomes 32.
Note that there is no scalar field in the graviton multiplet for
${\cal N} < 4$. If the maximum spin is to be two, the number ${\cal N}$ can not be
larger than 8.
The scalar fields lie in a coset space $G/H$ where
$G$ is some non-compact group and $H$ its maximal compact subgroup.
Group $H$ is a local
symmetry of the whole action while group $G$
is a global
symmetry of the equations of motion only(not the action) because
it acts on the spin-1 fields through duality transformations.
However, there exists a
non-compact subgroup $L$ of $G$, which is a global(rigid) symmetry of
the action.
One can gauge subgroup $K$ of the global symmetry group
$L$ of the action where the dimension of $K$ cannot exceed
the number of vector fields in the model.
To gauge the theory, one adds minimal Yang-Mills
couplings for $K$ both to the Lagrangian ${\cal L}_0$, which is the
Lagrangian of the ungauged theory, and to the supersymmetry
transformation rules of the ungauged theory with the vector fields
of the theory acting as gauge connections.
One should add coupling constant
dependent terms to both the action and supersymmetry transformation laws in
such a way that local supersymmetry is restored and gauge invariance
is maintained. Then one obtains a theory with Lagrangian
${\cal L}= {\cal L}_0 + {\cal L}_g$ where ${\cal L}_g$ consists of
minimal gauge couplings with coupling constant $g$,
fermionic bilinear terms proportional to $g$,
and a scalar potential proportional to $g^2$.
The minimal couplings and scalar potential break the symmetry $G$ of
the equations of motion and the symmetry $L$ of the action down to $K$
while leaving the local symmetry $H$ unchanged. Then the gauge theory
has both $H \times K$ local gauge symmetry and ${\cal N}$-extended local
supersymmetry.

The ungauged ${\cal N}=8$ supergravity(in this paper, we restrict
ourselves to
${\cal N}=8$ theory) has a symmetry
$G \times H={E_{7}}_{\mbox{global}}
\times SU(8)_{\mbox{local}}$ of the equations of
motion where the 28 vectors correspond to a global Abelian symmetry
between particles. Motivated by the fact that
realistic theories of fundamental interactions are based on
local, non-Abelian symmetries,
de Wit and Nicolai \cite{dn3,dn2} gauged the subgroup $K=SO(8)$ of
(the $L=SL(8,{\bf R})$ subgroup of) $E_7$ that is a global symmetry
of ${\cal L}_0$, and obtained a theory with a local $K\times H=
SO(8) \times SU(8)$
symmetry. The gauge group $K \subset L$ is a local symmetry:
${\cal L} \rightarrow {\cal L}$ under $K$
while
the remainder $L \backslash K$
of the non-compact group $L$ is a global symmetry
of ${\cal L}_0$ but not of ${\cal L}_g$:${\cal L} \rightarrow {\cal L'}=
{\cal L}_0 + {\cal L}'_g$ under  $L \backslash K$.
In other words,
acting with $L \backslash K$
changes the gauge covariantizations, fermion bilinear terms,
and scalar potential in ${\cal L}_g$ while keeping the ${\cal L}_0$
unaffected.
This is an invertible field-redefinition
for finite value of $t$ which appears in (\ref{et})
that leads to an equivalent theory, invariant
under the local supersymmetries and local gauge symmetry.

The contraction procedure
involves a sequence of change of
basis transformations depending on the parameter \cite{gilmore}.
Although the transformation becomes singular in the zero limit of
a parameter, the Lie bracket exists and is well defined in this singular
limit. The original and contracted algebras are not isomorphic.
Note that non-singular changes of bases can never lead to new algebras
because under such a transformation the new structure
constant tensor possesses exactly as much information as the original.
Let us consider a sequence of non-singular elements $E(\xi)$
of $L$  with $\xi$ real parameter and $E(1)=1$, and an identity transformation,
whose limit point $E(0)$
is singular and not in $L$.
As long as $E(\xi)$ remains nonsingular$(\xi \neq 0)$, the structure constants
have the usual tensor properties.
Acting on the Lagrangian with $E(\xi)$
yields a sequence of Lagrangian:$ {\cal L} \rightarrow
{\cal L}^{\prime}(\xi)={\cal L}_0 +
{{\cal L}_g}^{\prime}(\xi)$. If one also rescales the coupling constant
$g$ to an $\xi$-dependent one through $g \rightarrow g^{\prime}(\xi)$ for some
choices of the sequence $E(\xi)$ in $L$, the limit of ${{\cal L}_g}^{
\prime}(\xi)$ as $\xi \rightarrow 0$($\equiv {{\cal L}_g}^{\prime}(0)$) exists
and is well defined(the new structure constants characterize a Lie algebra)
so that ${\cal L}^{\prime}(0)={\cal L}_0 + {{\cal L}_g}^{\prime}(0)$ gives
the Lagrangian for a gauge-invariant supersymmetric theory.
The gauge group corresponding to
${\cal L}'(0)$ is not $K=SO(8)$ itself but
 an Inonu-Wigner \cite{iw}
contraction of $K$ denoted by $CSO(p, q)^{+}$ with $p+q=8$ 
\cite{hullplb2,hullcqg}.
A new(different from the de Wit-Nicolai compact gauged supergravity theory)
gauging, inequivalent to the original one,
is obtained by a singular, noninvertible field redefinition.
One can also continue the Lagrangian ${\cal L}^{\prime}(\xi)$ to negative
values of $\xi$.
In this case, ${\cal L}'(-1) $ is the Lagrangian
for another gauging and the gauge group is non-compact $SO(p,q)^{+}$ with
$p+q=8$ \cite{hullplb2,hullcqg}.

In section 2.1, starting with the action of $L=SL(8, {\bf R})$ element
on the de Wit-Nicolai
theory, we present a
superpotential which is an eigenvalue of $A_1$ tensor, that is,
a partially contracted T-tensor.
In section 2.2, with explicit $\xi$-dependence on the
T-tensor, one recovers
more general scalar potetential which will be reduced to the one in
section 2.1 when we put
$\xi=0$ and obtains
a more general superpotential.
In section 2.3, we present other cases, $CSO(p, q)^{+}$ and
$SO(p, q)^{+}$ gaugings
where $p=6, 5, 4, 3, 2, 1$ and $q=8-p$ and review their critical points
in a scalar potential.
In section 2.4, we construct
a domain-wall solution from an energy-functional.
Finally in section 2.5,  as an aside, we will concentrate
on the construction of a scalar potential for
the vacuum expectation value given
in terms of real, anti-self-dual(not self-dual), totally
anti-symmetric tensor. The parametrization for
this singlet-space is invariant under the $SO(p)^{-} \times SO(q)^{-}$
where $p+q=8$.

\subsection{ Superpotential in $CSO(7,1)^{+}=ISO(7)^{+}$ Gauging
\cite{hullprd}  }

Following the procedure we have introduced, the action of
the non-compact part of $SL(8, {\bf R})$, $L \backslash K$,
on the theory can be
used to other gauged ${\cal N}=8$ supergravity.
Let us consider the acting with the $L=SL(8, {\bf R}) \subset
E_{7(+7)}$
element
\bea
E(t) =
\mbox{exp} \left(
\begin{array}{cc}
0 &  \ t X^{+IJKL}  \\
 t X_{IJKL}^{+} & 0
\end{array} \right),
\label{et}
\eea
on the de Wit-Nicolai theory where $t$ is a real parameter proportional
to $- \ln \xi$ where $\xi$ was introduced before  and
$X^{+IJKL}$ is some real and self-dual
totally antisymmetric tensor that satisfies
\bea
X^{+IJKL}= {\overline{X}}_{IJKL}^{+} =
\frac{\eta}{24}\epsilon^{IJKLMNPQ}X^{+MNPQ}.
\nonu
\eea
Since $E(t)$ is in the real $SL(8, {\bf R})$ subgroup of $E_{7(+7)}$,
the ungauged Cremmer-Julia action ${\cal L}_0$ remains unchanged but
the $g$-dependent part ${\cal L}_g$ is modified nontrivially(changes
the minimal couplings and rotates the $A_1$ and $A_
{2} $ tensors into one another).
This gives one-parameter family of Lagrangian related to
the de Wit-Nicolai theory($t=0$ where $E(0)=1$, identity
transformation, or equivalently $\xi=1$ and
$E(\xi=1)=1$)
by the $SL(8, {\bf R})$ field-redefinition given by $E(t)$.
For all finite values of $t$, this yields a theory
which is equivalent to the de Wit-Nicolai theory by field-redefinition.
However, a different gauging might be found in the limit
$t \rightarrow \infty$(equivalent to $\xi \rightarrow 0 $)
if it exists. For many choices of the four-form
$X^{+IJKL}$, the limit does not exist.
The simplest and special choice for which this limit
exists \cite{hullprd} is\footnote{
We emphasize that the way we have chosen $X^{+IJKL}$ here
is different from the one in \cite{aw} in the sense that
in \cite{aw} the $SU(2)$ matrix of $SU(8)$ appears in the last
$2 \times 2$ block
diagonal. However, in this paper, we take it as the first
$2 \times 2$ block diagonal  matrix. The nonzero-component of
$X^{+IJKL}$ is either $1/2$ or $-1/2$ as in \cite{hullprd}.    }
\bea
X^{+IJKL}=Y^{IJKL}+\frac{\eta}{24}\epsilon^{IJKLMNPQ}Y^{MNPQ},
\label{xijkl}
\eea
where
\bea
Y^{IJKL}=\frac{1}{2} \left(
\delta^{IJKL}_{1\;2\;3\;4}+\delta^{IJKL}_{1\;2\;5\;6}+
\delta^{IJKL}_{1\;2\;7\;8}+\delta^{IJKL}_{1\;3\;7\;5}+
\delta^{IJKL}_{1\;3\;6\;8}
+\delta^{IJKL}_{1\;4\;5\;8}+\delta^{IJKL}_{1\;4\;6\;7} \right).
\nonu
\eea
Here $\eta=+1$ for $SO(7)^{+}$-invariant $X^{+IJKL}$ and
$\delta^{IJKL}_{MNPQ}$ has 1 when $I,J,K$ and $L$ form an
even permutation of $M,N,P,Q$ and $-1$ when they form an odd permutation
of $M,N,P,Q$ and vanishes.
We will come to the case in which $\eta=-1$ later in section 2.5 which holds for
$SO(7)^{-}$-invariant $X^{-IJKL}$.
The four-form tensor $X^{+IJKL}$ is closely
related to the torsion parallelizing
seven-sphere $\S^7$ \cite{dn1,englert,castetal,englertetal,warner}
and invariant under
the $SO(7)^{+}$-subgroup of $SO(8)$.
Turning on the vacuum expectation value
proportional to $X^{+IJKL}$ in the de Wit-Nicolai theory
gives rise to spontaneous symmetry breaking of $SO(8)$ into
$SO(7)^{+}$.
Regarded  as $28 \times 28$ matrices, $X^{+IJKL}$ has
21 eigenvalues of $-1$ and 7 eigenvalues of $+3$. Introducing
the projector $P_{+}$ into the 21-dimensional
eigenspace($P_{+}$ projects the generators of $SO(8)$ onto those of
$SO(7)^{+}$ while $P_{-}$ projects the generators of $SO(8)$ into the
remainder $SO(8) \backslash SO(7)^{+}$), \footnote{
Note that although the subscript minus sign in $P_{-}$ is nothing to
do with the anti-self dual part $SO(7)^{-}$ of $SO(8)$,
we will follow the same notation as in the previous
literature \cite{hullprd}.  In section 2.5, we take those
projectors as $P_1$ and $P_2$.} they are given in terms of $X^{+IJKL}$
\bea
P_{+}^{IJKL} =
\frac{3}{4} \left(  \delta^{IJ}_{KL} - \frac{1}{3} X^{+IJKL} \right),
\nonu
\eea
and\footnote{
$\delta^{IJ}_{KL}$ is defined as $\delta^{IJ}_{KL}=\frac{1}{2! 2!}
\left( \de_{K}^{I} \de_{L}^{J} - \de_{L}^{I} \de_{K}^{J} -
\de_{K}^{J} \de_{L}^{I} +\de_{L}^{J} \de_{K}^{I} \right)= \frac{1}{2} \left(
 \de_{K}^{I} \de_{L}^{J} - \de_{L}^{I} \de_{K}^{J} \right)$.}
\bea
P_{-}^{IJKL} =   \delta^{IJ}_{KL} -P_{+}^{IJKL} =
\frac{1}{4} \left(  \delta_{IJ}^{KL} +   X^{+IJKL} \right).
\nonu
\eea
Therefore, one has
\bea
X^{+IJKL}=  -P_{+IJKL} + 3 P_{-IJKL}.
\label{71x}
\eea
One can easily check that the projectors have the following
properties\footnote{In terms of $X^{+IJKL}$, we have the following
relation, $\left(  \delta^{IJ}_{KL} - \frac{1}{3} X^{+IJKL} \right)
\left(  \delta_{IJ}^{KL} +   X^{+IJKL} \right)=0$.  }
which will be used throughout
this paper
\bea
P_{\pm}^2 = P_{\pm}, \qquad P_{\pm} P_{\mp} =0.
\nonu
\eea
Here the product $P_{\pm}^2 $ is that of $28 \times 28$ matrices,
$(P_{\pm}^2)^{IJKL} = P_{\pm}^{IJMN} P_{\pm}^{MNKL}$.
The 28 $SO(8)$ generators $\La^{IJ}$ are projected onto
a 21-dimensional subspace by $P_{+}$, $\La_{+}^{IJ} = P_{+}^{IJKL}
\La^{KL}$ and this subspace is the Lie algebra for the
$SO(7)^{+}$-subgroup of $SO(8)$; in other words, the subgroup
stabilizes a right-handed  side $SO(8)$ spinor(See the appendix B).
Similarly, the remaining 7 generators
are generated by $\La_{-}^{IJ} = P_{-}^{IJKL} \La^{KL}$.

The change of the minimal couplings under supersymmetry gives
a net change of the action under an infinitesimal local supersymmetry
that can be parametrized by a  T-tensor \cite{hullprd}.
An expression for the  T-tensor, $T^{\prime\;jkl}_{i}$
can be obtained by realizing that a variation of $A_{\mu}^{IJ}$ leads to
a variation of the $SU(8)$-connection
${\cal B}_{\mu i}^{\;\;\;j}$
\bea
T^{\prime\;jkl}_{i} &  = &
\left( \overline{u}^{kl}_{\;\;\;IJ}+ \overline{v}^{klIJ}\right)
 \left[ M_{IJKL}\left( u_{im}^{\;\;\;\;KM}\overline{u}^{jm}_{\;\;\;\;
LM}-v_{imKM}\overline{v}^{jmLM} \right) \right. \nonumber\\
 & & \left. +N_{IJ}^{\;\;\;\;KLMN} \left( v_{imKL}
\overline{u}^{jm}_{\;\;\;\;MN}-u_{im}^{\;\;\;\;KL}\overline{v}^{jmMN}
\right) \right]
\label{newttensor}
\eea
where $M_{IJKL}$ and $ N_{IJ}^{\;\;\;\;KLMN}$ are defined in terms of
projectors
\bea
M_{IJKL}  & = & P_{+IJKL}+\frac{1}{2}P_{-IJKL},
\nonu \\
N_{IJ}^{\;\;\;\;KLMN} &= & \frac{1}{2}P_{-\;\;\;\;\;
[P}^{IJ[K}\delta^{L]}_{\;\;\;Q]} \left( P_{-}^{PQMN} -
P_{+}^{PQMN}  \right).
\nonu
\eea
The supersymmetry of the theory is restored by
adding ${\cal L}_g^{\prime}$
to the ungauged action ${\cal L}_0$ and
to the supersymmetry transformation rules
where $A_1, A_2$ tensors, that appear in ${\cal L}_g^{\prime}$,
 have a functional dependence
on the scalar field but with $T'$ tensor.
That is, for example,
\bea
A_{1}^{\prime \;ij}=-\frac{4}{21}T_{m}^{\prime\;\;ijm},\;\;\;
A_{2l}^{\prime\;\;\;ijk}=-\frac{4}{3}T_{l}^{\prime\;[ijk]}.
\label{A1A2}
\eea
The parametrization for the $SO(7)^{+}$-singlet space\footnote{The
35-dimensional fourth rank self-dual antisymmetric tensor
representation of $SO(8)$
splits into the $SO(7)^{+}$ representation ${\bf 35}
\rightarrow {\bf 27} +{\bf 7} +
{\bf 1}$ where the singlet $\bf 1$ is nothing but
$SO(7)^{+}$-invariant tensor $X^{+IJKL}$.}
that is
an invariant subspace under a particular $SO(7)^{+}$ subgroup of
$SO(8)$ becomes
\bea
\phi_{IJKL}=4\sqrt{2}sX_{IJKL}^{+}
\nonu
\eea
where $s$ is a real scalar field.

Therefore, 56-beins ${\cal V}(x)$ can be written as a $ 56\times 56$ matrix whose
elements are the function of scalar $s$
by exponentiating
the vacuum expectation value $\phi_{IJKL}$.
On the other hand, 28-beins $u_{ij}^{\;\;\;KL}$
and $v_{ijKL}$ are
elements of this ${\cal V}(x)$.
One can explicitly construct 28-beins $u_{ij}^{\;\;\;KL}$ and $v_{ijKL}$
in terms of scalar $s$
and they are given in the appendix E (\ref{so7so1}).
Now the complete expression for $A_1^{\prime}$ and
$A_{2}^{\prime}$ tensors are given in terms of $s$ using
(\ref{newttensor}) and (\ref{A1A2}).
It turns out from (\ref{A1A2}) that
$A_1^{\prime }$ tensor has  a single real
eigenvalue, $z_1$
with degeneracies 8 and has the following form
\bea
A_{1}^{\prime\;\;ij}=\mbox{diag}\left(z_{1},z_{1},z_{1},z_{1},
z_{1},z_{1},z_{1},z_{1}\right), \qquad
z_{1}=\frac{7}{8}e^{s}.
\label{a1cso71}
\eea
Similarly, $A_{2}^{\prime}$
tensor can be obtained from the triple product of
$u_{ij}^{\;\;\;KL}$ and $v_{ijKL}$ fields, that is, from (\ref{A1A2}).
They can be written as
\bea
A_{2i}^{\prime \;\;\;jkl}=\frac{1}{4}e^{s}X^{+ijkl}.
\label{a2cso71}
\eea
Finally,
the scalar potential together with new $A_1^{\prime}$
and $A_2^{\prime}$ tensors
can be written, by combining all the components of
$A_1^{\prime }, A_{2}^{\prime} $
tensors,  as \cite{hullprd,hullphysica}
\bea
V_{7,1,\xi=0}
=  -g^{2}\left(
\frac{3}{4}|A_{1}^{\prime\;ij}|^{2}-\frac{1}{24}|A_{2i}^{\prime\;\;\;jkl}|^2
\right) =
-\frac{35}{8} g^2 e^{2s}
\label{potentialcso}
\eea
which implies that there is no $SO(7)^{+}$-invariant critical point of
potential by differentiating this scalar potential with repect to
a field $s$.
The eigenvalue $z_1$ provides a superpotential which will
be analyzed in detail in section 2.3-2.4. The scalar
potential can be written as
\bea
V_{7,1,\xi=0} = g^2 \left[ \frac{2}{7} (\pa_s z_1)^2 -
6 z_1^2 \right] =
g^2 \left[ 4 (\pa_{\widetilde{s}} z_1)^2 -
6 z_1^2 \right]
\nonu
\eea
where $\widetilde{s} = \sqrt{14} s$.
The theory \cite{hullprd}
constitutes a gauging of the 28-dimensional, non-compact
$ISO(7)^{+}$ symmetry of the Cremmer-Julia action ${\cal L}_0$. The theory
has ${\cal N}=8$ local supersymmetry and $H \times K_{\xi=0,p=7,q=1}=
SU(8) \times ISO(7)^{+}$ local gauge
symmetry where $ISO(7)^{+}$ is the isometry group of
Euclidean 7 space, ${\bf R}^7$.
In the symmetric gauge the diagonal $SO(7)^{+}$ subgroup
is manifest.
The non-compact gauged ${\cal N}=8$ supergravity theories can
be obtained by compactification of 11-dimensional supergravity
on hyperboloids of a constant negative curvature. The contracted version
corresponds to a limit in which the hyperboloid degenerates to an infinite
cylinder \cite{hullwarner3}. Thus, the $ ISO(7)^{+}$ theory corresponds to
a compactification on the cylinder ${\bf S}^6 \times {\bf R}^1$ that can be
replaced by ${\bf S}^6 \times {\bf S}^1$ because the near-horizon limit of
the D2-brane is different from that of M2-brane \cite{bst}.
As near-horizon limits of the $k$ torus $T^k$ reduction of the M2-brane, one
expects that the corresponding theory is
$CSO(8-k, k)^{+}$ gauged ${\cal N}=8$, $d=4$ supergravity.

\subsection{Superpotential in Non-compact $SO(7,1)^{+}$ Gauging
\cite{hullplb1}  }

A suitable one-parameter family (that cannot be more
than 28-dimensional) where $\xi$ is a real parameter
of 28-dimensional subgroups of $L=SL(8, {\bf R})$ each parametrized
by some real antisymmetric generator,
is generated with nonzero $\xi$.
The commutation relations of the generators are given in
the previous subsection
and the only difference is that there exists another nonzero commutator.
Then the 21 $\La_{+}$ generate
$SO(7)^{+}$ group in which the seven linearly independent
$\La_{-}$ transforms as a ${\bf 7}$ representation. When $\xi >0$,
the algebra is that of $SO(8)$ and normalization is obtained when
$\xi =1$. When $\xi < 0$, one obtains $SO(7,1)^{+}$, the normalization being
obtained when $\xi=-1$. When $\xi =0$, it gives $ISO(7)^{+}$ as done
previously.
Then the one parameter family of gauged
${\cal N}=8$ supergravities can be described by inserting
$\xi$-dependent terms where the T-tensor is given by \cite{hullplb1}
\bea
T_{i}^{\prime \;\;jkl}\left( \xi \right) & =
& T_{i}^{\;\;jkl}-\left( 1-\xi \right)
\left(\overline{u}^{kl}_{\;\;\;\;IJ} + \overline{v}^{klIJ} \right)
\nonumber \\ & & \times \left[ \frac{1}{2} P_{-}^{IJKL} \left(
u_{im}^{\;\;\;KM}\overline{u}_{\;\;\;LM}^{jm} -
v_{imKM}\overline{v}^{jmLM} \right) \right. \nonumber \\ & &
\left. + N_{IJ}^{\;\;\;\;KLMN} \left(
v_{imKL}\overline{u}^{jm}_{\;\;\;MN} -
u_{im}^{\;\;\;KL}\overline{v}^{jmMN} \right) \right].
\label{71t}
\eea

When $\xi =1$, one gets the de Wit-Nicolai model with $SU(8) \times SO(8)$
gauge symmetry. When $\xi=0$, one has $SU(8) \times ISO(7)^{+}$ gauge symmetry.
Moreover, when $\xi =-1$, a different
 theory with $SU(8) \times SO(7,1)^{+}$ gauge
symmetry was obtained.
All the $\xi < 0$ theories are equivalent to
$\xi =-1$ model related to the $SL(8, {\bf R})$ transformation and
all the $\xi > 0$ theories are
equivalent to
$\xi =1$ de Wit-Nicolai theory related to the $SL(8, {\bf R})$ transformation.
Moreover, the $\xi=0$ theory was obtained by limiting either
$\xi > 0$ or $\xi < 0$ models, under which $SO(8)$ or $SO(7,1)^{+}$
is transformed from an Inonu-Wigner contraction to $ISO(7)^{+}$.
As we have done before, we can describe 28-beins in terms of $s$.
It turns out
that the $A_{1}^{\prime}$ tensor has a single eigenvalue $z_1$
with a multiplicity of 8 which will provide a superpotential
of a scalar potential and has the following form generalizing
(\ref{a1cso71})
\bea
A_{1}^{\prime\;\;ij}=\mbox{diag}\left(z_{1},z_{1},z_{1},z_{1},
z_{1},z_{1},z_{1},z_{1}\right), \qquad
z_{1}=\frac{1}{8}\left(7e^{s} + \xi e^{-7s}\right).
\label{71z1}
\eea
Additionally, we can construct an $A_{2}^{\prime}$
tensor generalizing (\ref{a2cso71})
which is the combination of the triple product of
28 beins, as given in
\bea
A_{2i}^{\prime\;\;\;jkl}=\frac{1}{4}\left(e^{s}-\xi
e^{-7s}\right)X^{+ijkl}.
\label{71A2}
\eea
Therefore,
the scalar potential generalizing (\ref{potentialcso})
in the $SO(7)^{+}$-invariant direction
by summing all the components of $A_1^{\prime}$ and $A_2^{\prime}$
tensors and counting
the degeneracies correctly is given by \cite{hullplb1,hullphysica}
\bea
V_{7,1,\xi}  =  -g^{2}\left(
\frac{3}{4}|A_{1}^{\prime\;ij}|^{2}-\frac{1}{24}|A_{2i}^{\prime\;\;\;jkl}|^2
\right)  =  \frac{1}{8} g^2 \left( -35e^{2s}- 14\xi
e^{-6s} + \xi^{2}e^{-14s}\right).
\nonu
\eea
This can be written as a superpotential:$
V_{7,1,\xi} = g^2 \left[ \frac{2}{7} (\pa_s z_1)^2 -
6 z_1^2 \right]= g^2 \left[ 4 (\pa_{\widetilde{s}} z_1)^2 -
6 z_1^2 \right] $
where $\widetilde{s} = \sqrt{14} s$.
It is easily determined that there are no $SO(7)^{+}$-invariant critical points.
The theory \cite{hullplb1}
has ${\cal N}=8$ local supersymmetry and $H \times K_{\xi=-1,p=7,q=1}=
SU(8) \times SO(7,1)^{+}$ local gauge
symmetry. $SO(7,1)^{+}$ gauge symmetry is broken down to
its compact subgroup.

\subsection{Superpotential in
Other $CSO(p,q)^{+}$ and $SO(p,q)^{+}$ Gaugings \cite{hullplb2,hullcqg} }

Starting from the $SO(8)$ gauging, the $ISO(7)^{+}$ and $SO(7,1)^{+}$
gaugings were obtained by exploiting the transformations generated
by the $SO(7)^{+}$-invariant fourth rank antisymmetric tensor.
Now if one uses the $SO(p)^{+} \times SO(8-p)^{+}$-invariant fourth rank tensor
to generate transformations, one expects an $SO(p,8-p)^{+}$ gauging and
a gauging of a certain contraction of $SO(p, 8-p)^{+}$ about its compact
subgroup $SO(p)^{+}$ \cite{hullplb2,hullcqg}.
Let us consider the $SO(p)^{+} \times SO(q)^{+}$ invariant generator
of $SL(8, {\bf R})$,
\bea
X_{ab} & = & \left(
\begin{array}{ccc} \alpha \textbf{1}_{p \times p} &
0 \\ 0 & \beta \textbf{1}_{q \times q}
\nonu
\end{array} \right)
\eea
with
\bea
\alpha p + \beta q = 0, \qquad p + q = 8
\nonu
\eea
where
${\bf 1}_{p \times p}$ is the $p \times p$ identity matrix.
The embedding of this $SL(8, {\bf R})$ in $E_7$ is such that
$X_{ab}$ corresponds to the $56 \times 56$ $E_7$ generator
which is a non-compact $SO(p)^{+} \times SO(q)^{+}$
invariant element of the $SL(8,{\bf R})$ subalgebra of $E_7$
\bea
\left(
\begin{array}{cc}
0 &  \  X^{+IJKL}  \\
  X_{IJKL}^{+} & 0
\end{array} \right),
\nonu
\eea
where the real, self-dual totally anti-symmetric $SO(p)^{+} \times SO(q)^{+}$
invariant four-form tensor
$X_{IJKL}^{+}$ can be written in terms of a symmetric,
trace-free, $8 \times 8$ matrix with $SO(8)$ right-handed spinor indices,
$X_{ab}$ using $SO(8)$ $\Ga$ matrices(See appendix B)
\bea
X_{IJKL}^{+} =-\frac{1}{8}\left(\Gamma_{IJKL}\right)^{ab}X_{ab}
\label{xgamma}
\eea
where $ \Gamma_{IJKL} = \Gamma_{[ I} \Gamma_{J} \Gamma_{K} \Gamma_{L ]}$ and
an arbitrary $SO(8)$ generator $L_{IJ}$ acts in the right-handed
spinor representation by $(L_{IJ} \Gamma_{IJ})^{ab}$.
When $p=7$ and $q=1$, this expression of (\ref{xgamma}) through
$\Gamma $ matrix coincides exactly with
the one in (\ref{xijkl}).
We also present (\ref{xgamma}) explicitly in Appendix A for various $p$
and $q$.

Regarded as a $28 \times 28$ matrix, $X^{+IJKL}$ has eigenvalues $\al, \be$
and $\ga=(\al+\be)/2 $ with degeneracies $d_{\al}, d_{\be}$ and $d_{\ga}$
respectively. Let it be recalled  that $SO(7)^{+}$-invariant four-form tensor
has eigenvalues of $-1$ and $+3$.
The eigenvalues and eigenspaces of the $SO(p)^{+} \times SO(q)^{+}$
invariant tensor are summarized in Table 1, including the
case of $(p, q)=(7,1)$.
By introducing projectors as done in previous cases,
$P_{\al}, P_{\be}$ and $P_{\ga}$
onto corresponding eigenspaces, we have a $28 \times 28$ matrix equation
that generalizes (\ref{71x}) to arbitrary $p$ and $q$
\bea
X^{+IJKL} = \alpha P_{\al}^{IJKL} +\be P_{\be}^{IJKL} + \ga P_{\ga}^{IJKL}.
\nonu
\eea
Projector
$P_{\al}(P_{\be})$ projects the $SO(8)$ Lie algebra onto its $SO(p)^{+}
(SO(q)^{+})$
subalgebra  while $P_{\ga}$ does onto the remainder
$SO(8)/(SO(p)^{+} \times SO(q)^{+})$.

\bea
\begin{array}{|c|c|c|c|c||c|c|c||c|}
\hline
p & q & \al & \be & \ga=(\al+\be)/2 & d_{\al}=p(p-1)/2 &
d_{\be}=q(q-1)/2 & d_{\ga}=p q & |X^{+}|^2 \nonu \\
\hline
 7 & 1 & -1 & 7 & 3 & 21 & 0 & 7 & 84 \nonu \\
\hline
6 & 2 & -1 & 3 & 1 & 15 & 1 & 12 & 36 \nonu \\
\hline
5 & 3 & -1 & 5/3 & 1/3 & 10 & 3 & 15 & 20 \nonu \\
\hline
4 & 4 & -1 & 1 & 0 & 6 & 6 & 16 & 12 \nonu \\
\hline
3 & 5 & -1 & 3/5 & -1/5 & 3 & 10 & 15 & 36/5 \nonu \\
\hline
2 & 6 & -1 & 1/3 & -1/3 & 1 & 15 &  12 & 4 \nonu \\
\hline
1 & 7 & -1 & 1/7 & -3/7 & 0 & 21 & 7 & 12/7 \nonu \\
\hline
\end{array}
\nonu
\eea
Table 1. \sl Eigenvalues and eigenspaces of the $SO(p)^{+}
\times SO(q)^{+}$
invariant tensor, $X^{+}$ where $|X^{+}|^2=d_{\al}|\al|^2+
d_{\be}|\be|^2 + d_{\ga}|\ga|^2$. We have taken this table from
\cite{hullplb2}. In \cite{freetal}, they displayed the signature
of the Killing-Cartan form by writing the numbers $n_{+}, n_{-}$
and $n_0$ of its positive, negative and zero eigenvalues. Here we
identify $d_{\al} + d_{\be}$ with $n_{+}$ and $d_{\ga}$
with $n_{-}$. \rm

Then the  $\xi$-dependent T-tensor \cite{hullplb2,hullcqg}
has a  much more complicated expression
that generalizes (\ref{71t})
\bea
T_{i}^{\prime \;\;jkl}\left( \xi \right) & =
& T_{i}^{\;\;jkl}-\left( 1-\xi \right)
\left(\overline{u}^{kl}_{\;\;\;\;IJ} + \overline{v}^{klIJ} \right)
\nonumber \\ & & \times \left[ \left( P_{\be}^{IJKL}
+ \frac{1}{2}  P_{\ga}^{IJKL} \right) \left(
u_{im}^{\;\;\;KM}\overline{u}_{\;\;\;LM}^{jm} -
v_{imKM}\overline{v}^{jmLM} \right) \right. \nonumber \\ & &
\left. + P_{\ga}^{IJRS} Z_{RS}^{KLMN} \left(
-v_{imKL}\overline{u}^{jm}_{\;\;\;MN} +
u_{im}^{\;\;\;KL}\overline{v}^{jmMN} \right) \right]
\label{tprime}
\eea
where we introduce the new quantity $Z_{IJKL}^{MN}$
\bea
Z_{IJKL}^{MN} = \frac{1}{2} \left[ \left(P_{\al} -P_{\be} \right)_{IJMP}
P_{\ga}^{NPKL} -P_{\ga}^{IJMP} \left(P_{\al} -P_{\be} \right)_{NPKL} \right].
\label{Zijklmn}
\eea
The 28-beins $u_{ij}^{\;\;\;KL}$ and $v_{ijKL}$
are given in Appendix E and the projectors
$P_{\si}^{IJKL} (\si=\al, \be, \ga)$ are given in Appendix F.
This $T'$ tensor \cite{hullplb2} defines new $A_1^{\prime}$ and
$A_2^{\prime}$ tensors. These models
 will have ${\cal N}=8$ local supersymmetry and local
$SU(8) \times K_{\xi, p, q}$ invariance.
The gauge groups are
\bea
SO(7,1)^{+}, \;\;\;SO(6,2)^{+}, \;\;\; SO(5,3)^{+} \;\;\; \mbox{and}
\;\;\; SO(4,4)^{+},
\nonu
\eea
when $\xi=-1(t=
i \pi/(\al-\be))$. When $\xi =0$($t=\infty$) there exist
the inhomogeneous groups
\bea
&& CSO(7,1)^{+}=ISO(7)^{+}, \;\;\; CSO(6,2)^{+},
\;\;\; CSO(5,3)^{+}, \;\;\;
CSO(4,4)^{+}, \nonu \\
&& CSO(3,5)^{+},\;\;\; CSO(2,6)^{+} \;\;\; \mbox{and} \;\;\; CSO(1,7)^{+}.
\nonu
\eea
Any other choice of $\xi >0(\xi <0)$ gives a model equivalent
to the $SO(8)(SO(p,q)^{+})$ gauging
by field-redefinition. The gauge symmetry $K_{\xi, p, q}$ is
broken down to its maximal
compact subgroup or some subgroup thereof. There are
three inequivalent distinct gaugings.
From the expression (\ref{tprime}) one gets a single eigenvalue
$z_1$ with degeneracies
8 which has the following form
\bea
A_{1}^{\prime\;\;ij}=\mbox{diag}\left(z_{1},z_{1},z_{1},z_{1},
z_{1},z_{1},z_{1},z_{1}\right), \qquad
z_{1}= \frac{1}{8} \left( p e^s + q \xi e^{-\frac{p}{q} s} \right)
\label{pqz1}
\eea
which include all the cases $p, q$ and $\xi$ and generalize (\ref{71z1}).
Similarly, one can construct $A_{2}^{\prime}$  generalizing
(\ref{71A2})
\bea
A_{2i}^{\prime\;\;\;jkl}=\frac{q}{4} \left( e^s-\xi
e^{-\frac{p}{q} s} \right) X^{+ijkl}.
\label{a2tensor}
\eea
Finally the $K_{\xi, p, q }$-invariant scalar potential as a
function of $p, q, \xi$ and $s$ by counting the degeneracies correctly
can be written as\footnote{
It is known \cite{hullwarner1} that for finite real $t$, the T-tensor
can be obtained from the old one, de Wit-Nicolai T-tensor
by replacing ${\cal V}$ with ${\cal V} E(t)^{-1}$ and scaling by a factor
of $e^{\al t}$:$T_{i}^{\prime \;\;jkl}({\cal V}) = e^{\al t}
T_{i}^{ \;\;jkl}({\cal V} E(t)^{-1})$. This can be used to give a simple
calculation of the potential in the $SO(p)^{+} \times SO(q)^{+}$ invariant
direction in the space of scalar field.  }
\bea
V  &=&  -g^{2}\left(
\frac{3}{4}|A_{1}^{\prime\;ij}|^{2}-\frac{1}{24}|A_{2i}^{\prime\;\;\;jkl}|^2
\right)  \nonu \\
&= &   -g^{2}\left(
\frac{3}{4}\times 8 \times
\left( \frac{1}{8} \left( p e^s + q \xi e^{-\frac{p}{q} s} \right) \right)^2
-\frac{1}{24} \times \left(\frac{q}{4} \left( e^s-\xi
e^{-\frac{p}{q} s} \right)  |X^{+ijkl}| \right)^2
\right)
\nonu
\eea
with
\bea
|X^{+}|^2= \frac{1}{2} p (p-1)|\al|^2+
\frac{1}{2} q(q-1)|\be|^2 +  p q|\ga|^2.
\nonu
\eea
The potentials $V_{p,q, \xi}$ for the $K_{\xi, p, q}$
gauging are given by \cite{hullcqg}
\bea
V_{7, 1, \xi} & = &  \frac{1}{8} g^2 \left( -35e^{2s}- 14\xi
e^{-6s} + \xi^{2}e^{-14s}\right),
\nonu \\
V_{6, 2, \xi} & = & -3 g^2 \left( e^{2s}+ \xi e^{-2s}\right),
\nonu \\
V_{5, 3, \xi} & = &  -\frac{3}{8} g^2 \left( 5e^{2s}+ 10\xi
e^{-2s/3} + \xi^{2}e^{-10s/3}\right),
\nonu \\
V_{4, 4, \xi} & = &  - g^2 \left( e^{2s}+4\xi
 + \xi^{2} e^{-2s}\right),
\nonu \\
V_{3, 5, \xi} & = &  -\frac{3}{8} g^2 \left( e^{2s}+ 10\xi
e^{2s/5} + 5\xi^{2} e^{-6s/5}\right),
\nonu \\
V_{2, 6, \xi} & = &  -3 g^2 \xi \left( e^{2s/3}+\xi
e^{-2s/3} \right),
\nonu \\
V_{1, 7, \xi} & = &  \frac{1}{8} g^2 \left( e^{2s}- 14\xi
e^{6s/7} -35 \xi^{2}e^{-2s/7}\right).
\label{potential}
\eea
Of course, the potential $V_{7, 1, \xi}$
is identical to the one in previous sections 2.1 and 2.2
and is obtained by putting $p=7$ and $q=1$
into the general expression of a scalar potential.
Note that for $\xi=-1$, the potentials for the $SO(p,q)^{+}$
gauging and the $SO(q,p)^{+}$ gauging coincide with each
other due to the fact that
the potential $V_{p,q, \xi}$ can be obtained from $V_{q,p, \xi}$ by rescaling
$s \rightarrow -p s /q$. However, this is not true for $\xi=0$ because
$V_{p, q, \xi=0} \neq  V_{q,p, \xi=0}$.

From the above effective non-trivial scalar potential one expects that the
superpotential $W$ maybe encoded in either the $A_1^{\prime}$ or
$A_2^{\prime}$ tensors.
It turns out that the eigenvalue of the $A_1^{\prime}$ tensor $z_1$ provides a
superpotential
and one can check that the scalar potential can be written in terms of
a superpotential as follows, observed newly in this paper
\bea
W_{p,q}(\xi; s)  & = & z_{1}=  \frac{1}{8} \left( p e^s + q \xi
e^{-\frac{p}{q} s}
\right)=  \frac{1}{8} \left( p e^{\sqrt{\frac{q}{2p}} \widetilde{s}} + q \xi
e^{-\sqrt{\frac{p}{2q}} \widetilde{s}}
\right) , \nonu \\
V_{p,q}(\xi; s) & = & g^2 \left[  \frac{2q}{p}
\left(\partial_{s} W_{p,q}(\xi;\widetilde{s}) \right)^2 - 6
W_{p,q}(\xi;\widetilde{s})^2 \right] \nonu \\
& = &
 g^2 \left[  4
\left(\partial_{\widetilde{s}} W_{p,q}(\xi;\widetilde{s})
\right)^2 - 6  W_{p,q}(\xi;\widetilde{s})^2 \right]
\label{superpotentialandpotential}
\eea
where $\widetilde{s}=\sqrt{\frac{2p}{q}} s$.
The scalar potential has critical points at 1) critical points of
superpotential and at 2) points for which
superpotential satisfies some differential equation.
By differentiating $W$ with respect to field $s$, one
finds that there are no critical points of superpotential
corresponding to supersymmetric critical points
except the trivial critical point which has ${\cal N}=8$ supersymmetry
 and whose cosmological constant
$\La =-6 g^2$ for which $W=1$. The other critical points of scalar
potential yield nonsupersymmetric
vacua that may or may not be stable. The superpotential has the
following values at the various
critical points.

\bea
\begin{array}{|c|c|c|c|c|c|c|c|}
\hline $\mbox{Gauge symmetry}$ & \cal N & p & q=8-p & \xi & s   &
W & V \nonu \\
\hline
   SO(8) & 8 &\mbox{any}  & \mbox{any} & 1
& 0
  & 1
 & -6 g^2 \nonu \\
\hline
   SO(7)^{+} \times SO(1)^{+} & 0 &7 & 1 & 1  &
 -\frac{1}{8} \ln 5
 & \frac{3}{2} \times 5^{-1/8} & - 2\times5^{3/4}g^2 \nonu \\
\hline
SO(5)^{+} \times SO(3)^{+} & 0   &5 &3 & -1
&  -\frac{3}{8} \ln 3   & -\frac{1}{2} \times 3^{-3/8}  &
2 \times 3^{1/4}g^2 \nonu \\
\hline
   SO(4)^{+} \times SO(4)^{+} & 0 &4 &4 & -1
&  0
& 0  &  2g^2 \nonu \\ \hline
SO(3)^{+} \times SO(5)^{+} & 0   &3 &5 & -1
&  \frac{5}{8} \ln 3   & \frac{1}{2} \times 3^{-3/8}  &
2 \times 3^{1/4}g^2 \nonu \\ \hline
 SO(2)^{+} \times U(1)^{+15} & 0 & 2 & 6 & 0
&  \mbox{any}   & e^s/4 & 0  \nonu
\\
\hline
   SO(1)^{+} \times SO(7)^{+} & 0 &1 & 7 & 1  &
 \frac{7}{8} \ln 5
 & \frac{3}{2} \times 5^{-1/8} & - 2\times5^{3/4}g^2 \nonu \\
\hline
\end{array}
\nonu
\eea
Table 2. \sl Summary of various critical points \cite{hullcqg}
in the context of
superpotential observed in this paper first : Gauge symmetry, supersymmetry,
vacuum expectation value of field, superpotential and
cosmological constants. For $SO(3)^{+} \times SO(5)^{+}$ case,
one can check it by the change of variable of $SO(5)^{+} \times SO(3)^{+}$ case,
$s \rightarrow - 3s/5$ that corresponding potential of
$SO(3)^{+} \times SO(5)^{+}$ is obtained while by change of variable,
$s \rightarrow - s/7$, the potential of $SO(1)^{+} \times SO(7)^{+}$
can be found from $SO(7)^{+} \times SO(1)^{+}$ case.
Although the corresponding superpotential of these two cases
may be different from the original ones, the scalar potentials are the same. \rm

$\bullet$ $SO(8)$ case: ${\cal N}=8$

By differentiating the scalar potential with respect to real scalar field
$s$, there exists a solution of $s =0$ when $\xi =1$ for all possible values of
$p$ and $q$. This is nothing but de Wit-Nicolai's $SO(8)$-invariant
critical point and vacuum, which is fully supersymmetric(because
in this case, $\pa_s W|_{s=0}=0$ implying that $V= -6g^2W^2$.
In other words, $|W|= \sqrt{-V/6g^2}$.
All the eight eigenvalues of the $A_1^{\prime}$ tensor give rise to
the number of supersymmetries)
and hence are stable.
All the scalar potential $V_{p,q,\xi}$ becomes $-6 g^2$ when $s=0$ for
$\xi=1$.

$\bullet$ $SO(7)^{+} \times SO(1)^{+}$ case: ${\cal N}=0$

This is exactly the $SO(7)^{+}$-invariant critical point of the
$SO(8)$ theory. As in Table 2, it has no supersymmetry and is unstable.

$\bullet$ $SO(5)^{+} \times SO(3)^{+}$ case: ${\cal N}=0$

In this case,   the value of the scalar potential gives
a positive cosmological constant  where
the eigenvalue of the $A_1^{\prime}$ tensor is $-\frac{1}{2} \times 3^{-3/8}$
and the $A_2^{\prime}$ tensor has a value of $2\times 3^{5/8} X^{+ijkl}$.
It is known to be unstable.

$\bullet$ $SO(4)^{+} \times SO(4)^{+}$ case: ${\cal N}=0$

At this critical point, the value of the scalar potential gives
a positive cosmological constant  where
the $A_1^{\prime}$ tensor vanishes and the $A_2^{\prime}$
tensor has a value of $4 X^{+ijkl}$.
It is known to be unstable.
The positivity of the cosmological constant from the analysis of
11-dimensional field equations for $SO(5,3)^{+}$ and $SO(4,4)^{+}$
theories was confirmed in \cite{hullwarner3}.

$\bullet$ $SO(2)^{+} \times U(1)^{+15}$ case: ${\cal N}=0$

When $\xi=0$, the potential vanishes implying that for any value
of $s$, there exists a zero cosmological constant critical point.
In addition, the potential is also flat in the
$SO(2)^{+} \times SO(6)^{+}$-invariant direction.
Nonetheless, global $SO(6)^{+}$ symmetry
remains unbroken by the vacuum. In this case,
the eigenvalue of the $A_1^{\prime}$ tensor is equal to $e^s/4$ and
the $A_2^{\prime}$ tensor
is $3 e^s X^{+ijkl}$.

\subsection{Domain Wall in
$CSO(p,q)^{+}$ and $SO(p,q)^{+}$ Gaugings \cite{hullplb2,hullcqg} }

Let us begin with the resulting Lagrangian of the
scalar-gravity sector by explicitly determining the scalar kinetic terms
appearing in the action in terms of $s $.
The scalar kinetic term is $- \frac{1}{96} \left| A_{\mu}^{\;\;ijkl}
\right|^2$ where the generalized $g$-dependent $A_{\mu}^{\;\; ijkl}$
can be obtained
\bea
A_{\mu}^{\;\; ijkl} & = & -2 \sqrt{2} \left( \overline{u}^{ij}_{\;\;IJ}
\partial_{\mu} \overline{v}^{klIJ} -
\overline{v}^{ijIJ} \partial_{\mu} \overline{u}^{kl}_{\;\;IJ} \right) \nonu \\
&& + 4\sqrt{2} (1-\xi) g A_{\mu IJ} \left[ \left( P_{\be}^{IJKL}
+ \frac{1}{2}  P_{\ga}^{IJKL} \right) \left( -\overline{u}^{ij}_{\;\;KM}
\overline{v}^{klLM} + \overline{v}^{ijKM} \overline{u}^{kl}_{\;\;LM}
\right) \right. \nonu \\
&& \left. +  P_{\ga}^{IJRS} Z_{RS}^{KLMN} \left(\overline{u}^{ij}_{\;\;KL}
\overline{u}^{kl}_{\;\;MN} - \overline{v}^{ijKL} \overline{v}^{klMN}
\right) \right].
\label{kineticpq}
\eea
By taking the product of $  A^{\;\;IJKL}_{\mu}$  and
its complex conjugation and taking into account the
multiplicity with vanishing $ A_{\mu IJ} $,
we arrive at the following expression for $(p, 8-p)$ where
$p= 7, 6, 5, 4, 3, 2, 1$
\bea
-\frac{1}{96} \left| A^{\;\;IJKL}_{\mu} \right|^2
 =  -\left(7 ,3 ,5/3 ,1 ,3/5 ,1/3 , 1/7\right) \pa^{\mu} s
\partial_{\mu} s.
\nonu
\eea
Let us define a new variable $\widetilde{s}$, in order to
have usual canonical kinetic terms, normalized by $1/2$, as
\bea
\widetilde{s} = \sqrt{\frac{2p}{q}} s.
\nonu
\eea
Therefore, the resulting Lagrangian of scalar-gravity sector takes the form:
\bea
\int d^4 x \sqrt{-g} \left( \frac{1}{2} R
- \frac{1}{2} \partial^{\mu} \widetilde{s}  \partial_{\mu} \widetilde{s}    -
V_{p, q}(\xi; \widetilde{s}) \right),
\label{action1}
\eea
together with (\ref{potential}) where $s$ replaced by $\widetilde{s}$.
Having established the holographic duals of both supergravity critical points,
and examined small perturbations around the corresponding fixed point
field theories, one can proceed the supergravity description.
The supergravity scalar whose vacuum expectation value leads to the new
critical point tells us what relevant operators in the dual field theory
would drive a flow to the fixed point in the IR.
To construct the kink corresponding  to
the supergravity description of the nonconformal
(in special case: RG) flow from one scale to two
other connecting critical points in $d=3$ field theories,
the form of a 3d Poincare invariant metric breaking
the full conformal group invariance takes the form \cite{lps}:
\bea
ds^2= e^{2A(r)} \eta_{\mu \nu} dx^{\mu} dx^{\nu} + e^{2B(r)} dr^2, \;\;\;
\eta_{\mu \nu}=(-,+,+),
\label{ansatz}
\eea
characteristic of space-time with a domain wall where $r$ is the
coordinate transverse to the wall(can be interpreted
 as an energy scale) and $A(r)$ is the scale factor
in the four-dimensional metric.

Our interest in domain wall space-times arises from
their relevance to dual field theories.
The distance from horizon $U=\infty$ corresponds
to long
distance in the bulk(UV in the dual field theory) and $U=0$(near
horizon
corresponds to short distances in the bulk(IR in the dual field theory).
We are looking for ``interpolating'' solutions.
We will show how supergravity can provide a description of the entire
flow from the maximal supersymmetric UV theory to the IR fixed point.
With the above  ansatz (\ref{ansatz})
the equations of motion for the scalars and the metric from (\ref{action1})
read
\bea
& &
 \pa_r^2 A -\pa_r A \pa_r B+ \frac{3}{2} (\pa_r A)^2 +
\frac{1}{4} (\pa_r \widetilde{s})^2   +
   \frac{1}{2} e^{2B} V_{p, q, \xi} = 0, \nonu \\
& & \pa_r^2 \widetilde{s} +
3 \pa_r A \pa_r \widetilde{s}  -
 \pa_r B \pa_r \widetilde{s} - e^{2B} \pa_{\widetilde{s}} V_{p, q, \xi}
=0.
\label{eom}
\eea

By substituting the domain-wall ansatz (\ref{ansatz})
into the Lagrangian (\ref{action1}),
the energy-density  $E[A, \widetilde{s}]$
\cite{st}, with the integration by parts
on the term of $\pa_r^2 A $, per
unit area transverse to $r$-direction is given by
\bea
E[A, \widetilde{s}]&  = &- \int_{-\infty}^{\infty} dr e^{3A+B}
\left[- 3 e^{-2B} \left( 2 (\pa_r A )^2 + \pa_r^2 A -\pa_r A \pa_r B \right) -
\frac{1}{2} e^{-2B} \left( \partial_r \widetilde{s} \right)^2 -
V_{p, q, \xi}(\widetilde{s})\right].
\nonu
\eea
We are looking for a nontrivial configuration along $r$-direction in order
to find out the first-order differential equations satisfying the domain-wall,
let us rewrite and reorganize
the energy-density by complete squares plus others due to
usual squaring-procedure as follows:
\bea
&& E[A, \widetilde{s}] =  \nonu \\
& &    \frac{1}{2} \int_{-\infty}^{\infty} dr e^{3A+B}
\left[ - 6 \left( e^{-B} \pa_r A + \sqrt{2} g W_{p,q}(\xi;\widetilde{s})
\right)^2 + \left( e^{-B} \pa_r \widetilde{s} -2 \sqrt{2} g
\pa_{\widetilde{s}} W_{p,q}(\xi;\widetilde{s}) \right)^2  \right. \nonu \\
&& \left. 12 \sqrt{2} g e^{-B} W_{p,q}(\xi;\widetilde{s}) \pa_r A +
4 \sqrt{2}  g
e^{-B}  \pa_r W_{p,q}(\xi;\widetilde{s}) \right]
\nonu
\eea
where superpotential $W_{p,q}(\xi;\widetilde{s})$
is given by (\ref{superpotentialandpotential}).
Then one can easily check the last two terms in the above can be
combined as
$4 \sqrt{2} g \pa_r ( e^{3A}  W_{p,q}(\xi;\widetilde{s})) $.
Therefore, one arrives at
\bea
& &  \frac{1}{2}
\int_{-\infty}^{\infty} dr e^{3A+B}
\left[ - 6 \left( e^{-B} \pa_r A + \sqrt{2} g W_{p,q}(\xi;\widetilde{s})
\right)^2 +
 \left( e^{-B} \pa_r \widetilde{s} -2\sqrt{2} g
\pa_{\widetilde{s}} W_{p,q}(\xi;\widetilde{s}) \right)^2 \right] \nonu \\
&  & + 2\sqrt{2}g \left(
e^{3A} W_{p,q}(\xi;\widetilde{s}) \right) |_{- \infty}^
{\infty}.
\nonu
\eea
Finally, we find BPS bound, inequality of the energy-density
\bea
E[A, \widetilde{s}]
 \geq  2\sqrt{2}g \left( e^{3A(\infty)} W_{p,q}(\xi;\widetilde{s})(\infty) -
e^{3A(-\infty)} W_{p,q}(\xi;\widetilde{s})(-\infty)\right).
\label{Ebound}
\eea

Then $E[A, \widetilde{s}]$ is extremized by the following
so-called BPS domain-wall solutions. The first order differential
equations for the scalar field are the gradient flow equations
of a superpotential defined on a restricted slice of
the scalar manifold and simply related to the potential of gauged supergravity
on this slice. The equations describing the flow are then
\bea
\partial_{r} \widetilde{s} & = &
\pm 2 \sqrt{2} e^B g \partial_{\widetilde{s}}
W_{p,q}(\xi;\widetilde{s}) ,\nonu
\\
\partial_{r} A & = & \mp \sqrt{2} e^{B} g W_{p,q}(\xi;\widetilde{s}).
\label{first}
\eea
There exists a supersymmetry \cite{st,hullrecent}
of the background with a nonvanishing metric and
a single scalar field, for each spinor satisfying the Killing
spinor condition. The background satisfying (\ref{first})
preserve half the supersymmetry.
It is straightforward to verify that any solutions
$\widetilde{s}(r), A(r)$ of (\ref{first})
satisfy the gravitational and
scalar equations of motion given by the
second order differential equations (\ref{eom}).
Embedding or consistent truncation means that
the flow is entirely determined by the
equations of motion of supergravity in four-dimensions and any solution
of the truncated theory can be lifted to a solution of untruncated
theory \cite{dn86}.
Using (\ref{first}), the monotonicity \cite{domainwall} of
$\pa_r A $  which is related to the local potential energy
of the kink leads to
$\pa_r^2 A \leq 0$ when $B$ is constant.
Note that the value of superpotential at either end of a kink may be thought
of as determining the topological sector.
The analytic solutions of (\ref{first})
for $(p, q)=(4,4)$ when $B$ is a constant become
\bea
\widetilde{s}(r) = \sqrt{2} \log \left[
\sqrt{\xi} \frac{ (e^{\sqrt{2\xi} g  (
c-r) }-1)}{(e^{\sqrt{2\xi} g  (
c-r) }+1)} \right], \qquad
A(r) = \left(1+\sqrt{2\xi} g \right) c + \log
\left[2\sinh\sqrt{2\xi} g(r-c) \right]
\nonu
\eea
where $c$ is some constant.
For other values of $(p,q)$, the analytic solutions exist only for $\xi=0$.

\subsection{$SO(7)^{-}$ Invariant Sector from
$SO(8)$ Gauging  }

The four-form tensor\footnote{The $SL(8, {\bf R})$ does act
on the vector potential and is generated by the $SO(8)$ and self-dual
part. The remainder of $E_7$ including the anti-self-dual part does not
act on the vector potentials but does act on the field strengths. Therefore,
contrary to the self-dual case we have discussed in previous sections,
the anti-self-dual case does not act on the vector potential. We thank
C.M. Hull for pointing out this to us.} $X^{-IJKL}$
is invariant under the $SO(7)^{-}$
subgroup of $SO(8)$. Turning on the vacuum expectation value
proportional to $X^{-IJKL}$ in the de Wit-Nicolai theory
gives rise to spontaneous symmetry breaking $SO(8)$ into
$SO(7)^{-}$.
Regarded  as a $28 \times 28$ matrix, $X^{-IJKL}$ has
21 eigenvalues of $1$ and 7 eigenvalues of $-3$. Introducing
the projector $P_{1}$ onto the 21-dimensional
eigenspace($P_{1}$ projects the generators of $SO(8)$ onto those of
$SO(7)^{-}$ while $P_{2}$ projects the generators of $SO(8)$ onto the
remainder $SO(8) \backslash SO(7)^{-}$), they are given in terms of
$X^{-IJKL}$
\bea
P_{1}^{IJKL} & = &
\frac{3}{4} \left(  \delta^{IJ}_{KL} + \frac{1}{3} X^{-IJKL} \right), \nonu \\
P_{2}^{IJKL} & = &  \delta^{IJ}_{KL} -P_{1}^{IJKL} =
\frac{1}{4} \left(  \delta_{IJ}^{KL} -   X^{-IJKL} \right).
\nonu
\eea
One can easily check that they satisfy
\bea
P_{1}^2 = P_{1}, \qquad P_{2}^2 =P_{2}, \qquad P_{1} P_{2} =P_{2}
P_{1} =0.
\nonu
\eea
The 28 $SO(8)$ generators $\La^{IJ}$ are projected onto
a 21-dimensional subspace by $P_{1}$, $\La_{1}^{IJ} = P_{1}^{IJKL}
\La^{KL}$ and this subspace is the Lie algebra for the
$SO(7)^{-}$ subgroup of $SO(8)$; in other words, the subgroup
stabilizes a left-handed  $SO(8)$ spinor(See the appendix B).
The remaining 7 generators
are $\La_{2}^{IJ} = P_{2}^{IJKL} \La^{KL}$. The usual commutation
relations for $SO(8)$ are given in terms of $\La_{1}^{IJ}$ and
$\La_{2}^{IJ}$.

Viewed as a $28 \times 28$ matrix, $X^{-IJKL}$ has eigenvalues $\al, \be$
and $\ga=(\al+\be)/2 $ with degeneracies $d_{\al}, d_{\be}$ and $d_{\ga}$
respectively(For the explicit construction of $X^{-IJKL}$ see the Appendix A).
The eigenvalues and eigenspaces of the $SO(p)^{-} \times SO(q)^{-}$
invariant tensor are summarized similarly.
By introducing projectors as done in previous cases,
$P_{\al}, P_{\be}$ and $P_{\ga}$
onto corresponding eigenspaces, we have a $28 \times 28$ matrix equation
to arbitrary $p$ and $q$.
The parametrization for the $SO(p)^{-} \times SO(q)^{-}$-singlet space that is
invariant subspace under a particular $SO(p)^{-} \times SO(q)^{-}$ subgroup of
$SO(8)$ becomes
\bea
\phi_{IJKL}=4\sqrt{2} i s X_{IJKL}^{-}
\nonu
\eea
where $s$ is a real scalar field.
Note the presence of imaginary number $i$.
As in the previous consideration, the
$A_{1}^{\prime}$ tensor we obtained is
a single complex eigenvalue with degeneracies 8
\bea
A_{1}^{\prime\;\;ij} & = & \mbox{diag}\left(z_{1},z_{1},z_{1},z_{1},
z_{1},z_{1},z_{1},z_{1}\right), \nonu \\
z_{1} & = & \frac{1}{16} (1+i)
\left( p e^s + q  e^{-\frac{p}{q} s} \right)+
\frac{1}{16} (1-i)
\left( p e^{-s} + q  e^{\frac{p}{q} s} \right).
\label{z1-}
\eea
For the $A_{2}^{\prime}$ tensor we get
\bea
A_{2i}^{\prime\;\;\;jkl}=\frac{p}{8} \left[ \left(1+i \right) \left(
e^{-\frac{p}{q} s} -e^s\right) + \left(1-i \right)\left(
 e^{\frac{p}{q} s} -e^{-s}
 \right) \right] X^{-ijkl}.
\nonu
\eea
Therefore, we are now ready to calculate the full expression of
a scalar potential and it turns out
\bea
V_{7, 1} & = &  \frac{1}{16} g^2 e^{-14s} \left( 1+e^{4s} \right)^5
\left( 1- 5 e^{4s} + e^{8s}\right),
\nonu \\
V_{6, 2} & = & -3 g^2 e^{-2s} \left( 1+  e^{4s}\right),
\nonu \\
V_{5, 3} & = &  -\frac{3}{16} g^2 e^{-10s/3} \left( 1+ e^{4s/3}\right)^5,
\nonu \\
V_{4, 4} & = &  - g^2 \left( 4 + e^{-2s}
 + e^{2s}\right),
\nonu \\
V_{3, 5} & = &  -\frac{3}{16} g^2 e^{-2s} \left( 1+ e^{4s/5}\right)^5,
\nonu \\
V_{2, 6} & = &  -3 g^2 e^{-2s/3} \left( 1+ e^{4s/3} \right),
\nonu \\
V_{1, 7} & = &  \frac{1}{16} g^2 e^{-2s}
\left( 1 + e^{4s/7} \right)^5 \left(1- 5
e^{4s/7} + e^{8s/7}\right).
\nonu
\eea
Note that the potential $V_{p, q}$ can be obtained from
$V_{q, p}$ by rescaling $s \rightarrow p s/q$.
The eigenvalue of the $A_1^{\prime}$ tensor $z_1$ provides a
superpotential
and one can check that the scalar potential can be written in terms of
superpotential:
\bea
W_{p,q}(s)  & = & |z_{1}|  , \nonu \\
V_{p,q}(s) & = & g^2 \left[  \frac{2q}{p}
\left(\partial_{s} W_{p,q}(s) \right)^2 - 6  W_{p,q}(s)^2 \right]
 = g^2 \left[  4
\left(\partial_{\widetilde{s}} W_{p,q}(s) \right)^2 - 6  W_{p,q}(s)^2 \right]
\nonu
\eea
where $\widetilde{s}=\sqrt{\frac{2p}{q}} s$ and
$z_1$ is given by (\ref{z1-}).
The kinetic terms are equivalent to the previous cases.
In this case, there are
no such first order differential equations for either a flow
between $SO(8)$ fixed point and $SO(7)^{-} \times SO(1)^{-}$ fixed point or
a flow between $SO(8)$ and $SO(1)^{-} \times SO(7)^{-}$,
contrary to the previous
$SO(p)^{+} \times SO(q)^{+}$ embedding case. The superpotential
has the following values at the two critical points.

\bea
\begin{array}{|c|c|c|c|c|c|c|}
\hline $\mbox{Gauge symmetry}$ & \cal N & p & q=8-p &  s   &
W & V \nonu \\
\hline
   SO(8) & 8 &\mbox{any}  & \mbox{any} & 1
& 0
 & -6 g^2 \nonu \\
\hline
   SO(7)^{-} \times SO(1)^{-} & 0 &7 & 1  &
 \frac{1}{2} \ln \frac{1}{2} (\pm 1 + \sqrt{5})
 & \frac{3 \times 5^{3/4}}{8} & - \frac{25 \sqrt{5}}{8}g^2 \nonu \\
\hline
   SO(1)^{-} \times SO(7)^{-} & 0 &1 &7
&  \frac{7}{2} \ln \frac{1}{2} (\pm 1 + \sqrt{5})
&  \frac{3 \times 5^{3/4}}{8} &  - \frac{25 \sqrt{5}}{8}g^2  \nonu \\ \hline
\end{array}
\nonu
\eea
Table 3. \sl Summary of critical points in the context of
superpotential : symmetry group, supersymmetry,
vacuum expectation values of field, superpotential, and
cosmological constants. For either case, it is exactly $SO(7)^{-}$-invariant
critical point of the $SO(8)$ theory. It has no supersymmetry and
is unstable.    \rm

\section{Domain Wall from
$SO(p)^{+} \times SO(q)^{+} \times SO(r)^{+}$
Sectors  of ${\cal N}=8$ Supergravity }

Let us consider a sequence of non-singular elements $E(\xi)$
of $L=SL(8, {\bf R})$
with $\xi$ real parameter and $E(1)=1$, identity transformation,
whose limit point $E(0)$ is singular and not in $L$.
As long as $E(\xi)$ remains nonsingular$(\xi \neq 0)$, the structure constants
have the usual tensor properties.
Acting on the Lagrangian with $E(\xi)$
yields a sequence of Lagrangian:$ {\cal L} \rightarrow
{\cal L}^{\prime}(\xi)={\cal L}_0 +
{{\cal L}_g}^{\prime}(\xi)$. If one also rescales the coupling constant
$g$ by a $\xi$-dependent one through $g \rightarrow g^{\prime}(\xi)$ for some
choices of the sequence $E(\xi)$ in $L$, a limit of ${{\cal L}_g}^{
\prime}(\xi)$ exists
and is well defined.
One can continue the Lagrangian ${\cal L}^{\prime}(\xi)$ to negative
values of $\xi$.
In this case, ${\cal L}'(-1) $ is the Lagrangian
for different gauging and the gauge group is non-compact $SO(p,q+r)^{+}$ with
$p+q+r=8$  which will be discussed in section 3.2.
One continues to consider a sequence of non-singular elements $F(\zeta)$
of $L$  with $\zeta$ real parameter and $F(1)=1$, identity transformation,
whose limit point $F(0)$ is singular and not in $L$.
As long as $F(\zeta)$
remains nonsingular$(\zeta \neq 0)$, the structure constants
have the usual tensor properties.
Acting on the Lagrangian ${\cal L}'$ with $F(\zeta)$
yields a sequence of Lagrangian:$ {\cal L}' \rightarrow
{\cal L}'' (\zeta)={\cal L}_0 +
{{\cal L}_g}''(\zeta, \xi)$.
 If one also rescales the coupling constant
$g'$ by a $\zeta$-dependent one through $g' \rightarrow g''(\zeta)$ for some
choices of the sequence $F(\zeta)$ in $L$, the limit of
${{\cal L}_g}''(\zeta)$ exists as $\zeta \rightarrow 0$
so that $ {\cal L}''(\zeta =0) = {\cal L}_0 +
{{\cal L}_g}''(\zeta=0)$ gives the Lagrangian.
The gauge group corresponding to ${\cal L}''(\zeta=0, \xi=-1)$
is an Inonu-Wigner
contraction of $K_{\xi, \zeta, p, q, r}$ denoted by $CSO(p, q, r)^{+}$ with
$p+q+r=8$ \cite{hullcqg}.

In section 3.1, we start with the most general gaugings which
generalize previous considerations by introducing
two parameters, $\xi$ and $\zeta$.
The gauging denoted by $CSO(p,q, r)^{+}$ preserves a metric
with $p$ positive eigenvalues, $q$ negative eigenvalues and
$r$ zero eigenvalues.
In section 3.2, by analyzing two successive $SL(8, {\bf R})$
transformations(repeating twice) in the context of $SO(p, q+r)^{+}$
and $SO(p+q, r)^{+}$ gaugings, we discover a $T^{\prime}$ tensor
which depends on these two parameters, $\xi$ and $\zeta$.
As done in previous sections, the $A_1$ and $A_2$ tensors can be
easily determined by realizing that 56-beins
are product of each 56-bein for each parametrization of the singlet-space.
In section 3.3
it turns out that one has a scalar potential which can be written as
a superpotential in very simple form and
in section 3.4, we find the domain-wall solutions.
In section 3.5, by starting with $SO(p)^{+} \times SO(q)^{+}
\times SO(r)^{+}$ invariant generator of $SL(8, {\bf R})$ directly,
one can construct the projectors corresponding to this invariant four-form
tensor, which we will compare with the approach given in section 3.2-3.4.

\subsection{Non-semi-simple and Non-compact Gaugings \cite{hullcqg} }

It is possible to gauge the 28-dimensional subgroup $K_{\xi, \zeta, p,
q, r}$ of
$L=SL(8, {\bf R})$ whose algebra
\bea
[ \La_{ab}, \La_{cd} ]_{\xi, \zeta} & = &
\La_{ad} \eta_{bc} -\La_{ac} \eta_{bd} -
\La_{bd} \eta_{ac} +\La_{bc} \eta_{ad}, \nonu \\
\eta_{ab} & = &
\left( \begin{array}{ccc}
{\bf 1}_{p \times p} & 0 & 0  \\
0 & \xi {\bf 1}_{q \times q} & 0 \\
0 & 0 &   \xi \zeta {\bf 1}_{r \times r}
\end{array} \right), \qquad p + q + r =8
\nonu
\eea
where $a, b = 1, \cdots, 8$ and $\La_{ab}=-\La_{ba}$. \\
$\bullet$ When $(\xi, \zeta)=(1, 1)$,
this leads to the algebra of $SO(8)$ and the
de Wit-Nicolai gauging is recovered. When  $(\xi, \zeta)=(1, 0)$ it will
give $CSO(p+q, r)^{+}$ algebra which was discussed in the previous section and the
maximal compact subgroup is $SO(p+q)^{+} \times U(1)^{+r(r-1)/2}$.
Moreover, when  $(\xi, \zeta)=(1, -1)$, one gets $SO(p+q, r)^{+}$ algebra
which was already considered and the maximal compact subgroup is
$SO(p+q)^{+} \times SO(r)^{+}$. \\
$\bullet$ When  $(\xi, \zeta)=(-1, 1)$, it will give non-compact
$SO(p,q+r)^{+}$ gauging whose maximal compact subgroup is $SO(p)^{+}
\times SO(q+r)^{+}$.  When  $(\xi, \zeta)=(-1, 0)$,
it gives a certain non-semi-simple
algebra of the Inonu-Wigner contraction of $SO(8)$
 about its $SO(p, q)^{+}$ subgroup, denoted by $CSO(p, q, r)^{+}$
\cite{hullcqg}.
The maximal compact subgroup is
$SO(p)^{+} \times SO(q)^{+} \times U(1)^{+r(r-1)/2}$. Note that
$CSO(p, q, 1)^{+}= ISO(p, q)^{+}$, the inhomogeneous group.
For $(\xi, \zeta)=(-1, -1)$, one gets $SO(p+r, q)^{+}$ algebra. \\
$\bullet$ When $\xi=0$, it gives Inonu-Wigner contraction
$CSO(p,q+r)^{+}$ which was already considered.

The $CSO(p,q, r)^{+}$ gauging initiated by Hull which preserves a metric
with $p$ positive eigenvalues, $q$ negative eigenvalues and
$r$ zero eigenvalues
can be obtained by group contractions of
$SO(8)$ as follows. One decomposes each $SO(8)$
generator $\La$ into the part $\La_{(\al)}$ in the $SO(p)^{+}$ sub-algebra,
the part $\La_{(\be)}$ in the $SO(q)^{+}$ sub-algebra,
the part $\La_{(\ga)}$ in the $SO(r)^{+}$ sub-algebra,
and the remainders $\La_{(\de)}, \La_{(\la)},$ and $\La_{(\rho)}$
where $\La =\La_{(\al)}+\La_{(\be)}+ \La_{(\ga)} +
\La_{(\de)}+\La_{(\la)}+ \La_{(\rho)} $. See also the discussion around in
(\ref{relation}).
One performs the rescaling as
\bea
\La \rightarrow \La_{(\al)}+ \xi \left(\La_{(\be)}+  \zeta
\La_{(\ga)} + \sqrt{\zeta} \La_{(\rho)}\right) +
\sqrt{\xi} \left( \La_{(\de)} + \sqrt{\zeta} \La_{(\la)} \right).
\nonu
\eea
The rescaled algebra can be represented as
\bea
&&[\La_{(\al)}, \La_{(\al)}] \approx \La_{(\al)}, \qquad
[\La_{(\al)}, \La_{(\de)}] \approx \La_{(\de)}, \qquad
[\La_{(\al)}, \La_{(\la)}] \approx \La_{(\la)}, \nonu \\
&&[\La_{(\be)}, \La_{(\be)}] \approx \xi \La_{(\be)}, \qquad
[\La_{(\be)}, \La_{(\de)}] \approx \xi \La_{(\de)}, \qquad
[\La_{(\be)}, \La_{(\rho)}] \approx \xi \La_{(\rho)}, \nonu \\
&&[\La_{(\ga)}, \La_{(\ga)} ] \approx \xi \zeta \La_{(\ga)}, \qquad
[\La_{(\ga)}, \La_{(\rho)}] \approx \xi \zeta \La_{(\rho)}, \qquad
[\La_{(\ga)}, \La_{(\la)}] \approx \xi \zeta \La_{(\la)}, \nonu \\
&&[\La_{(\de)}, \La_{(\de)}] \approx \xi \La_{(\al)} + \La_{(\be)}, \qquad
[\La_{(\de)}, \La_{(\rho)]} \approx \xi \La_{(\la)}, \qquad
[\La_{(\de)}, \La_{(\la)}] \approx \La_{(\rho)}, \nonu \\
&&[\La_{(\rho)}, \La_{(\rho)}] \approx
 \xi \zeta \La_{(\be)} + \xi \La_{(\ga)}, \qquad
[\La_{(\rho)}, \La_{(\la)} ] \approx \xi \zeta \La_{(\de)}, \qquad
[\La_{(\la)}, \La_{(\la)}] \approx \xi \zeta \La_{(\al)} + \La_{(\ga)}, \nonu
\eea
with other commutators vanishing.
By taking the contraction, $\zeta \rightarrow 0$,
the $SO(r)^{+}$ subgroup generated by $\La_{(\ga)}$ collapses to an abelian
group $U(1)^{+r(r-1)/2}$ and the maximal compact subgroup of
$CSO(p, q, r)^{+}$ is $SO(p)^{+} \times SO(q)^{+} \times U(1)^{+r(r-1)/2}$.
The generators $\La_{(\ga)}$ commute  all the generators except appearing
on the right hand sides of $[ \La_{(\la)}, \La_{(\la)} ]$ and
$[ \La_{(\rho)}, \La_{(\rho)} ]$.
The methods described in previous section will
be used to obtain a $CSO(p, q, r)^{+}$ gaugings.

\subsection{ T-tensor in
$CSO(p, q, r)^{+}$ Gaugings \cite{hullcqg}}

The $CSO(p, q, r)^{+}$ gaugings can be obtained by acting on the
$SO(p, q+r)^{+}$ gauging first. Some idea in this direction
was already given in the paper of \cite{hullwarner3}.
Let us consider the $SO(p)^{+} \times SO(q+r)^{+}$ invariant generator
of $SL(8, {\bf R})$ we have discussed in previous section,
\bea
X_{ab} & = & \left(
\begin{array}{ccc} \alpha \textbf{1}_{p \times p} &
0 \\ 0 & \beta \textbf{1}_{(q+r) \times (q+r)}
\label{xabt}
\end{array} \right)
\eea
with
\bea
\alpha p + \beta (q+r) = 0, \qquad p + q + r = 8.
\label{alphat}
\eea
Regarded as a $28 \times 28$ matrix, real, self-dual totally
anti-symmetric  $SO(p)^{+} \times SO(q+r)^{+}$-invariant
four-form tensor $X^{+IJKL}_t$ has eigenvalues $\al, \be$
and $\ga=(\al+\be)/2 $ with degeneracies $d_{\al}, d_{\be}$ and $d_{\ga}$
respectively.
The eigenvalues and eigenspaces of the $SO(p)^{+} \times SO(q+r)^{+}$
invariant tensor are summarized in Table 1.
By introducing projectors, $P_{\al,t}, P_{\be,t}$ and $P_{\ga,t}$
onto corresponding eigenspaces, we have a $28 \times 28$ matrix equation.
Projector
$P_{\al,t}(P_{\be,t})$ projects the $SO(8)$ Lie algebra onto its $SO(p)^{+}
(SO(q+r)^{+})$
subalgebra  while $P_{\ga,t}$ does onto the remainder
$SO(8)/(SO(p)^{+} \times SO(q+r)^{+})$. Note that
$q$ over there is replaced by $q+r$ here. The projectors
can be constructed from
$X^{+IJKL}_t$.
The combination $g A_{\mu}^{IJ}$ in the minimal couplings will
be finite as $t \rightarrow \infty$ if $g$ is rescaled to
\bea
g(t) = g e^{\al t}
\nonu
\eea
for constant $\al$ which we have chosen as $-1$ so that
\bea
g(t) A_{\mu}^{IJ}(t) & = & g \left(   A_{\mu (\al)}^{IJ} +
e^{(\al-\be) t} A_{\mu (\be)}^{IJ} + e^{(\al-\ga) t} A_{\mu (\ga)}^{IJ}
\right)  \nonu \\
& = &  g \left( A_{\mu (\al)}^{IJ} + \xi  A_{\mu (\be)}^{IJ} +
\sqrt{\xi}  A_{\mu (\ga)}^{IJ} \right),
\label{gA}
\eea
where $\xi =e^{(\al -\be) t}$ as before.
One finds that on taking the limit $t \rightarrow \infty(\xi \rightarrow 0)$
one obtains a gauging with gauge group contraction of $SO(8)$ about its
$SO(p)^{+}$ subgroup. If, instead, one analytically continues to $t =i
\pi/(\al-\be)$, one obtains a gauging of $SO(p, q+r)^{+}$.

Let us consider the additional, second $SL(8, {\bf R})$ transformation
using the $SO(p+q)^{+} \times SO(r)^{+}$ invariant generator
of $SL(8, {\bf R})$,
\bea
X_{ab} & = & \left(
\begin{array}{ccc} \alpha' \textbf{1}_{(p+q) \times (p+q)} &
0 \\ 0 & \beta' \textbf{1}_{r \times r}
\label{xabs}
\end{array} \right)
\eea
with
\bea
\alpha' (p+q) + \beta' r = 0, \qquad p + q + r = 8.
\label{alphas}
\eea
Regarded as a $28 \times 28$ matrix,
real, self-dual totally
anti-symmetric  $SO(p+q)^{+} \times SO(r)^{+}$-invariant
four-form tensor
$X^{+IJKL}_s$ has eigenvalues $\al', \be'$
and $\ga'=(\al'+\be')/2 $ with degeneracies $d_{\al'}, d_{\be'}$ and
$d_{\ga'}$
respectively.
The eigenvalues and eigenspaces of the $SO(p+q)^{+} \times SO(r)^{+}$
invariant tensor are summarized in Table 1.
By introducing projectors, $P_{\al',s}, P_{\be',s}$ and $P_{\ga',s}$
onto corresponding eigenspaces, we have a $28 \times 28$ matrix equation.
Projector
$P_{\al',s}(P_{\be',s})$ projects the $SO(8)$ Lie algebra onto its
$SO(p+q)^{+}
(SO(r)^{+})$
subalgebra  while $P_{\ga',s}$ does onto the remainder
$SO(8)/(SO(p+q)^{+} \times SO(r)^{+})$. Note that $p$ over there
is replaced by
$p+q$ here.
The projectors
can be constructed from
$X^{+IJKL}_s$ similarly.
The combination $g A_{\mu}^{IJ}$ in the minimal couplings will
be finite as $s \rightarrow \infty$ if $g$ is rescaled to
\bea
g(s) = g e^{\al' s}
\nonu
\eea
for some constant $\al'$(taken as $-1$) so that
by acting $[\exp(-s X_s^{+})]^{IJKL}$ on the right hand side
of (\ref{gA})
\bea
g(s,t) A_{\mu}^{IJ}(s,t) & = & g \left(   P_{\al',s}^{IJKL} +
e^{(\al'-\be') s} P_{\be',s}^{IJKL} +
e^{(\al'-\ga') s} P_{\ga',s}^{IJKL}
\right)   \nonu \\
&&  \times \left(   A_{\mu (\al)}^{KL} +
e^{(\al-\be) t} A_{\mu (\be)}^{KL} + e^{(\al-\ga) t} A_{\mu (\ga)}^{KL}
\right) \nonu \\
& = &  g \left( A_{\mu (\al' \al)}^{IJ} + \xi  A_{\mu (\al' \be)}^{IJ} +
\sqrt{\xi}  A_{\mu (\al' \ga)}^{IJ}+
\zeta \xi A_{\mu (\be' \be)}^{IJ} +
\sqrt{\zeta} \xi  A_{\mu (\ga' \be)}^{IJ} +
\sqrt{\zeta \xi}  A_{\mu (\ga' \ga)}^{IJ}
 \right) \nonu
\eea
where $\zeta =e^{(\al' -\be') s}$ as before.
Here we used the fact that
\bea
P_{\be',s} P_{\al,t} = P_{\be',s} P_{\ga,t} =
P_{\ga',s} P_{\al,t} =0
\label{property}
\eea
which can be shown by the explicit expression of
projectors given in Appendix F
and we denote the simplified notations for
$A_{\mu (\si' \si)}^{IJ}$, where
$\si'=\al', \be', \ga', \si=\al, \be, \ga$ as follows:
\bea
&&A_{\mu (\al' \al)}^{IJ}  \equiv
\left(P_{\al',s} P_{\al,t}\right)^{IJMN} A_{\mu}^{MN},
\qquad
A_{\mu (\al' \be)}^{IJ}   \equiv
\left(P_{\al',s} P_{\be,t}\right)^{IJMN} A_{\mu}^{MN},
\nonu \\
&& A_{\mu (\al' \ga)}^{IJ}   \equiv
\left( P_{\al',s} P_{\ga,t} \right)^{IJMN} A_{\mu}^{MN},\qquad
A_{\mu (\be' \be)}^{IJ}   \equiv
\left( P_{\be',s}  P_{\be,t} \right)^{IJMN} A_{\mu}^{MN},
\nonu \\
&& A_{\mu (\ga' \be)}^{IJ}   \equiv
\left( P_{\ga',s} P_{\be,t} \right)^{IJMN} A_{\mu}^{MN},
\qquad
A_{\mu (\ga' \ga)}^{IJ}  \equiv
\left( P_{\ga',s} P_{\ga,t} \right)^{IJMN} A_{\mu}^{MN}.
\nonu
\nonu
\eea
Now we can think of the product of these projectors,
$P_{\si',s}^{IJKL} P_{\si,t}^{KLMN}$, as a single projector.
Therefore, let us define them, to satisfy the usual property of projectors, as
\bea
&&P_{\al',s} P_{\al,t} \equiv P_{\al}, \qquad
P_{\al',s} P_{\be,t} \equiv P_{\be}, \qquad
P_{\al',s} P_{\ga,t} \equiv P_{\de}, \nonu \\
&&P_{\be',s} P_{\be,t} \equiv P_{\ga}, \qquad
P_{\ga',s} P_{\be,t} \equiv P_{\rho}, \qquad
P_{\ga',s} P_{\ga,t} \equiv P_{\la}.
\label{relation}
\eea
We will see that $\de=(\al+\be)/2, \la=(\al+\ga)/2$ and
$\rho =(\be+\ga)/2$ and $\al$ and $\be$ are related to $\al$'s in
(\ref{alphat}) and (\ref{alphas}).
Projector
$P_{\al}(P_{\be})[P_{\ga}]$
projects the $SO(8)$ Lie algebra onto its $SO(p)^{+}
(SO(q)^{+})[SO(r)^{+}]$
subalgebra  while $P_{\de}(P_{\la})[P_{\rho}]$ projects onto the remainder
\bea
\frac{SO(8)}{SO(p)^{+} \times SO(q)^{+}}
\left(\frac{SO(8)}{SO(p)^{+} \times SO(r)^{+}} \right) \left[
\frac{SO(8)}{SO(q)^{+} \times SO(r)^{+}} \right]
\nonu
\eea
which will be discussed
in next section 3.5.
One obtains these projectors explicitly
from the relation (\ref{relation}) where the projectors in
$SO(p)^{+} \times SO(q)^{+}$-invariant sector are given in
the appendix F.
In terms of these new projectors, one can write
the combination $g(s,t) A_{\mu}^{IJ}(s, t)$ as
\bea
A_{\mu (\al)}^{IJ}+ \xi \left(A_{\mu (\be)}^{IJ}+  \zeta
A_{\mu (\ga)}^{IJ} + \sqrt{\zeta} A_{\mu (\rho)}^{IJ} \right) +
\sqrt{\xi} \left( A_{\mu (\de)}^{IJ} +
\sqrt{\zeta} A_{\mu (\la)}^{IJ} \right).
\label{result}
\eea
By expanding
$ g(t,s)  D (A_{\mu}, \xi, \zeta)$ with respect to both
$t$ and $s$, there exist many terms that seem to diverge
as $t \rightarrow \infty$ or
$s \rightarrow \infty$. However, by exploiting some identities
of the generators given in appendix D,
it implies that those divergent terms vanish
identically and therefore
a limit of $t \rightarrow \infty$ or $s \rightarrow \infty$
exists.

By simplifying the expressions appearing in $ g(t,s)
D (A_{\mu}, \xi, \zeta)$, one gets, for example,
the first $28 \times 28$ block diagonal terms given by
\bea
&& A_{(\al) \mu} + P_{\al} A_{(\de) \mu} P_{\de} +
 P_{\de} A_{(\de) \mu} P_{\be}
+ P_{\la} A_{(\de) \mu} P_{\rho}  +
 P_{\la} A_{(\la) \mu} P_{\ga} +
 + P_{\al} A_{(\la) \mu } P_{\la} +
P_{\de} A_{(\la) \mu } P_{\rho}
\nonu \\
&&  + \xi \left(  A_{(\be) \mu} + P_{\de} A_{(\de) \mu } P_{\al} +
 P_{\be} A_{(\de) \mu } P_{\de} +  P_{\rho} A_{(\de) \mu } P_{\la} +
 P_{\be} A_{(\rho) \mu } P_{\rho} + P_{\de} A_{(\rho) \mu } P_{\la}
 \right. \nonu \\
&& \left. + P_{\rho} A_{(\rho) \mu } P_{\ga}
\right)  + \xi \zeta
\left( A_{(\ga) \mu} +  P_{\rho} A_{(\rho) \mu } P_{\be} +
 P_{\la} A_{(\rho) \mu } P_{\de}
+ P_{\ga} A_{(\rho) \mu } P_{\rho} +
 P_{\la} A_{(\la) \mu } P_{\al}
  \right. \nonu \\
&& \left.   + P_{\rho} A_{(\la) \mu } P_{\de}
+  P_{\ga} A_{(\la) \mu } P_{\la} \right)
\label{block}
\eea
where
we used the properties between projectors and vector fields:
\bea
 P_{\al} A_{(\rho) \mu } P_{\al} =
P_{\ga} A_{(\de) \mu } P_{\ga} =
P_{\be} A_{(\la) \mu } P_{\be} =0.
\nonu
\eea
One can prove that (\ref{block}) becomes the one we have considered
for $SO(p, q+r)^{+}$ gauging
when $\zeta=1$ by combining $\xi$-dependent terms with
$\xi \zeta$-dependent terms\footnote{
When $\zeta=1$, (\ref{block}) becomes,
\bea
\underline{A}_{(\al) \mu }
+ P_{\al,t} \underline{A}_{(\ga) \mu } P_{\ga,t}+
P_{\ga,t} \underline{A}_{(\ga) \mu } P_{\be,t} +
\xi \left(
\underline{A}_{(\be) \mu }  +
P_{\be,t} \underline{A}_{(\ga) \mu } P_{\ga,t}+
P_{\ga,t} \underline{A}_{(\ga) \mu } P_{\al,t} \right).
\nonu
\eea
} and removing
the projectors $P_{\si',s}(\si'=\al', \be', \ga')$
with (\ref{relation}) under the extensive manipulation of properties of
projectors. On the other
hand, when $\xi =1$, it becomes the one in $SO(p+q, r)^{+}$ gauging by
combining the $\xi, \zeta$-independent terms with $\xi$-depedent terms and
removing the projectors $P_{\si,t}(\si=\al, \be, \ga) $.
In this case, we can write it similarly\footnote{
When $\xi =1$, (\ref{block}) becomes
\bea
\underline{A}_{(\al') \mu }
+ P_{\al',s} \underline{A}_{(\ga') \mu } P_{\ga',s}+
P_{\ga',s} \underline{A}_{(\ga') \mu } P_{\be',s} +
\zeta \left(
\underline{A}_{(\be') \mu }  +
P_{\be',s} \underline{A}_{(\ga') \mu } P_{\ga',s}+
P_{\ga',s} \underline{A}_{(\ga') \mu } P_{\al',s} \right).
\nonu
\eea}.

Collecting all other terms by simplifying other three $28 \times 28$
blocks we get
\bea
&& g(t,s)  D (A_{\mu}, \xi, \zeta)
=  g D(A_{\mu})  \nonu \\
& & -(1-\xi) g
\left(
\begin{array}{cc}
\underline{A}_{(\be) \mu }  + \frac{1}{2} \left(
\underline{A}_{(\de) \mu }+ \underline{A}_{(\rho) \mu }
\right),
 &  Z_{(\rho) IJKL}^{MN} A_{(\rho) \mu }^{MN} -
Z_{(\de) IJKL}^{MN} A_{(\de) \mu }^{MN}  \\
Z_{(\rho) IJKL}^{MN} A_{(\rho) \mu }^{MN} -
Z_{(\de) IJKL}^{MN} A_{(\de) \mu }^{MN},
&
\underline{A}_{(\be) \mu }  + \frac{1}{2} \left(
\underline{A}_{(\de) \mu }+ \underline{A}_{(\rho) \mu }
\right)
\end{array} \right) \nonu \\
&& -(1- \xi \zeta ) g
\left(
\begin{array}{cc}
\underline{A}_{(\ga) \mu }  + \frac{1}{2} \left(
\underline{A}_{(\la) \mu }+ \underline{A}_{(\rho) \mu }
\right),
 &  Z_{(\la) IJKL}^{MN} A_{(\la) \mu }^{MN} -
Z_{(\rho) IJKL}^{MN} A_{(\rho) \mu }^{MN}   \\
Z_{(\la) IJKL}^{MN} A_{(\la) \mu }^{MN} -
Z_{(\rho) IJKL}^{MN} A_{(\rho) \mu }^{MN},
&
\underline{A}_{(\ga) \mu }  + \frac{1}{2} \left(
\underline{A}_{(\la) \mu }+ \underline{A}_{(\rho) \mu }
\right)
\end{array} \right)
\nonu
\eea
where $Z_{(\si) IJKL}^{MN}$ are quadratic forms of projectors
\bea
Z_{(\de)IJKL}^{MN} & = &
\frac{1}{2} \left[ \left(P_{\al} -P_{\be} \right)_{IJMP}
P_{\de}^{NPKL} -P_{\de}^{IJMP} \left(P_{\al} -P_{\be} \right)_{NPKL} \right.
\nonu \\
& & \left. -
\left( P_{\rho IJMP} P_{\la}^{NPKL} -  P_{\la IJMP} P_{\rho}^{NPKL}
\right) \right],
\label{Zgeneral}
\eea
and $Z_{(\la)IJKL}^{MN}$ can be written
by performing the change of the above indices in (\ref{Zgeneral}) as
$\al \rightarrow \ga, \be \rightarrow \al, \de \rightarrow \la,
\rho \rightarrow \de, \la \rightarrow \rho$ and
$Z_{(\rho)IJKL}^{MN}$ which can be expressed by
changing the indices in (\ref{Zgeneral}) as
$\al \rightarrow \be, \be \rightarrow \ga, \de \rightarrow \rho,
\rho \rightarrow \la, \la \rightarrow \de$.
Then  our $SU(8)$ $T'$ tensor encoding the structure of
the scalar sector of the ${\cal N}=8$ supergravity
 can be read off and one arrives at the
final complicated expression:
\bea
T_{i}^{\prime \;\;jkl}\left( \xi, \zeta \right) & =
& T_{i}^{\;\;jkl}-\left( 1-\xi \right)
\left(\overline{u}^{kl}_{\;\;\;\;IJ} + \overline{v}^{klIJ} \right)
\nonumber \\ & & \times \left[ \left( P_{\be}^{IJKL}
+ \frac{1}{2} \left( P_{\de}^{IJKL} +  P_{\rho}^{IJKL} \right) \right) \left(
u_{im}^{\;\;\;KM}\overline{u}_{\;\;\;LM}^{jm} -
v_{imKM}\overline{v}^{jmLM} \right) \right. \nonumber \\ & &
\left. + \left( P_{\de}^{IJRS} Z_{(\de) RS}^{KLMN} -
P_{\rho}^{IJRS} Z_{(\rho) RS}^{KLMN}
\right) \left(
-v_{imKL}\overline{u}^{jm}_{\;\;\;MN} +
u_{im}^{\;\;\;KL}\overline{v}^{jmMN} \right) \right]
\nonu \\
&&
 -\left( 1-\xi \zeta \right)
\left(\overline{u}^{kl}_{\;\;\;\;IJ} + \overline{v}^{klIJ} \right)
\nonumber \\
& & \times \left[ \left( P_{\ga}^{IJKL}
+ \frac{1}{2} \left( P_{\la}^{IJKL} +  P_{\rho}^{IJKL} \right) \right) \left(
u_{im}^{\;\;\;KM}\overline{u}_{\;\;\;LM}^{jm} -
v_{imKM}\overline{v}^{jmLM} \right) \right. \nonumber \\ & &
\left. + \left( P_{\rho}^{IJRS} Z_{(\rho) RS}^{KLMN} -
P_{\la}^{IJRS} Z_{(\la) RS}^{KLMN}
\right) \left(
-v_{imKL}\overline{u}^{jm}_{\;\;\;MN} +
u_{im}^{\;\;\;KL}\overline{v}^{jmMN} \right) \right].
\label{tprimepqr}
\eea
Let us examine the structure of $T^{\prime}$-tensor\footnote{
It is more convenient to use the $SL(8, {\bf R})$ basis in order to
compare operators in the dual CFT. This is nothing but
a triality rotation of the $SU(8)$ basis we have considered in which
the two representations of $SO(8)$ are converted to each other using
gamma matrices. In an  $SL(8, {\bf R})$ basis, the expression of
T-tensor was already given as Eq. (21) in \cite{hullcqg}.  }. When $\xi =1$, it
consists of $\zeta$-independent part plus $\zeta$-dependent part. One can
prove that by plugging $P_{\si}(\si=\al, \be, \ga, \de, \la, \rho)$
into the product of
$P_{\si', s}(\si'=\al', \be', \ga')$ and $P_{\si,t}(\si=\al, \be,
\ga)$. According
to (\ref{relation}), the expressions of projectors proportional to
$1-\zeta$ are identical to those in (\ref{tprime}) for $SO(p+q)^{+}
\times SO(r)^{+}$-invariant sector.
On the other hand, when $\zeta=1$,
the above (\ref{tprimepqr}) will
consist of $\xi$-independent part plus $\xi$-dependent part.
By substituting $P_{\si}(\si=\al, \be, \ga, \de, \la,
\rho)$ back into $P_{\si', s}(\si'=\al', \be', \ga')$
and $P_{\si,t}(\si=\al, \be, \ga)$. According
to (\ref{relation}), the expressions of projectors proportional to
$1-\xi$ are the same as those in (\ref{tprime}) for $SO(p)^{+}
\times SO(q+r)^{+}$-invariant sector. Therefore,
the expressions of projectors proportional to
$1-\xi$ in (\ref{tprimepqr}) are the difference between
the one in $(p, q+r)$ and the one in $(p+q, r)$.
One can easily see that
the expressions of projectors proportional to
$1-\xi \zeta$ in (\ref{tprimepqr}) are
the one in $(p+q, r)$. This implies that one can use the projectors
in (\ref{tprimepqr}) for those in $SO(p)^{+} \times SO(q)^{+}$
invariant sector. Alternatively, one can exploit those projectors from (\ref{productproj})
directly.

When $(\xi, \zeta)=(1, 1)$,
this leads to the algebra of $SO(8)$ and one obtains
de Wit-Nicolai gauging with $SU(8) \times SO(8)$ gauge symmetry.
When  $(\xi, \zeta)=(1, 0)$ one
has $CSO(p+q, r)^{+}$ algebra with $SU(8) \times CSO(p+q, r)^{+}$
gauge symmetry.
Moreover, when  $(\xi, \zeta)=(1, -1)$, one gets $SO(p+q, r)^{+}$ algebra
with $SU(8) \times SO(p+q, r)^{+}$ gauge symmetry.
When  $(\xi, \zeta)=(-1, 1)$, it will give non-compact
$SO(p,q+r)^{+}$ gauging with $SU(8) \times SO(p, q+r)^{+}$ gauge symmetry.
When  $(\xi, \zeta)=(-1, 0)$,
it yields a nontrivial non-semi-simple
algebra of the Inonu-Wigner contraction of $SO(8)$
about its $SO(p, q)^{+}$ subgroup, denoted by $CSO(p, q, r)^{+}$ with
$SU(8) \times CSO(p, q, r)^{+}$ gauge symmetry.
For $(\xi, \zeta)=(-1, -1)$, one gets $SO(p+r, q)^{+}$ algebra with
gauge symmetry $SU(8) \times SO(p+r, q)^{+}$.
Finally, when $\xi=0$, it gives Inonu-Wigner contraction
$CSO(p,q+r)^{+}$ with gauge symmetry $SU(8) \times CSO(p, q+ r)^{+}$.
The gauge group will be spontaneously broken to its maximal
compact subgroup.

\subsection{ Superpotential and Scalar Potential in
$CSO(p, q, r)^{+}$ Gaugings \cite{hullcqg}}

The parametrization for the $SO(p)^{+} \times SO(q)^{+}
\times SO(r)^{+}$-singlet space
that is invariant subspace under a particular $SO(p)^{+} \times SO(q)^{+}
\times SO(r)^{+}$ subgroup of $SO(8)$ becomes
\bea
\phi_{IJKL} = 4 \sqrt{2} \left(m X^{+}_{IJKL, s} + n X^{+}_{IJKL, t} \right)
\nonu
\eea
where $m, n$ are two real scalar fields.
The two scalar fields parametrize an
$SO(p)^{+} \times SO(q)^{+}
\times SO(r)^{+}$-invariant subspace of the
full scalar manifold $E_{7(7)}/SU(8)$.
The 56-beins ${\cal V}$ can be written as a $56 \times 56$
matrix by exponentiating the vacuum expectation value $\phi_{IJKL}$.
One can construct 28-beins
$u_{ij}^{\;\;\;KL}$ and $v_{ijKL}$
in terms of scalars $m, n$
explicitly, which can be given in terms of the products
of $u_{ij,t}^{\;\;\;KL}$,
$v_{ijKL,t}$, $u_{ij,s}^{\;\;\;KL}$ and $v_{ijKL,s}$,
as given in the Appendix E\footnote{One can express
$ u_{IJ}^{\;\;\;KL} $ and $ v_{IJKL}$ in terms of sum of
product of $4 \times 4$ matrices as follows:
\bea
u_{IJ}^{\;\;\;KL}  & = & \mbox{diag} \left(
u_{1,t} u_{1,s} + v_{1,t} \overline{v}_{1,s},
u_{2,t} u_{2,s} + v_{2,t} \overline{v}_{2,s},
u_{3,t} u_{3,s} + v_{3,t} \overline{v}_{3,s}, \right. \nonu \\
&& \left. u_{4,t} u_{4,s} + v_{4,t} \overline{v}_{4,s},
u_{5,t} u_{5,s} + v_{5,t} \overline{v}_{5,s},
u_{6,t} u_{6,s} + v_{6,t} \overline{v}_{6,s},
u_{7,t} u_{7,s} + v_{7,t} \overline{v}_{7,s} \right), \nonu \\
v_{IJKL}  & = & \mbox{diag} \left(
u_{1,t} v_{1,s} + v_{1,t} \overline{u}_{1,s},
u_{2,t} v_{2,s} + v_{2,t} \overline{u}_{2,s},
u_{3,t} v_{3,s} + v_{3,t} \overline{u}_{3,s}, \right. \nonu \\
&& \left. u_{4,t} v_{4,s} + v_{4,t} \overline{u}_{4,s},
u_{5,t} v_{5,s} + v_{5,t} \overline{u}_{5,s},
u_{6,t} v_{6,s} + v_{6,t} \overline{u}_{6,s},
u_{7,t} v_{7,s} + v_{7,t} \overline{u}_{7,s} \right)
\nonu
\eea
where each $u_{i,t}$ and $u_{i, s}$ corresponds to seven $4 \times 4$
block diagonal matrices for $u_{IJ,t}^{\;\;\;KL}$  and $u_{IJ,s}^{\;\;\;KL}$
respectively as in Appendix E and  $v_{i,t}$ and $v_{i, s}$ for
$v_{IJKL,t}$ and $ v_{IJKL,s}$ respectively. Their complex conjugations
hold similarly.}.
Now the full expression for $A_1^{\prime }$ and
$A_{2}^{\prime}$ tensors are given in terms of $m, n$ using
(\ref{A1A2}) and (\ref{tprimepqr}) with  $T^{\prime}$ tensor.
\bea
{\cal V}(x) & = &
 \mbox{exp}\left(
\begin{array}{ccc}
0  & -\frac{1}{2\sqrt{2}}\;\phi_{IJPQ} \\
-\frac{1}{2\sqrt{2}}\;\overline{\phi}^{MNKL} &
0
\end{array}\right) \nonu \\
& = &
 \mbox{exp}\left(
\begin{array}{ccc}
0  & -\frac{1}{2\sqrt{2}}\;\phi_{IJPQ,t} \\
-\frac{1}{2\sqrt{2}}\;\overline{\phi}^{MNKL,t} &
0
\end{array}\right) \times
 \mbox{exp}\left(
\begin{array}{ccc}
0  & -\frac{1}{2\sqrt{2}}\;\phi_{IJPQ,s} \\
-\frac{1}{2\sqrt{2}}\;\overline{\phi}^{MNKL,s} &
0
\end{array}\right) \nonu \\
& = & \left( \begin{array}{ccc} u_{ij,t}^{\;\;\;IJ}
& v_{ijKL,t} \\
 \overline{v}^{klIJ}_t & \overline{u}^{kl}_{\;\;\;KL,t}  \end{array} \right)
\times
\left( \begin{array}{ccc} u_{ij,s}^{\;\;\;IJ}
& v_{ijKL,s} \\
 \overline{v}^{klIJ}_s & \overline{u}^{kl}_{\;\;\;KL,s}  \end{array} \right)
= \left( \begin{array}{ccc} u_{ij}^{\;\;\;IJ}
& v_{ijKL} \\
 \overline{v}^{klIJ} & \overline{u}^{kl}_{\;\;\;KL}  \end{array} \right)
\label{uvgeneral}
\eea
where $\phi_{IJKL, s}= 4\sqrt{2} m X^{+}_{IJKL, s}$ and
 $\phi_{IJKL, t}= 4\sqrt{2} n X^{+}_{IJKL, t}$, which commute each
other.
It turns out from (\ref{tprimepqr})
that  the $A_1^{\prime }$ tensor has a single real eigenvalue,
$z_1$, with degeneracies 8 that has the following form
\bea
A_1^{\prime ij} & = &
\mbox{diag} (z_1, z_1, z_1, z_1, z_1, z_1,z _1, z_1), \nonu
\\
z_1 & = &  \frac{1}{8} \left( p e^{m+n} + q e^{m -\frac{p}{(q+r)} n} \xi +
r e^{-\frac{p+q}{r} m - \frac{p}{(q+r)} n} \xi \zeta  \right).
\label{a1pqr}
\eea
It is now straightforward to verify that this yields (\ref{pqz1}) for
$(p+q, r)$ gauging when $\xi=1$ and $n=0$. However, when $\zeta =1$ and $m=0$ it
becomes (\ref{pqz1}) for $(p, q+r)$ gauging.
In particular the superpotential, $W$, for the flow is found
as one of the eigenvalues of the this symmetric tensor.
Additionally, we can construct the $A_{2}^{\prime}$
tensor from (\ref{tprimepqr})
which are the combinations of the triple product of
28-beins as given in
\bea
A_{2i}^{\prime\;\;\;jkl} & = & \frac{1}{4} ( q+r) e^m \left(e^{n}-\xi
e^{-\frac{p}{q+r} n}\right)X^{+ijkl}_{t} +
\frac{1}{4} r \xi e^{-\frac{p}{q+r} n} \left(e^{m}-\zeta
e^{-\frac{p+q}{r} m}\right)X^{+ijkl}_{s} \nonu \\
&= & e^m (A_{2, t})_{i}^{\;\;jkl} +
\xi e^{-\frac{p}{q+r} n} (A_{2, s})_{i}^{\;\;jkl}
\label{a2pqr}
\eea
where $(A_{2, t})_{i}^{\;\;jkl}$ is the same as the one (\ref{a2tensor}) for
$SO(p)^{+} \times SO(q+r)^{+}$ sector and $(A_{2, s})_{i}^{\;\;jkl}$
for  $SO(p+q)^{+} \times SO(r)^{+}$.
Moreover $X^{+ijkl}_{t}$ is $\sum_{\si = \al, \be, \ga} \si P_{\si, t}$
while $X^{+ijkl}_{s}$ is $\sum_{\si' = \al', \be', \ga'} \si'
P_{\si', s}$\footnote{
One can prove $A_1^{\prime}$ and $A_2^{\prime}$ can be obtained by
analytic continuation. The $T^{\prime}$ tensor we obtained is
$T^{\prime \;\;jkl}_i ( E(-n) \times F(-m) , \xi, \zeta) $. By considering
only the $SL(8, {\bf R})$ transformation by $\xi$, this can be reduced to
$e^{\al t} T^{\prime \;\;jkl}_i ( E(t+n)^{-1} \times F(-m) , 0, \zeta)$.
Moreover, this becomes
$e^{\al (t-n)}
T^{\prime \;\;jkl}_i ( E(t)^{-1} \times F(-m) , e^{(\al-\be)n}, \zeta)$.
Now we arrive at the following intermediate expression:
$e^{-\al n}
T^{\prime \;\;jkl}_i ({\bf 1} \times F(-m) , \xi e^{(\al-\be)n}, \zeta)$.
Next we apply $SL(8, {\bf R})$ transformation by $\zeta$. Then by doing
a similar procedure we arrive at the final expression:
\bea
T^{\prime \;\;jkl}_i ( E(-n) \times F(-m) , \xi, \zeta)=
e^{-\al' m} e^{-\al n}
T^{\prime \;\;jkl}_i ({\bf 1}, \xi e^{(\al-\be)n},
\zeta e^{(\al'-\be')m}).
\nonu
\eea
At the origin, $\phi_{IJKL}=0, {\cal V}={\bf 1}$,
the $T^{\prime}$ tensor is from (\ref{tprimepqr})
\bea
T^{\prime \;\;jkl}_i ({\bf 1}, \xi, \zeta)  =
\frac{3}{2} [ 1 - (1-\xi) a_1^{-1} - \xi(1-\zeta)
a_2^{-1} ] \de_{ij}^{kl} -\frac{3}{2} (1-\xi) a_1^{-1} X^{ijkl}_t
- \frac{3}{2} \xi(1-\zeta) a_2^{-1} X^{ijkl}_s.
\label{analytic}
\eea
Finally we possess all the information of
$T^{\prime \;\;jkl}_i ( E(-n) \times F(-m) , \xi, \zeta)$ because
by transforming $\xi \rightarrow \xi e^{(\al-\be)n},
\zeta \rightarrow \zeta e^{(\al'-\be')m}$ as in (\ref{analytic})
we get  $T^{\prime \;\;jkl}_i ({\bf 1}, \xi e^{(\al-\be)n},
\zeta e^{(\al'-\be')m})$.
From this, one can obtain
$A_1^{\prime}$ tensor which is $\frac{4}{21}
T^{\prime \;\;jkl}_i ( E(-n) \times F(-m) , \xi, \zeta)$.
It turns out that it coincides with the one in (\ref{a1pqr}).
We used the numerical values: $\al'=-1=\al, \be=\frac{p}{q+r}, \be'=
\frac{p+q}{r}$ and $a_1 = \frac{p}{q+r} +1$, $a_2 =\frac{p+q}{r}+1$. Additionally,
we have checked that
$ A_{2,i}^{\prime \;\;jkl} = - \frac{4}{3}
T^{\prime \;\;[jkl]}_i ( E(-n) \times F(-m) , \xi, \zeta)
$ is identical to the one in (\ref{a2pqr}).  }.
Finally the $K_{\xi,\zeta, p, q, r}$-invariant scalar potential
as a function of $p, q, r, \xi,\zeta$ and $m,n$ by
combining all the components
can be written as
\bea
V_{p,q,r}(\xi,\zeta; m, n)  &= &
-g^{2}\left(
\frac{3}{4}|A_{1}^{\prime\;ij}|^{2}-\frac{1}{24}|A_{2i}^{\prime\;\;\;jkl}|^2
\right) \nonu \\
&= & \frac{1}{1536}  e^{-\frac{2(p+q)m}{r}-\frac{2pn}{q+r}}
\left( V_1 + V_2 \xi +V_3 \xi \zeta +V_4 \xi^2 + V_5 \xi^2 \zeta
+ V_6 \xi^2 \zeta^2  \right) \nonu
\eea
where we introduce intermediate functions $V_i$'s
as the coefficients of the $\xi$
and $\zeta$
\bea
V_1 & = & e^{\frac{16m}{r}+\frac{16n}{q+r}}
 p \left(p^{2}(q+r)+(-2+q+r)(q+r)^{2}
\right.  \nonumber
 \\ & &
  \left. +2p(-72+q^{2}-r+r^{2}+q(-1+2r) \right),  \nonu
 \\
V_2  & = &  -
 2e^{\frac{16m}{r}+\frac{8n}{q+r}}pq \left(144+p^{2}+
q^{2}+2q(-1+r)-2r+r^{2}+2p(-1+q+r) \right), \nonu \\
V_3 & = & -
 2e^{\frac{8m}{r}+\frac{8n}{q+r}}pr \left(144+p^{2}+q^{2}+
2q(-1+r)-2r+r^{2}+2p(-1+q+r) \right),
  \nonumber
 \\
V_4  & = &
 e^{\frac{16m}{r}}q \left(p^{3}+q^{2}r+(-2+r)r^{2}+
p^{2}(-2+2q+3r)+2q(-72-r+r^{2})
 \right. \nonumber
 \\ & & \left.
 +p(q^{2}+r(-4+3r)+q(-2+4r)) \right), \nonumber
 \\
V_5 & = & -
 2e^{\frac{8m}{r}}qr \left(144+p^{2}+q^{2}+2q(-1+r)-2r+
r^{2}+2p(-1+q+r) \right),
  \nonumber
 \\
V_6 & = &
 r \left(p^{3}+q^{3}+2q^{2}(-1+r)-144r+q(-2+r)r+p^{2}(-2+3q+2r) \right.
\nonumber
 \\ & & \left.
 +p(3q^{2}+4q(-1+r)+(-2+r)r) \right).
\nonu
\eea

By looking at the form of scalar potential, it is easy to see that
$V_{r, q, p}(\xi=-1, \zeta=-1; m, n)$ can be obtained
from $V_{p, q, r}(\xi=-1, \zeta=-1; -\frac{r}{p+q} n, -\frac{q+r}{p} m)$.
Under the change of real fields, they are equivalent to each other.
Moreover, the potential $V_{r, q, p}(\xi=-1, \zeta=1; m, n)$
can be obtained from $V_{p, q, r}(\xi=1, \zeta=-1;  -\frac{r}{p+q} n ,
-\frac{q+r}{p} m)$.
On this basis, the kinetic terms are not
the usual ones but
there exists a cross term, $\pa^{\mu} m \pa_{\mu} n$
which makes it difficult to find first-order differential equations
for domain-wall solutions. Now we have to change the basis for which one has
usual kinetic terms. We calculated all the quantities for 21 possible cases
of $CSO(p, q, r)^{+}$ gaugings
and summarized them in Appendix G:kinetic terms in terms of old
fields\footnote{
One can generalize the kinetic terms (\ref{kineticpq})
of $SO(p)^{+} \times SO(q)^{+}$-invariant sector to write down
\bea
&& A_{\mu}^{\;\; ijkl}  =  -2 \sqrt{2} \left( \overline{u}^{ij}_{\;\;IJ}
\partial_{\mu} \overline{v}^{klIJ} -
\overline{v}^{ijIJ} \partial_{\mu} \overline{u}^{kl}_{\;\;IJ} \right) \nonu \\
&& + 4\sqrt{2} (1-\xi) g A_{\mu IJ} \left[
\left( P_{\be}^{IJKL}
+ \frac{1}{2} \left( P_{\de}^{IJKL} +  P_{\rho}^{IJKL} \right) \right)
\left( -\overline{u}^{ij}_{\;\;KM}
\overline{v}^{klLM} + \overline{v}^{ijKM} \overline{u}^{kl}_{\;\;LM}
\right) \right. \nonu \\
&& \left. +
\left( P_{\de}^{IJRS} Z_{(\de) RS}^{KLMN} -
P_{\rho}^{IJRS} Z_{(\rho) RS}^{KLMN}
\right)
 \left(\overline{u}^{ij}_{\;\;KL}
\overline{u}^{kl}_{\;\;MN} - \overline{v}^{ijKL} \overline{v}^{klMN}
\right) \right] \nonu \\
&& + 4\sqrt{2} (1-\xi \zeta) g A_{\mu IJ} \left[
\left( P_{\ga}^{IJKL}
+ \frac{1}{2} \left( P_{\la}^{IJKL} +  P_{\rho}^{IJKL} \right) \right)
\left(
 -\overline{u}^{ij}_{\;\;KM}
\overline{v}^{klLM} + \overline{v}^{ijKM} \overline{u}^{kl}_{\;\;LM}
\right) \right. \nonu \\
&& \left.
\left( P_{\rho}^{IJRS} Z_{(\rho) RS}^{KLMN} -
P_{\la}^{IJRS} Z_{(\la) RS}^{KLMN}
\right)
\left(
\overline{u}^{ij}_{\;\;KL}
\overline{u}^{kl}_{\;\;MN} - \overline{v}^{ijKL} \overline{v}^{klMN}
 \right) \right].
\nonu
\eea
Of course, in this case we put $A_{\mu}^{IJ}$ to zero because
we are interested in the scalar plus gravity parts of the Lagrangian.},
change of variables, superpotential, and
scalar potential as new fields.
From the results in Appendix G, one can describe
a superpotential and scalar potential in terms of new real scalar
fields $\widetilde{m}$ and $\widetilde{n}$ that are related to
the old fields $m$ and $n$ as follows:
\bea
m = -\frac{r\sqrt{pq(p+q)}}{4q(p+q)} \widetilde{m} -
\frac{\sqrt{2r(p+q)}}{2(p+q)} \widetilde{n}, \qquad
n = \frac{(q+r) \sqrt{pq(p+q)}}{4pq} \widetilde{m}.
\nonu
\eea
Then in terms of new fields the superpotential can be written as
\bea
W_{p,q, r}(\xi, \zeta; \widetilde{m}, \widetilde{n})
 =
 \frac{1}{8} \left( p e^{2\sqrt{\frac{q}{p(p+q)}} \widetilde{m}+
\sqrt{\frac{r}{2(p+q)}} \widetilde{n}}  + q e^{-2 \sqrt{\frac{p}{q(p+q)}}
\widetilde{m} -
\sqrt{\frac{r}{2(p+q)}} \widetilde{n}} \xi +
r e^{\sqrt{\frac{p+q}{2r}} \widetilde{n}} \xi \zeta  \right),
\label{generalsuperpotential}
\eea
and the supergravity potential is then given by
\bea
V_{p,q,r}(\xi, \zeta; \widetilde{m}, \widetilde{n}) &  = &   g^2 \left[  4
\left(\partial_{\widetilde{m}} W_{p,q, r}(\xi, \zeta; \widetilde{m},
\widetilde{n}) \right)^2 +
4 \left(\partial_{\widetilde{n}} W_{p,q, r}(\xi, \zeta;
\widetilde{m}, \widetilde{n}) \right)^2 \right. \nonu \\
&& \left. - 6  W_{p,q, r}(\xi, \zeta; \widetilde{m}, \widetilde{n})^2 \right].
\label{pqrsuperpotential}
\eea
Note that the coefficients, 4 and 4, in the first and second terms
in the above are a simple generalization of
(\ref{superpotentialandpotential}) for two scalar fields.

The superpotential has the following values at the various critical
points in Table 4.

$\bullet$
The first row corresponds to the maximal supersymmetric case of
de Wit-Nicolai's $SO(8)$-invariant trivial critical point.

$\bullet$
The second row corresponds to the $SO(7)^{+}$-invariant critical point of
the $SO(8)$ theory that is equivalent to the second one in Table 2.
We find $ V_{r, q, p}(\xi=1, \zeta=1; m, n)=
V_{p, q, r}(\xi=1, \zeta=1;  -\frac{r}{p+q} n ,
-\frac{q+r}{p} m)$.
This means that by a change of variables, the solutions
of $(p,q,r)=(4,1,3), (5,1,2), (6,1,1)$ are obtained from
$(p,q,r)=(3,1,4), (2,1,5), (1,1,6)$, respectively.

$\bullet$
The third row is the gauging of $CSO(p+q, r)^{+}=CSO(2,6)^{+}$
that corresponds to
the sixth in Table 2.

$\bullet$
The fourth row implies $SO(p+q, r)^{+}=
SO(5,3)^{+}$ that
corresponds to negative superpotential, being equivalent to
the third one in Table 2, or $
SO(p+q, r)^{+}=SO(3,5)^{+}$ that corresponds to positive superpotential, being equal to
the fifth in Table 2. In each case, the potentials are the same, although
the superpotentials are different at the critical points.

$\bullet$
The fifth row implies $SO(p+q, r)^{+}=SO(4,4)^{+}$, being equivalent to
the fourth one in Table 2.

$\bullet$
For the sixth row one has $SO(p, q+r)^{+}=
SO(3, 5)^{+}$ gauging with positive superpotential and $SO(p, q+r)^{+}=
SO(5, 3)^{+}$
with negative superpotential.
According to the symmetry betwen the potential, one can see that
all the critical points in this row can be obtained from those in fourth row:
$ V_{r, q, p}(\xi=-1, \zeta=1; m, n)=
V_{p, q, r}(\xi=1, \zeta=-1;  -\frac{r}{p+q} n ,
-\frac{q+r}{p} m)$.
This implies that within a given class with the same scalar potential,
these solutions are coming from the same critical point in the fourth row
but are viewed along different directions in scalar space.

$\bullet$
For the seventh row, we have $SO(p, q+r)^{+}=
SO(4,4)^{+}$ gauging.
Additionally, in this case, all the critical points
are similarly obtained from those in the fifth row.
These solutions are coming from the same critical point in the fifth row
but are viewed along different directions in scalar space.

$\bullet$
For the eighth row, one has either $SO(p+r, q)^{+}=
SO(5,3)^{+}$
gaugings with negative superpotential
or $SO(p+r, q)^{+}=SO(3, 5)^{+}$ with positive superpotential.

$\bullet$
In the ninth row $SO(p+r, q)^{+}=SO(4,4)^{+}$
gauging.

$\bullet$
Finally the last
row corresponds to $CSO(p, q+r)^{+}=CSO(2, 6)^{+}$ gauging.

For the eighth and ninth  rows we have the following symmetry
in the potential:
$V_{r, q, p}(\xi=-1, \zeta=-1; m, n)=
V_{p, q, r}(\xi=-1, \zeta=-1; -\frac{r}{p+q} n, -\frac{q+r}{p} m)$.
In other words, the solutions corresponding to
$(p,q,r)=(2,5,1),(3,3,2),(4,3,1)$ can be obtained from
$(p,q,r)=(1,5,2),(2,3,3),(1,3,4)$ respectively in the eighth row. Similarly,
in the ninth row the solution of $(p,q,r)=(1,4,3)$ is related to $(p,q,r)=(3,4,1)$.

\subsection{ Domain Wall in
$CSO(p, q, r)^{+}$ Gaugings \cite{hullcqg}}

The resulting Lagrangian of scalar-gravity sector takes
\bea
\int d^4 x \sqrt{-g} \left( \frac{1}{2} R
- \frac{1}{2} \partial^{\mu} \widetilde{m}  \partial_{\mu} \widetilde{m}
- \frac{1}{2} \partial^{\mu} \widetilde{n}  \partial_{\mu} \widetilde{n}
- V_{p, q, r}(\xi, \zeta; \widetilde{m}, \widetilde{n}) \right).
\label{action2}
\eea
With the ansatz (\ref{ansatz}) the equations of motion for
the scalars and metric read
\bea
& &
 \pa_r^2 A -\pa_r A \pa_r B+ \frac{3}{2} (\pa_r A)^2 +
\frac{1}{4} (\pa_r \widetilde{m})^2   +
\frac{1}{4} (\pa_r \widetilde{n})^2
  + \frac{1}{2} e^{2B} V_{p, q, r} (\xi, \zeta; \widetilde{m},
\widetilde{n}) = 0, \nonu \\
& & \pa_r^2 \widetilde{m} +
3 \pa_r A \pa_r \widetilde{m}  -
 \pa_r B \pa_r \widetilde{m} - e^{2B} \pa_{\widetilde{m}}
V_{p, q, r} (\xi, \zeta; \widetilde{m},
\widetilde{n})=0, \nonu \\
 & & \pa_r^2 \widetilde{n} +
3 \pa_r A \pa_r \widetilde{n}  -
 \pa_r B \pa_r \widetilde{n} - e^{2B} \pa_{\widetilde{n}}
V_{p, q, r} (\xi, \zeta; \widetilde{m},
\widetilde{n})=0.
\label{secondeq}
\eea
By plugging the domain-wall ansatz (\ref{ansatz})
into the Lagrangian (\ref{action2}), the energy-density
per unit area transverse to $r$-direction
with the integration by parts on the term of $\pa_r^2 A$
can be expressed similarly
 and after rewriting and recombining the
energy-density by summation of complete squares plus other terms,
one gets
\bea
&& E[A, \widetilde{m}, \widetilde{n}] =
\frac{1}{2} \int_{-\infty}^{\infty} dr e^{3A+B}
\left[ - 6 \left( e^{-B} \pa_r A + \sqrt{2} g
W_{p,q,r}(\xi,\zeta;\widetilde{m},\widetilde{n})
\right)^2 \right. \nonu \\
&& + \left( e^{-B} \pa_r \widetilde{m} - 2 \sqrt{2} g
\pa_{\widetilde{m}} W_{p,q,r}(\xi,\zeta;\widetilde{m},\widetilde{n})
 \right)^2
 +\left( e^{-B} \pa_r \widetilde{n} - 2 \sqrt{2} g
\pa_{\widetilde{n}} W_{p,q,r}(\xi,\zeta;\widetilde{m},\widetilde{n})
 \right)^2  \nonu \\
&& \left. +
 12 \sqrt{2} g e^{-B} W_{p,q,r}(\xi,\zeta;\widetilde{m},\widetilde{n})
\pa_r A + 4 \sqrt{2} g
e^{-B}  \pa_r W_{p,q,r}(\xi,\zeta;\widetilde{m},\widetilde{n}) \right].
\nonu
\eea
By recognizing that the last two terms can be written as
$4 \sqrt{2} g \pa_r ( e^{3A}  W_{p,q,r}(\xi,\zeta;\widetilde{m},
\widetilde{n}))$ we arrive at
\bea
& &  \frac{1}{2}
\int_{-\infty}^{\infty} dr e^{3A+B}
\left[ - 6 \left( e^{-B} \pa_r A + \sqrt{2} g
W_{p,q,r}(\xi,\zeta;\widetilde{m},\widetilde{n})
\right)^2 \right. \nonu \\
&& \left. + \left( e^{-B} \pa_r \widetilde{m} -2 \sqrt{2} g
\pa_{\widetilde{m}}
W_{p,q,r}(\xi,\zeta;\widetilde{m},\widetilde{n}) \right)^2 +
\left( e^{-B} \pa_r \widetilde{n} -2 \sqrt{2} g
\pa_{\widetilde{n}}
W_{p,q,r}(\xi,\zeta;\widetilde{m},\widetilde{n})
 \right)^2 \right] \nonu \\
&& + 2\sqrt{2}g \left(
e^{3A}
W_{p,q,r}(\xi,\zeta;\widetilde{m},\widetilde{n})
\right) |_{- \infty}^
{\infty}.
\nonu
\eea

Therefore, one finds the energy-density bound
\bea
E[A, \widetilde{m}, \widetilde{n}]
 \geq  2\sqrt{2}g \left( e^{3A(\infty)}
W_{p,q,r}(\xi,\zeta;\widetilde{m},\widetilde{n})(\infty) -
e^{3A(-\infty)}
W_{p,q,r}(\xi,\zeta;\widetilde{m},\widetilde{n})(-\infty)\right).
\nonu
\eea
This $E[A, \widetilde{m}, \widetilde{n}]$ is extremized by
domain-wall solutions and the first-order differential equations
for the scalar fields one finds are
the gradient flow equations of the superpotential
(\ref{generalsuperpotential}):
\bea
\partial_{r} \widetilde{m} & = &
\pm 2\sqrt{2} e^B g \partial_{\widetilde{m}}
W_{p,q,r}(\xi,\zeta;\widetilde{m},\widetilde{n})
 ,\nonu \\
\partial_{r} \widetilde{n} & = &
\pm 2\sqrt{2} e^B g \partial_{\widetilde{n}}
W_{p,q,r}(\xi,\zeta;\widetilde{m},\widetilde{n}) ,\nonu \\
\partial_{r} A & = & \mp \sqrt{2}  e^{B} g
W_{p,q,r}(\xi,\zeta;\widetilde{m},\widetilde{n}).
\label{solution}
\eea

\bea
\begin{array}{|c|c|c|c|c|c|c|c|c|c|}
\hline
 \cal N & p & q & r & \xi & \zeta & m & n   &
W & V \nonu \\
\hline
   8 &\mbox{any}  & \mbox{any} &  \mbox{any} & 1
 &1 & 0
 & 0 & 1
 & -6 g^2 \nonu \\
\hline
       &1 & 1 & 6 &  &   &
 \frac{3 }{4}\ln 5 & -\frac{7 }{8} \ln 5
 &  &  \nonu \\
      &2 & 1 & 5 &  &
  & \frac{5 }{8}\ln 5 &-\frac{3 }{4}\ln 5
 &  &  \nonu \\
     0 &3 & 1 & 4 & 1 & 1
 & \frac{1}{2} \ln 5 & -\frac{5 }{8}\ln 5
 & \frac{3}{2} \times 5^{-1/8} & - 2\times 5^{3/4}g^2 \nonu \\
      &4 & 1 & 3 &  &
  & \frac{3}{8} \ln 5 &-\frac{1}{2} \ln 5
 &  &  \nonu \\
     &5 & 1 & 2 &  &
 & \frac{1}{4} \ln 5 &- \frac{3}{8} \ln 5
 &  &  \nonu \\
      &6 & 1 & 1 &  &
 & \frac{1}{8} \ln 5 &  -\frac{1}{4} \ln 5
 &  &  \nonu \\
\hline
     0 &1 & 1 & 6 & 1 & 0  &
 \mbox{any} & 0
 & e^{m}/4 & 0 \nonu \\
\hline
      &1 & 2 & 5 &  &   &
 \frac{5}{8} \ln 3  &
 &  \frac{1}{2} \times 3^{-3/8} &  \nonu \\
      &1 & 4 & 3 &  &   &
 -\frac{3}{8} \ln 3&
 &  -\frac{1}{2} \times 3^{-3/8} &  \nonu \\
   0   &2 & 1 & 5 & 1 & -1  &
 \frac{5}{8} \ln 3  & 0
 & \frac{1}{2} \times 3^{-3/8}  &  2 \times 3^{1/4} g^2 \nonu \\
      &2 & 3 & 3 &  &   &
 -\frac{3}{8} \ln 3 &
 &  -\frac{1}{2} \times 3^{-3/8} &  \nonu \\
      &3 & 2 & 3 &  &   &
 -\frac{3}{8} \ln 3 &
 & -\frac{1}{2} \times 3^{-3/8}  &  \nonu \\
      &4 & 1 & 3 &  &   &
 -\frac{3}{8} \ln 3 &
 &  -\frac{1}{2} \times 3^{-3/8} &  \nonu \\
\hline
      &1 & 3 & 4 &  &   &
   &
 &  &  \nonu \\
     0 &2 & 2 & 4 & 1 & -1  &
 0 & 0
 & 0 & 2g^2  \nonu \\
      &3 & 1 & 4 &  &   &
  &
 &  &  \nonu \\
\hline
      &3 & 1 & 4 &  &   &
   & \frac{5}{8} \ln 3
 & \frac{1}{2} \times 3^{-3/8}  &  \nonu \\
      &3 & 2 & 3 &  &   &
  & \frac{5}{8} \ln 3
 &  \frac{1}{2} \times 3^{-3/8} &  \nonu \\
   0   &3 & 3 & 2 & -1 & 1  &
 0  & \frac{5}{8} \ln 3
 &\frac{1}{2} \times 3^{-3/8}  & 2 \times 3^{1/4} g^2 \nonu \\
      &3 & 4 & 1 &  &   &
  & \frac{5}{8} \ln 3
 & \frac{1}{2} \times 3^{-3/8}  &  \nonu \\
      &5 & 1 & 2 &  &   &
  & -\frac{3}{8} \ln 3
 & -\frac{1}{2} \times 3^{-3/8}  &  \nonu \\
      &5 & 2 & 1 &  &   &
  & -\frac{3}{8} \ln 3
 &  -\frac{1}{2} \times 3^{-3/8} &  \nonu \\
\hline
      & 4 & 1 & 3 &  &  &
   &
 &  &  \nonu \\
     0 &4 & 2 & 2 & -1 & 1  &
 0 & 0
 & 0  & 2g^2  \nonu \\
      &4 & 3 & 1 &  &   &
   &
 &  &  \nonu \\
\hline
      &1 & 3 & 4 &  &   &
 \frac{1}{2} \ln 3  & -\frac{7}{8} \ln 3
 & -\frac{1}{2} \times 3^{-3/8} &  \nonu \\
      &1 & 5 & 2 &  &   &
 -\frac{1}{4} \ln 3  & \frac{7}{8} \ln 3
 & \frac{1}{2} \times 3^{-3/8} &  \nonu \\
     0 &2 & 3 & 3 & -1 & -1  &
 \frac{3}{8} \ln 3 & -\frac{3}{4} \ln 3
 & -\frac{1}{2} \times 3^{-3/8} & 2 \times 3^{1/4} g^2 \nonu \\
      &2 & 5 & 1 &  &   &
 -\frac{1}{8} \ln 3 & \frac{3}{4} \ln 3
 & \frac{1}{2} \times 3^{-3/8} &  \nonu \\
      &3 & 3 & 2 &  &   &
 \frac{1}{4} \ln 3 & -\frac{5}{8} \ln 3
 & -\frac{1}{2} \times 3^{-3/8} &  \nonu \\
     &4 & 3 & 1 &  &   &
 \frac{1}{8} \ln 3 & -\frac{1}{2} \ln 3
 & -\frac{1}{2} \times 3^{-3/8} &  \nonu \\
\hline
      &1 & 4 & 3 &  &   &
   &
 &  &  \nonu \\
     0 &3 & 4 & 1 & -1 & -1  &
 0  & 0
 & 0 &  2g^2 \nonu \\
      &2 & 4 & 2 &  &   &
  &
 &  &  \nonu \\
\hline
     &2 & 1 & 5 &  &   &
  &
 &  &  \nonu \\
      &2 & 2 & 4 &  &   &
   &
 &  &  \nonu \\
     0 &2 & 3 & 3 & 0 & 1,0,-1  &
  \mbox{any} &    \mbox{any}
 & e^{m+n}/4 & 0 \nonu \\
      &2 & 4 & 2 &  &   &
  &
 &  &  \nonu \\
      &2 & 5 & 1 &  &   &
   &
 &  &  \nonu \\
\hline
\end{array}
\nonu
\eea
Table 4. \sl Summary of various critical points in the context of
superpotential : supersymmetry,
vacuum expectation values of fields, superpotential and
cosmological constants.
There is {\it no} $SO(p)^{+} \times SO(q)^{+} \times U(1)^{+r(r-1)/2}$
critical point of potential for $\xi =-1$ and $\zeta=0$.
The nontrivial $CSO(p,q,r)^{+}$ gauging in this section does not provide
any new extra critical points.
   \rm

It is easy to check whether solutions $\widetilde{m}(r), \widetilde{n}(r)$
and $A(r)$ of (\ref{solution}) satisfy the gravitational and scalar
equations of motion in the second order differential equations
(\ref{secondeq}).
The analytic solutions of (\ref{solution})
for $\xi=0$ when $B$ is a constant become
\bea
\widetilde{m}(x) & = & -\frac{2 q \left(\sqrt{\frac{2r}{p+q}} c_1 +
2 \log \left[ \frac{g(8q+p r) x +c_2}{4 \sqrt{2}(p+q)} \right]  \right)}
{\sqrt{\frac{q}{p(p+q)}} \left(8q+p r \right)}, \nonu \\
\widetilde{n}(x) & = &  \frac{8 q \sqrt{\frac{r}{p+q}} c_1 -\sqrt{2}
p r \log \left[ \frac{g(8q+p r) x +c_2}{4 \sqrt{2}(p+q)} \right]  }
{\sqrt{\frac{r}{(p+q)}} \left(8q+p r \right)} \nonu  \\
 A(x) & = & c_1 + \frac{p(p+q) \log \left[ \frac{g(8q+p r) x +
c_2}{4 \sqrt{2}(p+q)} \right] }{8q+p r}
\nonu
\eea
where
we change variable $r$ into $x$ in order not to
confuse it with the integer $r$ and $c_1$ and $c_2$ are constant.
One expects to have nontrivial analytic solutions for nonzero $\xi$ for
particular $(p, q, r)$ and $\zeta$, as in section 2.
In particular, for $CSO(1,1,6)^{+}$ gauging where we fix $\xi=-1$ and
$\zeta=0$, we
have
\bea
\widetilde{m}(x) & = & \frac{2}{7} \left(-\sqrt{3} c_1 +  \sqrt{2} \log
\left( \tan \left[ \frac{-7 g x +c_2}{4 \sqrt{2}} \right] \right)
\right), \nonu \\
\widetilde{n}(x) & = &  \frac{2}{14} \left(4 c_1 + \sqrt{6}
\log \left( \tan \left[ \frac{-7 g x +c_2}{4 \sqrt{2}} \right] \right)
\right), \nonu  \\
 A(x) & = & \frac{1}{7} \left( 7 c_1
 \tan
\left[ \frac{-7 g x +c_2}{4 \sqrt{2}} \right]
 + 1 \right) \log \left( \frac{1}{2} \sin
\left[ \frac{-7 g x +c_2}{2 \sqrt{2}} \right] \right). \nonu
\eea
Similarly, one also has an analytic solution of $CSO(3,3,2)^{+}$ gauging where
$\xi=-1$ and $\zeta=0$.

\subsection{ $CSO(p, q, r)^{+}$ Gaugings from $SO(8)$ Gaugings }

Thus far, the values of $p, q$ and $r$ are greater than or equal to 1.
If we allow those values to have zero, then   one can classify
them as follows:
1) $CSO(p, 0, 0)^{+}=SO(p)^{+}$,
2) $CSO(p, q, 0)^{+}=SO(p, q)^{+}$,
3) $CSO(p, 0, r)^{+}=CSO(p, r)^{+}$,
and 4) $CSO(p, q, r)^{+}$.
In this section, we take a different route from previous the case.
Some idea in this direction
was already given in the paper of \cite{hullwarner3}.
Let us consider the $SO(p)^{+} \times SO(q)^{+} \times SO(r)^{+}$
invariant generator
of $SL(8, {\bf R})$,
\bea
X_{ab} & = & \left(
\begin{array}{cccc} \alpha \textbf{1}_{p \times p} &
0 & 0 \\ 0 & \beta \textbf{1}_{q \times q} & 0 \\
0 & 0 & \gamma \textbf{1}_{r \times r}
\label{xabgeneral}
\end{array} \right)
\eea
with
\bea
\alpha p + \beta q + \gamma  r = 0, \qquad p + q + r = 8
\nonu
\eea
where
${\bf 1}_{p \times p}$ is $p \times p$ identity matrix.
The embedding of this $SL(8, {\bf R})$ in $E_7$ is such that
$X_{ab}$ corresponds to the $56 \times 56$ $E_7$ generator
which is a non-compact $SO(p)^{+} \times SO(q)^{+} \times SO(r)^{+}$
invariant element of the $SL(8,{\bf R})$ subalgebra of $E_7$
\bea
\left(
\begin{array}{cc}
0 &  \  X^{+IJKL}  \\
  X_{IJKL}^{+} & 0
\end{array} \right),
\nonu
\eea
where the real, self-dual totally anti-symmetric $SO(p)^{+}
\times SO(q)^{+} \times SO(r)^{+}$
invariant four-form tensor
$X_{IJKL}^{+}$ can be written in terms of a symmetric,
trace-free, $8 \times 8$ matrix with $SO(8)$ right-handed spinor indices,
$X_{ab}$ using $SO(8)$ $\Ga$ matrices(See Appendix B)
\bea
X_{IJKL}^{+} =-\frac{1}{8}\left(\Gamma_{IJKL}\right)^{ab}X_{ab}
\label{xgeneral}
\eea
where $ \Gamma_{IJKL} = \Gamma_{[ I} \Gamma_{J} \Gamma_{K} \Gamma_{L ]}$ and
an arbitrary $SO(8)$ generator $L_{IJ}$ acts in the right-handed
spinor representation by $(L_{IJ} \Gamma_{IJ})^{ab}$.
One can show that
$X^{+IJKL}$ (\ref{xgeneral}) can be
decomposed into $X^{+IJKL}_t$ and $X^{+IJKL}_s$:
\bea
X^{+IJKL} = X^{+IJKL}_t + X^{+IJKL}_s
\nonu
\eea
where the real, self-dual totally anti-symmetric $SO(p)^{+}
\times SO(q+r)^{+}$ invariant four-form tensor
$X^{+IJKL}_t$ was expressed in the previous subsection as
$\Ga$ matrices with (\ref{xabt}) and
$SO(p+q)^{+}
\times SO(r)^{+}$ invariant four-form tensor
$X^{+IJKL}_s$ with (\ref{xabs}).
Moreover, $\al$ and $\be$ in (\ref{xabgeneral}) consist of
$\al_t$ that was defined as (\ref{xabt}) and (\ref{alphat})(we replace
$\al$ by $\al_t$) and $\al_s$ as (\ref{xabs}) and (\ref{alphas}).
We also replace $\al^{\prime}$ with $\al_s$.
Therefore we have
\bea
\al = \al_t +\al_s, \qquad \be = \al_s -\frac{p}{q+r} \al_t.
\nonu
\eea

Regarded as a $28 \times 28$ matrix, $X^{+IJKL}$ has eigenvalues $\al, \be,
\ga,$ $\de=(\al +\be)/2, \la=(\al+\ga)/2, \rho=(\be +\ga)/2$
with degeneracies $d_{\al}, d_{\be}, d_{\ga},
d_{\de}, d_{\la}$ and $d_{\rho}$
respectively.
The eigenvalues and eigenspaces of the $SO(p)^{+} \times SO(q)^{+}
\times SO(r)^{+}$
invariant tensor are summarized in Table 5.
By introducing projectors, $P_{\al}, P_{\be}, P_{\ga}, P_{\de}, P_{\la}$
and $P_{\rho}$
onto corresponding eigenspaces, we have a $28 \times 28$ matrix equation
\bea
X^{+IJKL} =\sum_{\si =\al, \be, \ga, \de, \la, \rho} \si P_{\si}^{IJKL}.
\nonu
\eea
Projector
$P_{\al}(P_{\be})[P_{\ga}]$
projects the $SO(8)$ Lie algebra onto its $SO(p)^{+}
(SO(q)^{+})[SO(r)^{+}]$
subalgebra  while $P_{\de}(P_{\la})[P_{\rho}]$ projects onto the remainder
$\frac{SO(8)}{SO(p)^{+} \times SO(q)^{+}}
(\frac{SO(8)}{SO(p)^{+} \times SO(r)^{+}})[
\frac{SO(8)}{SO(q)^{+} \times SO(r)^{+}}]$.
The projectors can be constructed from
$X^{+IJKL}$,
\bea
P_{\si} & = & \prod_{\si' \neq \si}
\frac{1}{(\si'-\si)} \left( \si'  \delta_{IJ}^{MN} -
 X^{+IJMN}  \right),
\qquad \mbox{for} \qquad \si =\al, \be, \ga, \de, \la, \rho
\label{productproj}
\eea
and it is easily checked that they satisfy
\bea
P_{\si}^2 = P_{\si}, \qquad
P_{\si} P_{\si'}
= 0( \si \neq \si') \qquad \mbox{where} \qquad \si, \si'=\al, \be, \ga,
\de, \la, \rho.
\label{projector1}
\eea
Then using the relation obtained by the properties of projectors above
\bea
\left[ \exp(-s X^{+}_s) \right]^{IJKL}
\left[ \exp(-t X^{+}_t) \right]^{KLMN}  =
\sum_{\si' =\al', \be', \ga'}e^{-\si' s} P_{\si'}^{IJKL}
\sum_{\si =\al, \be, \ga}e^{-\si t} P_{\si}^{KLMN}
\nonu
\eea
one gets
\bea
g(s, t) A_{\mu}^{IJ}(s,t) & \equiv &
g e^{\al t} e^{\al' s} \left[ \exp(-s X^{+}_s) \right]^{IJKL}
\left[ \exp(-t X^{+}_t) \right]^{KLMN}
 A_{\mu}^{MN}  \nonu \\
& = & g e^{\al t} e^{\al' s}
\sum_{\si' =\al', \be', \ga'}e^{-\si' s} P_{\si', s}^{IJKL}
\sum_{\si =\al, \be, \ga}e^{-\si t} P_{\si, t}^{KLMN}
A_{\mu}^{MN} \nonu
\eea
which will be the same as  (\ref{result}) together with
$A_{\mu (\si)}^{IJ} \equiv  P_{\si}^{IJKL} A_{\mu}^{KL}$
for $ \si =\al, \be, \ga, \de, \la, \rho$.
In this section, the main difference with the previous section
is that we started with
projectors directly constructed from $SO(p)^{+}
\times SO(q)^{+} \times SO(r)^{+}$ invariant four-form tensor.
Of course, these projectors are very complicated expressions because
they are fifth power of $X^{+IJKL}$ or $\de_{IJ}^{KL}$ given in
(\ref{productproj}). In the previous section, according to (\ref{relation}),
we identified
the product of projectors in $SO(p, q+r)^{+}$ and $SO(p+q, r)^{+}$
with a single projector (\ref{productproj}) in this section.

\section{Conclusion }

In summary,

$\bullet$ the main results in section 2 is described by
(\ref{first}). There are BPS domain-wall solutions interpolating between
a maximally supersymmetric $SO(8)$ critical point and various
nonsupersymmetric ones.

$\bullet$
The analytic solution is available for only $p=q=4$
with general $\xi$. For $\xi=0$, we also have solutions for general
$(p,q)$. That is,  for $SO(4,4)^{+}$ and $CSO(p,8-p)^{+}$ gaugings
where $p=1, \cdots, 7$ there exist  analytic solutions. Among these
gaugings, only $SO(4,4)^{+}$ and $CSO(2,6)^{+}$ cases contain critical points
according to Table 2. Note that the presence of domain-wall solutions
do not have any critical points.

$\bullet$
In section 3, the crucial part is to obtain a  T-tensor as found in
(\ref{tprimepqr}). Although it is rather complicated and involved,
all the components of a T-tensor can be obtained from the information on
both the projectors and 28-beins established by $SO(p)^{+} \times SO(q)^{+}
\times SO(r)^{+}$-singlet space. The only nontrivial gauging comparable to
that in Section 2 corresponds to both $\xi=-1$ and $\zeta=0$ that gives rise to
$CSO(p,q,r)^{+}$ gauging. The other values of $\xi$ and $\zeta$
gave us the previous gaugings discussed in Section 2.

$\bullet$
Finally, we arrived at
(\ref{generalsuperpotential}) and (\ref{pqrsuperpotential}) that is
a general expression, our new findings, for two scalar fields as the one
(\ref{superpotentialandpotential}) for one scalar field.

$\bullet$ Moreover,
similar domain-wall solutions are described by (\ref{solution}).
Although the scalar potential for this case looks different from the case of
$SO(p)^{+} \times SO(q)^{+}$, the structure of the critical points
are reduced to those in $SO(p)^{+} \times SO(q)^{+}$-invariant sector.
We emphasize that although $CSO(p,q,r)^{+}$ gaugings(in this case $\xi=-1$
and $\zeta=0$)
do not have any
critical points analyzed in section 3.3, they {\it do}
have domain wall solutions
and even possess analytic solutions for particular
$(p,q,r)$ combinations.

\bea
\begin{array}{|c|c|c|c|c|c||c|c|c|c|c|c||c|c|c|c|}
\hline
p & q & r & \al & \be & \ga & \de & \la & \rho & d_{\al} &
d_{\be} & d_{\ga} & d_{\de}& d_{\la}& d_{\rho} & |X^{+}|^2 \nonu \\
\hline
 1 & 1 & 6 & -2 & 6/7 & 10/21 & -10/7 & -16/21 &-4/21
&0 & 0  & 15 & 1  & 6 &6 & 64/7 \nonu \\
\hline
1 & 2 &5 &-2  & -6/7 &  26/35 & -10/7 &-22/35  & -2/35
&0 &1 &10 &2 &5 &10 & 432/35 \nonu \\
\hline
2 & 1 &5 & -2 & -2/3  & 14/15 & -4/3 & -8/15 &  2/15
&1 &0 &10 & 2 & 10&5 & 96/5 \nonu \\
\hline
1 &3  &4 & -2 & -6/7 & 8/7  & -10/7 & -3/7 &  1/7
& 0 & 3&6 &3 &4 &12 & 120/7 \nonu \\
\hline
2 & 2 &4 & -2 & -2/3 & 2  & -4/3 & 0 &  2/3
&1  &1 &6 &4 & 8& 8&32 \nonu \\
\hline
3 & 1 &4 & -2 & -2/5 & 8/5 & -6/5 & -1/5 &  3/5
&3 &0 &6 & 3 &12 & 4& 168/5 \nonu \\
\hline
1 & 4 &3 & -2 &-6/7  & 38/21 &-10/7  &-2/21  &  10/21
& 0 &6 &3 & 4 & 3&12 & 176/7 \nonu \\
\hline
 2 & 3 & 3 & -2 & -2/3 &2  & -4/3 & 0 & 2/3
& 1 &3 &3 &6 &6 &9  & 32 \nonu \\
\hline
3 & 2 &3 & -2 & -2/5 & 34/15 & -6/5 & 2/15 & 14/15
& 3 &1 &3 &6 &9 &6 & 208/5  \nonu \\
\hline
4 & 1 &3 &-2  & 0 & 8/3  & -1 & 1/3  &  4/3
& 6& 0&3 & 4 &12 &3 & 56 \nonu \\
\hline
1 &5  &2 &-2  & -6/7 & 22/7 & -10/7 & 4/7 &  8/7
& 0 &10 &1 &5 &2 &10 & 288/7\nonu \\
\hline
2 & 4 &2 & -2 & -2/3 & 10/3  &-4/3  & 2/3 &  4/3
& 1 & 6& 1& 8& 4& 8& 48 \nonu \\
\hline
3 & 3 &2 & -2 & -2/5 & 18/5 & -6/5 & 4/5 &   8/5
& 3& 3& 1&9  &6 & 6& 288/5  \nonu \\
\hline
4 & 2 &2 & -2 &0 & 4  & -1 & 1 &  2
&   6& 1& 1 &8 &8 &4&72 \nonu \\
\hline
 5 & 1 & 2 & -2 & 2/3 & 14/3  & -2/3 & 4/3 & 8/3
& 10 & 0 &1 & 5& 10& 2 & 96 \nonu \\
\hline
1 & 6 &1 &-2  & -6/7 & 50/7  & -10/7 & 18/7 &  22/7
& 0& 15&0 &6 &1 &6 & 624/7  \nonu \\
\hline
2 & 5 &1 &-2  & -2/3 & 22/3 & -4/3 & 8/3 &  10/3
&1 & 10& 0&10  &2 & 5& 96 \nonu \\
\hline
3 &4  &1 & -2 & -2/5 & 38/5 &-6/5  & 14/5 &  18/5
& 3 &6 &0 &12 &3 &4 & 528/5 \nonu \\
\hline
4 & 3 &1 & -2 & 0 & 8 & -1 & 3 &  4
& 6 &3 & 0& 12& 4& 3& 120 \nonu \\
\hline
5 & 2 &1 & -2 & 2/3 & 26/3 & -2/3 & 10/3 &   14/3
& 10&1 &0 & 10 & 5&2 & 144 \nonu \\
\hline
6 & 1 &1 & -2 & 2 & 10 & 0 & 4 &  6
&   15& 0& 0 &6 & 6& 1 & 192 \nonu \\
\hline
\end{array}
\nonu
\eea
Table 5. \sl Eigenvalues and eigenspaces of the $SO(p)^{+}
\times SO(q)^{+} \times SO(r)^{+}$
invariant tensor, $X^{+}$ where $|X^{+}|^2= \sum_{\si =\al,
\be, \ga, \de, \la, \rho} d_{\si}|\si|^2$.
The degeneracies are given in
$d_{\al} =p(p-1)/2, d_{\be}=q(q-1)/2, d_{\ga}=r(r-1)/2, d_{\de}=pq, d_{
\la}=pr$ and $d_{\rho}=qr$.
In \cite{freetal}, they displayed the signature
of the Killing-Cartan form by writing the numbers $n_{+}, n_{-}$
and $n_0$ of its positive, negative and zero eigenvalues. Here we
identify $d_{\al} + d_{\be}$ with $n_{+}$, $d_{\de}$
with $n_{-}$ and $d_{\ga}+ d_{\la} +d_{\rho}$
with $n_{0}$.
\rm

Recently, 11-dimensional embedding \cite{cpw,ai} of supersymmetric vacua
of compact-gauged supergravity was found. For solutions with varying
scalars(due to the $r$-dependence of vacuum expectation values), the
ansatz for the field strength was more complicated. In this direction
it was crucial to know about the 11-dimensional analog of superpotential,
so-called geometric superpotential, in order to achieve the M-theory
lift of the RG flow. Provided that the $r$-dependence
of the vevs is controlled by the RG flow equations, an exact solution
to the 11-dimensional Einstein-Maxwell equations was obtained.
As mentioned in
the introduction, the 11-dimensional origin of $SO(p,q)^{+}$ and $CSO(p,q)^{+}$
gaugings was found in \cite{hullwarner3} for constant scalars.
In this paper, we describe explict $r$-dependence on the vevs by domain-wall
solutions. It is natural to ask whether 11-dimensional embedding
of various vacua we have considered of non-compact and non-semi-simple
gauged supergravity can be obtained. In a recent paper \cite{hullgibbons},
the metric on the 7-dimensional internal space and domain wall in
11-dimensions was found. However, they did not provide an ansatz for
an 11-dimensional
three-form gauge field. It would be interesting to study the
geometric superpotential,
11-dimensional analog of superpotential we have obtained.
We expect that the nontrivial $r$-dependence of vevs makes
Einstein-Maxwell equations consistent not only at the critical points
but also along the supersymmetric RG flow connecting two critical points.

In \cite{warner}, all critical points of the scalar potential of
the ${\cal N}=8$ supergravity with $SO(8)$
gauge symmetry that break the local $SO(8)$ down to a solution with
symmetry that is at least
some specified subgroup of $SO(8)$ were found. One considers only
those scalars which are
singlets of that subgroup and searches critical points of the
potential restricted to be a function  only of
the singlets. Schurr's lemma tells us that any critical point of
restricted potential will be a critical point
of the original complete scalar potential. Then the problem of
finding critical points of the potential
is reduced to the simpler one of finding critical points of the
restricted potential which is a singlet
sector.
In this paper,  we  applied similar techniques to the non-compact
and non-semi-simple gauged
supergravities and
the subgroup is to be $SO(p)^{+} \times SO(q)^{+}$ for the $SO(p, q)^{+}$
gaugings and $CSO(p,q)^{+}$ gaugings while that will be
$SO(p)^{+} \times SO(q)^{+} \times SO(r)^{+}$ for the
$CSO(p,q, r)^{+}$ gaugings.

In \cite{warner}, the specified subgroup
$H$ was taken to be $SU(3)$ for $SO(8)$ gauged supergravity.
One can think of the $H$ subgroup as a compact subgroup of
$SO(p, q)^{+}$ gauged model because this is necessary to the validity of
Schurr's lemma. When 56-beins commute the $SL(8, {\bf R})$ transformation
$E(t)$, it is rather easy to calculate the scalar potential. However,
it may happen that for the noncommutativity of 56-beins
${\cal V}$ and $E(t)$, it will
be rather complicated to find the scalar potential because of the
presence of additional Baker-Hausdorff terms appearing in the
calculations of exponentials of matrices.
According to \cite{hullwarner2}, it was found that no
$G_2$-invariant critical points exist for $SO(7,1)^{+}$ gauging, no
$SU(3)$-invariant critical points for $SO(6,2)^{+}$ gauging
and a $SO(5)$-invariant critical point with positive cosmological
constant, and no supersymmetry  for $SO(5,3)^{+}$ gauging. It would be
interesting to investigate whether there exist any critical points
of the potential restricted to the $H$-singlet sector for the most
general $CSO(p, q, r)^{+}$ gaugings we have considered in this paper.
Here group $H$ is a compact subgroup of this model.

\section{Appendix A: Four-form (Anti)Self-dual
Tensors in $28 \times 28$ Matrices  }

Let us consider the $SO(p)^{-} \times SO(q)^{-}$ invariant generator
of $SL(8, {\bf R})$,
\bea
X_{\dot{a} \dot{b}} & = & \left(
\begin{array}{ccc} \alpha \textbf{1}_{p \times p
} & 0 \\ 0 & \beta \textbf{1}_{q \times q}
\end{array} \right) \;\;\; \mbox{with} \;\;\;
\alpha p + \beta q = 0, \qquad p + q = 8,
\nonu
\eea
where
${\bf 1}_{p \times p}$ is a $p \times p$ identity matrix.
The embedding of this $SL(8, {\bf R})$ in $E_7$ is such that
$X_{\dot{a} \dot{b}}$
corresponds to the $56 \times 56$ $E_7$ generator with $X^{-IJKL}$
\bea
\left(
\begin{array}{cc}
0 &  \  X^{-IJKL}  \\
  X_{IJKL}^{-} & 0
\end{array} \right),
\nonu
\eea
where the real, anti-self-dual totally anti-symmetric tensor
$X^{-IJKL}$ is given by the following form through the $\widetilde{\Ga}$
matrix
\bea
X_{IJKL}^{-} =-\frac{1}{8}\left(
{\widetilde{\Gamma}}_{IJKL}\right)^{\dot{a} \dot{b}}X_{\dot{a} \dot{b}}
\label{xgammatilde}
\eea
where $ {\widetilde{\Gamma}}_{IJKL} =
{\widetilde{\Gamma}}_{[ I} {\widetilde{\Gamma}}_{J}
{\widetilde{\Gamma}}_{K} {\widetilde{\Gamma}}_{L ]}$ as
in section 2.5  and
an arbitrary $SO(8)$ generator $L_{IJ}$ acts in the left-handed
spinor representation(See Appendix B for this representation)
by $(L_{IJ} {\widetilde{\Gamma}}_{IJ})^{\dot{a} \dot{b}}$.
When $p=7$ and $q=1$, one can see that
this expression of (\ref{xgammatilde}) through
$\widetilde{\Gamma} $ matrix coincides exactly with
the one in section 2.6 or $X_{7,1}^{-IJKL}$ presented
below explicitly.

We have seen real (anti) self-dual tensors in the $SU(8)$-basis through
$\Ga$ matrices in (\ref{xgamma}) and (\ref{xgammatilde}).
Now one can express them as the following
forms which will be a useful and illuminating description,
viewed as a $28 \times 28$
matrix representation,
after doing the $\Ga$ matrix algebra
\bea
X^{\pm IJKL}_{p,q}=Y^{IJKL}_{p,q}+\frac{\eta}{24}
\epsilon^{IJKLMNPQ}Y^{MNPQ}_{p,q},
\nonu
\eea
where self-duality $+$ corresponds to $\eta =1$ and anti-self-duality
$-$ corresponds to $\eta=-1$ and $Y^{IJKL}_{p, q}$
tensors are given for each
$p$ and $q$ in
\bea
Y^{IJKL}_{7,1} & = & \frac{1}{2} \left(
\delta^{IJKL}_{1\;2\;3\;4}+\delta^{IJKL}_{1\;2\;5\;6}+
\delta^{IJKL}_{1\;2\;7\;8}+\delta^{IJKL}_{1\;3\;7\;5}+
\delta^{IJKL}_{1\;3\;6\;8}
+\delta^{IJKL}_{1\;4\;5\;8}+\delta^{IJKL}_{1\;4\;6\;7} \right),
\nonu \\
Y^{IJKL}_{6,2} & = & \frac{1}{2} \left(
\delta^{IJKL}_{1\;2\;3\;4}+\delta^{IJKL}_{1\;2\;5\;6}+
\delta^{IJKL}_{1\;2\;7\;8} \right),
\nonu \\
Y^{IJKL}_{5,3} & = & \frac{1}{6}\left(
3 \delta^{IJKL}_{1\;2\;3\;4}+\delta^{IJKL}_{1\;2\;5\;6}+
\delta^{IJKL}_{1\;2\;7\;8}+\delta^{IJKL}_{1\;5\;3\;7}+
\delta^{IJKL}_{1\;3\;6\;8}
+\delta^{IJKL}_{1\;5\:4\;8}+\delta^{IJKL}_{1\;6\;4\;7} \right),
\nonu \\
Y^{IJKL}_{4,4} & = & \frac{1}{2}
\delta^{IJKL}_{1\;2\;3\;4},
\nonu \\
Y^{IJKL}_{3,5} & = & \frac{1}{10}\left(
3 \delta^{IJKL}_{1\;2\;3\;4}+\delta^{IJKL}_{1\;5\;2\;6}+
\delta^{IJKL}_{1\;2\;7\;8}+\delta^{IJKL}_{1\;3\;5\;7}+
\delta^{IJKL}_{1\;3\;6\;8}
+\delta^{IJKL}_{1\;4\;5\;8}+\delta^{IJKL}_{1\;6\;4\;7} \right),
\nonu \\
Y^{IJKL}_{2,6} & = & \frac{1}{6}\left(
\delta^{IJKL}_{1\;2\;3\;4}+\delta^{IJKL}_{1\;5\;2\;6}+
\delta^{IJKL}_{1\;2\;7\;8} \right),
\nonu \\
Y^{IJKL}_{1,7} & = & \frac{1}{14}\left(
\delta^{IJKL}_{1\;2\;3\;4}+\delta^{IJKL}_{1\;5\;2\;6}+
\delta^{IJKL}_{1\;2\;7\;8}+\delta^{IJKL}_{1\;3\;5\;7}+
\delta^{IJKL}_{1\;3\;6\;8}
+\delta^{IJKL}_{1\;5\;4\;8}+\delta^{IJKL}_{1\;4\;6\;7} \right).
\nonu
\eea
Actually the case of $X^{\pm IJKL}_{5,3}$ can be identified with
$SO(5)^{\pm}$-singlets among six scalars \cite{romans}
when restricted to equal real
parameters($\phi_{IJKL}$ depends on only three real parameters because
of $SO(3)^{\pm}$ rotation).

\section{Appendix B: $SO(8)$ $\Ga$ Matrices and Its Representations   }

The 28 $SO(8)$ generators are denoted by
$\La_{MN}$ where $M, N=1, 2, \cdots, 8$ and
they can be decomposed into
$\La_{MN} = (\La_{mn}, \La_{m1})$.
Here $\La_{mn}= - \La_{nm}$ where $m, n= 2, 3,
\cdots, 8$ are the 21 generators of $SO(7)$. Then the $8 \times 8$ $SO(7)$
gamma matrices satisfy $\{ \Ga_m, \Ga_n \} = -2 \de_{mn}$ and the generators
act on the 8-dimensional spinor representation of $SO(7)$
by $\frac{1}{4} \La^{mn} \Ga_{mn}$.
Then the $16 \times 16$ $SO(8)$ gamma matrices
have the following form, $\ga_{MN}= \mbox{diag} ((\Ga_{MN})^{ab},
({\widetilde{\Ga}}_{MN})^{\dot{a} \dot{b}})$ where
\bea
\Ga_{MN} = {\widetilde{\Ga}}_{MN} =\Ga_{mn}, \qquad M, N= 2, 3, \cdots, 8,
\qquad \Ga_{M1} = - {\widetilde{\Ga}}_{M1} =  \Ga_m
\nonu
\eea
and $a,b$ are right-handed spinor indices and $\dot{a}, \dot{b}$ are
left-handed spinors.
The $SO(8)$ has three different eight-dimensional representations:
the vector representation ${\bf 8}_v$ generated by $\La_{MN}$,
the right-handed spinor representation  ${\bf 8}_s$ generated by
$\frac{1}{4} \La^{mn} \Ga_{mn}$, and
the left-handed spinor representation  ${\bf 8}_c$ generated by
$\frac{1}{4} \La^{mn} {\widetilde{\Ga}}_{mn}$.
This induces three inequivalent $SO(7)$ subgroups of $SO(8)$.
That is,
the stability group of the vector, $SO(7)$ is generated by $\La_{MN}$,
$M,N=2, 3, \cdots, 8$, the stabilizer of a right -handed spinor,
$SO(7)^{+}$ is generated by  $\La^{MN} \Ga_{MN}$, and
the stabilizer of a left-handed spinor,
$SO(7)^{-}$ is generated by  $\La^{MN} {\widetilde{\Ga}}_{MN}$.
The $SO(7)^{+}$-singlet under the branching rule of 35-dimensional
fourth rank self-dual antisymmetric tensor representation  of $SO(8)$
into $SO(7)^{+}$ corresponds to the $SO(7)^{+}$-invariant tensor
$X^{+IJKL}$ given in Section 2.3.
Moreover, we present explicit realizations of $\Gamma$ matrices
we are using here as follows \cite{gso,cremmer}:
\bea
&&\Gamma^2=\left(
\begin{array}{ccc}
\al^3 & 0 \\ 0 & -\al^3
\nonu
\end{array} \right),
\Gamma^3=\left(
\begin{array}{ccc}
 \al^2& 0 \\ 0 & -\al^2
\nonu
\end{array} \right),
\Gamma^4=\left(
\begin{array}{ccc}
\al^1 & 0 \\ 0 & -\al^1
\nonu
\end{array} \right), \nonu \\
&&\Gamma^5=\left(
\begin{array}{ccc}
0 & {\bf 1} \\ -{\bf 1} & 0
\nonu
\end{array} \right),
\Gamma^6=\left(
\begin{array}{ccc}
0 & -\be^3 \\ -\be^3 & 0
\nonu
\end{array} \right),
\Gamma^7=\left(
\begin{array}{ccc}
0  & \be^2 \\ \be^2 & 0
\nonu
\end{array} \right),
\Gamma^8=\left(
\begin{array}{ccc}
0 & \be^1 \\ \be^1 & 0
\nonu
\end{array} \right),
\eea
where $\al^i$'s and $\be^i$'s are given in terms of usual $2 \times 2$
Pauli matrices $\si^i$'s
\bea
&&\al^1 = \left(
\begin{array}{ccc}
0 & \si^1 \\ -\si^1 & 0
\nonu
\end{array} \right),
\al^2= \left(
\begin{array}{ccc}
0  & -\si^3 \\ \si^3 & 0
\nonu
\end{array} \right),
\al^3 = \left(
\begin{array}{ccc}
i \si^2 &  0 \\ 0 & i \si^2
\nonu
\end{array} \right), \nonu \\
&&\be^1 = \left(
\begin{array}{ccc}
0 &  i\si^2 \\ i \si^2 & 0
\nonu
\end{array} \right),
\be^2 = \left(
\begin{array}{ccc}
0 &  1 \\ -1 & 0
\nonu
\end{array} \right),
\be^3 = \left(
\begin{array}{ccc}
-i \si^2 &  0 \\ 0 & i \si^2
\nonu
\end{array} \right).
\eea
\section{Appendix C: Some Identities between Invariant Generators and
Projectors in $SO(p)^{+} \times SO(q)^{+}$ Sectors }

For any $SO(p)^{+} \times SO(q)^{+}$ generator $\La_{(\al)}^{IJ}$,
 the invariance of $X^{+IJKL}$ under the $SO(p)^{+}$ implies
\bea
E(t)^{-1} \left(
\begin{array}{ccc}
\underline{\La}_{(\al)} &  0 \\ 0 & \underline{\La}_{(\al)}
\nonu
\end{array} \right) E(t) =
\left(
\begin{array}{ccc}
\underline{\La}_{(\al)} &  0 \\ 0 & \underline{\La}_{(\al)}
\nonu
\end{array} \right) ,
\nonu
\eea
which is equivalent to
\bea
[P_{\al}, \underline{\La}_{(\al)}] =[P_{\be}, \underline{\La}_{(\al)}]=
[P_{\ga}, \underline{\La}_{(\al)}] = 0.
\nonu
\eea
Similarly,
the invariance of $X^{+IJKL}$ under the $SO(q)^{+}$ implies
\bea
E(t)^{-1} \left(
\begin{array}{ccc}
\underline{\La}_{(\be)} &  0 \\ 0 & \underline{\La}_{(\be)}
\nonu
\end{array} \right) E(t) =
\left(
\begin{array}{ccc}
\underline{\La}_{(\be)} &  0 \\ 0 & \underline{\La}_{(\be)}
\nonu
\end{array} \right) ,
\nonu
\eea
which will lead to vanishing of commutators between $P_{\al, \be, \ga}$
and $\underline{\La}_{(\be)}$
\bea
[P_{\al}, \underline{\La}_{(\be)}] =[P_{\be}, \underline{\La}_{(\be)}]=
[P_{\ga}, \underline{\La}_{(\be)}] =0.
\nonu
\eea
One gets
the following identities
\bea
&&P_{\al} \underline{\La}_{(\al)} P_{\ga} =
P_{\be} \underline{\La}_{(\al)} P_{\ga}=
P_{\al} \underline{\La}_{(\be)} P_{\ga}=
P_{\be} \underline{\La}_{(\be)} P_{\ga}= 0,\nonu \\
&& P_{\ga} \underline{\La}_{(\al)} P_{\al} =
P_{\ga} \underline{\La}_{(\al)} P_{\be}=
P_{\ga} \underline{\La}_{(\be)} P_{\al}=
P_{\ga} \underline{\La}_{(\be)} P_{\be}= 0, \nonu \\
&& P_{\be} \underline{\La}_{(\al)} P_{\al} =
P_{\be} \underline{\La}_{(\be)} P_{\al}=
P_{\al} \underline{\La}_{(\al)} P_{\be}=
P_{\al} \underline{\La}_{(\be)} P_{\be}= 0.
\label{iden1}
\eea
Moreover, one gets for the $SO(8)/(SO(p)^{+} \times SO(q)^{+})$
generator $\La_{(\ga)}$
\bea
&&
P_{\al} \underline{\La}_{(\ga)} P_{\al} =
P_{\be} \underline{\La}_{(\ga)} P_{\al}=
P_{\al} \underline{\La}_{(\ga)} P_{\be}=
P_{\be} \underline{\La}_{(\ga)} P_{\be}=
P_{\ga} \underline{\La}_{(\ga)} P_{\ga}= 0.
\label{ga}
\eea
With $1 =P_{\al} + P_{\be} + P_{\ga}$, the combinations of
(\ref{ga}) will give us
\bea
P_{\al} \underline{\La}_{(\ga)} P_{\ga} & = & P_{\al} \underline{\La}_{(\ga)},
\qquad
P_{\ga} \underline{\La}_{(\ga)} P_{\al} = \underline{\La}_{(\ga)} P_{\al},
\nonu \\
P_{\be} \underline{\La}_{(\ga)} P_{\ga} & =& P_{\be} \underline{\La}_{(\ga)},
\qquad
P_{\ga} \underline{\La}_{(\ga)} P_{\be} = \underline{\La}_{(\ga)} P_{\be}.
\label{identities}
\eea
By combining the first(second) and third(fourth) relations of
(\ref{identities}) respectively and using (\ref{ga})
it is easily checked that
\bea
\left( P_{\al} + P_{\be} \right)  \underline{\La}_{(\ga)} =
\underline{\La}_{(\ga)} P_{\ga},\qquad
\underline{\La}_{(\ga)} \left( P_{\al} + P_{\be} \right) =
 P_{\ga} \underline{\La}_{(\ga)}.
\nonu
\eea

\section{Appendix D: Some Identities between Invariant Generators and
Projectors  in $SO(p)^{+} \times SO(q)^{+} \times SO(r)^{+}$ Sectors}

For any $SO(p)^{+} \times SO(q)^{+} \times SO(r)^{+}$
generator $\La_{(\al)}^{IJ}$,
the invariance of $X^{+IJKL}$ under the $SO(p)^{+}$ implies
\bea
E(t)^{-1} \left(
\begin{array}{ccc}
\underline{\La}_{(\al)} &  0 \\ 0 & \underline{\La}_{(\al)}
\nonu
\end{array} \right) E(t) =
\left(
\begin{array}{ccc}
\underline{\La}_{(\al)} &  0 \\ 0 & \underline{\La}_{(\al)}
\nonu
\end{array} \right) ,
\nonu
\eea
which is equivalent to
\bea
[P_{\si}, \underline{\La}_{(\al)}] =
0 \qquad \mbox{for} \qquad \si =\al, \be, \ga, \de, \la, \rho.
\nonu
\eea
Similarly,
the invariance of $X^{+IJKL}$ under the $SO(q)^{+}$ implies
\bea
E(t)^{-1} \left(
\begin{array}{ccc}
\underline{\La}_{(\be)} &  0 \\ 0 & \underline{\La}_{(\be)}
\nonu
\end{array} \right) E(t) =
\left(
\begin{array}{ccc}
\underline{\La}_{(\be)} &  0 \\ 0 & \underline{\La}_{(\be)}
\nonu
\end{array} \right) ,
\nonu
\eea
which will lead to
\bea
[P_{\si}, \underline{\La}_{(\be)}] =
0 \qquad \mbox{for} \qquad \si=\al, \be, \ga, \de, \la, \rho.
\nonu
\eea
Similarly,
the invariance of $X^{+IJKL}$ under the $SO(r)^{+}$ implies
\bea
E(t)^{-1} \left(
\begin{array}{ccc}
\underline{\La}_{(\ga)} &  0 \\ 0 & \underline{\La}_{(\ga)}
\nonu
\end{array} \right) E(t) =
\left(
\begin{array}{ccc}
\underline{\La}_{(\ga)} &  0 \\ 0 & \underline{\La}_{(\ga)}
\nonu
\end{array} \right) ,
\nonu
\eea
which will lead to
\bea
[P_{\si}, \underline{\La}_{(\ga)}] =
0 \qquad \mbox{for} \qquad \si=\al, \be, \ga, \de, \la, \rho.
\nonu
\eea
Using the relations (\ref{projector1}), one gets
the following identities
\bea
P_{\si} \underline{\La}_{(\al)} P_{\si'} =
P_{\si} \underline{\La}_{(\be)} P_{\si'} =
P_{\si} \underline{\La}_{(\ga)} P_{\si'} = 0, \qquad
\mbox{for } \qquad \si, \si' = \al, \be, \ga, \de, \la , \rho
\qquad \si \neq \si'.
\nonu
\eea
Moreover, one gets for the $SO(8)/(SO(p)^{+} \times SO(q)^{+})$
generator $\La_{(\de)}^{IJ}$
\bea
P_{\al} \underline{\La}_{(\de)} P_{\si} & = & 0,
\qquad \si= \al, \be, \ga, \la,
\rho \nonu \\
P_{\be} \underline{\La}_{(\de)} P_{\si}& = & 0,
\qquad \si= \al, \be, \ga, \la,
\rho \nonu \\
P_{\ga} \underline{\La}_{(\de)} P_{\si} & = & 0,
\qquad \si= \al, \be, \de, \la,
\rho \nonu \\
P_{\de} \underline{\La}_{(\de)} P_{\si} & = & 0,
\qquad \si= \ga, \de, \la,
\rho \nonu \\
P_{\la} \underline{\La}_{(\de)} P_{\si} & = & 0,
\qquad \si= \al, \be, \ga, \de,
\la \nonu \\
P_{\rho} \underline{\La}_{(\de)} P_{\si} & = & 0,
\qquad \si= \al, \be, \ga, \de, \rho.
\label{property1}
\eea
With $1 =\sum_{\si = \al, \be, \ga, \de, \la, \rho}
P_{\si} $, the combinations of
(\ref{property1}) will give us
\bea
&&P_{\al} \underline{\La}_{(\de)} P_{\de}  =  P_{\al} \underline{\La}_{(\de)},
\qquad
P_{\be} \underline{\La}_{(\de)} P_{\de}  = P_{\be} \underline{\La}_{(\de)},
\qquad
P_{\ga} \underline{\La}_{(\de)} P_{\ga}  = P_{\ga} \underline{\La}_{(\de)},
\nonu \\
&&P_{\de} \underline{\La}_{(\de)} P_{\al} = \underline{\La}_{(\de)} P_{\al},
\qquad
P_{\de} \underline{\La}_{(\de)} P_{\be} = \underline{\La}_{(\de)} P_{\be},
\qquad
P_{\la} \underline{\La}_{(\de)} P_{\rho}  = P_{\la} \underline{\La}_{(\de)},
\nonu \\
&&
P_{\rho} \underline{\La}_{(\de)} P_{\la}  = P_{\rho} \underline{\La}_{(\de)}.
\label{identity1}
\eea
Moreover, one gets for the $SO(8)/(SO(p)^{+} \times SO(r)^{+})$
generator $\La_{(\la)}^{IJ}$
\bea
P_{\al} \underline{\La}_{(\la)} P_{\si} & = & 0,
\qquad \si= \al, \be, \ga, \de,
\rho \nonu \\
P_{\be} \underline{\La}_{(\la)} P_{\si}& = & 0,
\qquad \si= \al, \ga, \de, \la,
\rho \nonu \\
P_{\ga} \underline{\La}_{(\la)} P_{\si} & = & 0,
\qquad \si= \al, \be, \ga, \de,
\rho \nonu \\
P_{\de} \underline{\La}_{(\la)} P_{\si} & = & 0,
\qquad \si= \al, \be, \ga, \de,
\la \nonu \\
P_{\la} \underline{\La}_{(\la)} P_{\si} & = & 0,
\qquad \si= \be, \de, \la,
\rho \nonu \\
P_{\rho} \underline{\La}_{(\la)} P_{\si} & = & 0,
\qquad \si= \al, \be, \ga, \la, \rho.
\label{property2}
\eea
With $1 =\sum_{\si = \al, \be, \ga, \de, \la, \rho}
P_{\si} $, the combinations of
(\ref{property2}) will give us
\bea
&&P_{\al} \underline{\La}_{(\la)} P_{\la}  =  P_{\al} \underline{\La}_{(\la)},
\qquad
P_{\be} \underline{\La}_{(\la)} P_{\be}  = P_{\be} \underline{\La}_{(\la)},
\qquad
P_{\ga} \underline{\La}_{(\la)} P_{\la}  = P_{\ga} \underline{\La}_{(\la)},
\nonu \\
&&P_{\de} \underline{\La}_{(\la)} P_{\rho}  = P_{\de} \underline{\La}_{(\la)},
\qquad
P_{\la} \underline{\La}_{(\la)} P_{\al} = \underline{\La}_{(\la)} P_{\al},
\qquad
P_{\la} \underline{\La}_{(\la)} P_{\ga} = \underline{\La}_{(\la)} P_{\ga},
\nonu \\
&&
P_{\rho} \underline{\La}_{(\la)} P_{\de}  = P_{\rho} \underline{\La}_{(\la)}.
\label{identity2}
\eea
Moreover, one gets for the $SO(8)/(SO(q)^{+} \times SO(r)^{+})$
generator $\La_{(\rho)}^{IJ}$
\bea
P_{\al} \underline{\La}_{(\rho)} P_{\si} & = & 0,
\qquad \si= \be, \ga, \de,
\la, \rho \nonu \\
P_{\be} \underline{\La}_{(\rho)} P_{\si}& = & 0,
\qquad \si= \al, \be, \ga, \de,
\la \nonu \\
P_{\ga} \underline{\La}_{(\rho)} P_{\si} & = & 0,
\qquad \si= \al, \be, \ga, \de, \la \nonu \\
P_{\de} \underline{\La}_{(\rho)} P_{\si} & = & 0,
\qquad \si= \al, \be, \ga, \de, \rho \nonu \\
P_{\la} \underline{\La}_{(\rho)} P_{\si} & = & 0,
\qquad \si= \al, \be, \ga, \la,
\rho \nonu \\
P_{\rho} \underline{\La}_{(\rho)} P_{\si} & = & 0,
\qquad \si= \al, \de, \la,
\rho.
\label{property3}
\eea
With $1 =\sum_{\si = \al, \be, \ga, \de, \la, \rho}
P_{\si} $, the combinations of
(\ref{property3}) will give us
\bea
&&P_{\al} \underline{\La}_{(\rho)} P_{\al}  =  P_{\al}
\underline{\La}_{(\rho)},
\qquad
P_{\be} \underline{\La}_{(\rho)} P_{\al}  = P_{\be} \underline{\La}_{(\rho)},
\qquad
P_{\ga} \underline{\La}_{(\rho)} P_{\rho}  = P_{\ga} \underline{\La}_{(\rho)},
\nonu \\
&&
P_{\de} \underline{\La}_{(\rho)} P_{\la}  = P_{\de} \underline{\La}_{(\rho)},
\qquad
P_{\la} \underline{\La}_{(\rho)} P_{\de}  = P_{\la} \underline{\La}_{(\rho)},
\qquad
P_{\rho} \underline{\La}_{(\rho)} P_{\ga} = \underline{\La}_{(\rho)} P_{\ga},
\nonu \\
&&
P_{\rho} \underline{\La}_{(\rho)} P_{\be} = \underline{\La}_{(\rho)} P_{\be}.
\label{identity3}
\eea
Using (\ref{identity1}), (\ref{identity2}) and (\ref{identity3})
it is easily checked that
\bea
&&\left( P_{\al} + P_{\be} \right)  \underline{\La}_{(\de)} =
\underline{\La}_{(\de)} P_{\de},\qquad
\underline{\La}_{(\de)} \left( P_{\al} + P_{\be} \right) =
 P_{\de} \underline{\La}_{(\de)}, \qquad
\left( P_{\al} + P_{\ga} \right)  \underline{\La}_{(\la)} =
\underline{\La}_{(\la)} P_{\la}, \nonu \\
&&\underline{\La}_{(\la)} \left( P_{\al} + P_{\ga} \right) =
 P_{\la} \underline{\La}_{(\la)}, \qquad
\left( P_{\be} + P_{\ga} \right)  \underline{\La}_{(\rho)} =
\underline{\La}_{(\rho)} P_{\rho},\qquad
\underline{\La}_{(\rho)} \left( P_{\be} + P_{\ga} \right) =
 P_{\rho} \underline{\La}_{(\rho)}, \nonu \\
&& P_{\ga}   \underline{\La}_{(\de)} = \underline{\La}_{(\de)} P_{\ga},
\qquad
 P_{\la}   \underline{\La}_{(\de)} = \underline{\La}_{(\de)} P_{\rho},
\qquad
 P_{\rho}   \underline{\La}_{(\de)} = \underline{\La}_{(\de)} P_{\la}, \nonu \\
&&  P_{\be}   \underline{\La}_{(\la)} = \underline{\La}_{(\la)} P_{\be},
\qquad
 P_{\de}   \underline{\La}_{(\la)} = \underline{\La}_{(\la)} P_{\rho},
\qquad
 P_{\rho}   \underline{\La}_{(\la)} = \underline{\La}_{(\la)} P_{\de}, \nonu \\
&& P_{\al}   \underline{\La}_{(\rho)} = \underline{\La}_{(\rho)} P_{\al},
\qquad
 P_{\de}   \underline{\La}_{(\rho)} = \underline{\La}_{(\rho)} P_{\la},
\qquad
 P_{\la}   \underline{\La}_{(\rho)} = \underline{\La}_{(\rho)} P_{\de}.
 \nonu
\eea

\section{Appendix E: 28-beins $ u^{IJ}_{\;\;\;KL}$ and
$v^{IJKL}$ for Each Invariant Sector }

The 28-beins  $ u_{IJ}^{\;\;\;KL}$ and
$v_{IJKL}$ fields can be obtained by exponentiating the
vacuum expectation values
$\phi_{IJKL}$. The nonzero components of those have the following
seven $4 \times 4$
block diagonal matrices respectively
\bea u_{IJ}^{\;\;\;KL} & = & \mbox{diag} \left(
u_{1},u_{2},u_{3},u_{4},u_{5},u_{6},u_{7} \right), \nonumber \\
v_{IJKL} & = & \mbox{diag}\left(
v_{1},v_{2},v_{3},v_{4},v_{5},v_{6},v_{7} \right). \nonu \eea
Each hermitian submatrix is a $4 \times 4$ matrix and we denote antisymmetric
indices explicitly for convenience. For simplicity, we make an
empty space corresponding
to lower triangle elements. We also denote $\varepsilon_{+} =1$(self-dual),
$\varepsilon_{-}=i$(anti-self-dual) and
$\eta=1$ corresponding to self-dual case or $-1$ anti-self dual case.
We write down here each hermitian matrices.

$\bullet$ $SO(7)^{\pm} \times SO(1)^{\pm}$ Invariant Sectors:


\bea && u_1 = \left(
\begin{array}{ccccc}
& \left[12 \right] & \left[34 \right] & \left[56 \right] &
\left[78 \right] \nonu \\
 \left[12 \right] &  A & \eta B & \eta B & \eta B \nonu \\
\left[34 \right]  &  & A & B &
B \nonu \\
 \left[ 56\right] &  &  & A &
B \nonu \\
 \left[ 78\right] &  &   &  & A \\
\end{array}
\right), u_2 = \left(
\begin{array}{ccccc}
& \left[13 \right] & \left[24 \right] & \left[57 \right] &
\left[68 \right] \nonu \\
 \left[13 \right] &  A & -\eta B & -\eta B & \eta B \nonu \\
\left[24 \right]  &  & A & B &
-B \nonu \\
 \left[ 57\right] &  &  & A &
-B \nonu \\
 \left[ 68\right] &  &   &  & A \\
\end{array}
\right), \nonu \\
&& u_3 = \left(
\begin{array}{ccccc}
& \left[14 \right] & \left[23 \right] & \left[58 \right] &
\left[67 \right] \nonu \\
 \left[14 \right] &  A & \eta B & \eta B & \eta B \nonu \\
\left[23 \right]  &  & A & B &
B \nonu \\
 \left[ 58\right] &  &  & A &
B \nonu \\
 \left[ 67\right] &  &  &  & A \\
\end{array}
\right), u_4 = \left(
\begin{array}{ccccc}
& \left[15 \right] & \left[26 \right] & \left[37 \right] &
\left[48 \right] \nonu \\
 \left[15 \right] &  A & -\eta B & \eta B &  -\eta B \nonu \\
\left[26 \right]  &  & A & -B &
B \nonu \\
 \left[ 37\right] &  &  & A &
-B \nonu \\
 \left[ 48\right] &  &  &  & A \\
\end{array}
\right), \nonu \\
&& u_5 = \left(
\begin{array}{ccccc}
& \left[16 \right] & \left[25 \right] & \left[38 \right] &
\left[47 \right] \nonu \\
 \left[16 \right] &  A & \eta B & -\eta B &  -\eta B \nonu \\
\left[25 \right]  &  & A & -B &
-B \nonu \\
 \left[ 38\right] &  &  & A &
B \nonu \\
 \left[ 47\right] &  &  &  & A \\
\end{array}
\right), u_6 = \left(
\begin{array}{ccccc}
& \left[17 \right] & \left[28 \right] & \left[35 \right] &
\left[46 \right] \nonu \\
 \left[17 \right] &  A & -\eta B & -\eta B &  \eta B \nonu \\
\left[28 \right]  &  & A & B &
-B \nonu \\
 \left[35 \right] &  &  & A &
-B \nonu \\
 \left[46 \right] &  &  &  & A \\
\end{array}
\right), \nonu \\
&& u_7 = \left(
\begin{array}{ccccc}
& \left[18 \right] & \left[27 \right] & \left[36 \right] &
\left[45 \right] \nonu \\
 \left[18 \right] &  A & \eta B & \eta B &  \eta B \nonu \\
\left[27 \right]  &  & A & B &
B \nonu \\
 \left[ 36\right] &  &  & A &
B \nonu \\
 \left[ 45\right] &  &  &  & A \\
\end{array}
\right),
 v_1 = -\varepsilon_{\pm} \left(
\begin{array}{ccccc}
& \left[12 \right] & \left[34 \right] & \left[56 \right] &
\left[78 \right] \nonu \\
 \left[12 \right] &  F & \eta G & \eta G & \eta G \nonu \\
\left[34 \right]  &   &  F &  G &
 G \nonu \\
 \left[ 56\right] &  &  &  F &
G \nonu \\
 \left[ 78\right] &  &  &  &  F \\
\end{array}
\right), \nonu \\
& & v_2 = -\varepsilon_{\pm}\left(
\begin{array}{ccccc}
& \left[13 \right] & \left[24 \right] & \left[57 \right] &
\left[68 \right] \nonu \\
 \left[13 \right] &  F & -\eta G & -\eta G & \eta G \nonu \\
\left[24 \right]  &  &  F &  G &
-G \nonu \\
 \left[ 57\right] &  &   &  F &
-G \nonu \\
 \left[ 68\right] &  &   &   &  F \\
\end{array}
\right),  v_3 = -\varepsilon_{\pm}\left(
\begin{array}{ccccc}
& \left[14 \right] & \left[23 \right] & \left[58 \right] &
\left[67 \right] \nonu \\
 \left[14 \right] &  F & \eta G & \eta G & \eta G \nonu \\
\left[23 \right]  &  &  F &  G &
 G \nonu \\
 \left[ 58\right] &  &  &  F &
 G \nonu \\
 \left[ 67\right] &  &  &  &  F \\
\end{array}
\right), \nonu \\
& & v_4 = -\varepsilon_{\pm}\left(
\begin{array}{ccccc}
& \left[15 \right] & \left[26 \right] & \left[37 \right] &
\left[48 \right] \nonu \\
 \left[15 \right] &  F & -\eta G & \eta G & -\eta G \nonu \\
\left[26 \right]  &  &  F &  -G &
 G \nonu \\
 \left[ 37\right] &  &   &  F &
 -G \nonu \\
 \left[ 48\right] &  &   &  &  F \\
\end{array}
\right), v_5 = -\varepsilon_{\pm}\left(
\begin{array}{ccccc}
& \left[16 \right] & \left[25 \right] & \left[38 \right] &
\left[47 \right] \nonu \\
 \left[16 \right] &  F & \eta G & -\eta G & -\eta G \nonu \\
\left[25 \right]  &  &  F & - G &
- G \nonu \\
 \left[ 38\right] &  &   &  F &
G \nonu \\
 \left[ 47\right] &  &   &  G &  F \\
\end{array}
\right), \nonu \\
& & v_6 = -\varepsilon_{\pm}\left(
\begin{array}{ccccc}
& \left[17 \right] & \left[28 \right] & \left[35 \right] &
\left[46 \right] \nonu \\
 \left[17 \right] &  F & -\eta G & -\eta G & \eta G \nonu \\
\left[28 \right]  &  &  F &  G &
- G \nonu \\
 \left[ 35\right] &  &   &  F &
 -G \nonu \\
 \left[ 46\right] &  &   &   &  F \\
\end{array}
\right),  v_7 = -\varepsilon_{\pm}\left(
\begin{array}{ccccc}
& \left[18 \right] & \left[27 \right] & \left[36 \right] &
\left[45 \right] \nonu \\
 \left[18 \right] &  F & \eta G & \eta G & \eta G \nonu \\
\left[27 \right]  &  &  F &  G &
 G \nonu \\
 \left[ 36\right] &  &   &  F &
 G \nonu \\
 \left[ 45\right] &  &   &  &  F \\
\end{array}
\right), \\
\label{so7so1}
\eea
where
\bea A & = & \cosh^{3} s,
\qquad B=\cosh s \sinh^{2} s,
\nonu \\
 F & = & \sinh^{3} s,
\qquad G=\sinh s \cosh^{2} s.
\nonu
\eea
From now on, we do not include the index pairs into the $4\times 4$
matrices $u_i$ and $v_i$, for simplicity. For example,
when we write $u_2 =u_3$ below, this implies that although the indices they
possess are
different, the corresponding matrix elements are identical.

$\bullet$
$SO(6)^{\pm} \times SO(2)^{\pm}$ Invariant Sectors:


\bea
&& u_1 = \left(
\begin{array}{ccccc}
   A & \eta B & \eta B & \eta B \nonu \\
  & A & B &
B \nonu \\
  &  & A &
B \nonu \\
   &   &  & A \\
\end{array}
\right), u_2 = C {\bf 1}_{4 \times 4}=u_3 = u_4 = u_5 = u_6 = u_7,
\nonu \\
&& v_1 = -\varepsilon_{\pm}\left(
\begin{array}{ccccc}
  F & \eta G & -\eta G & \eta G \nonu \\
  & F & G &
G \nonu \\
   &  & F &
-G \nonu \\
   &  &  & F \\
\end{array}
\right), \nonu \\
&&v_2 = \varepsilon_{\pm}\left(
\begin{array}{ccccc}
  0& \eta H & 0 & 0 \nonu \\
  & 0 & 0 & 0
 \nonu \\
   &  & 0 &
H \nonu \\
   &  &  & 0 \\
\end{array}
\right)
=-v_3 = v_4 = -v_5 =
v_6 = -v_7
\nonu
\eea
where
\begin{eqnarray}
A & = & \cosh^{3} s, \qquad B=\cosh s \sinh^{2} s, \qquad C=\cosh s,
\nonu \\
 F & = &  \sinh^{3} s, \qquad G=\sinh s \cosh^{2} s, \qquad H=\sinh s.
\nonu
\end{eqnarray}

$\bullet$
$SO(5)^{\pm} \times SO(3)^{\pm}$ Invariant Sectors:


\bea && u_1 = \left(
\begin{array}{ccccc}
   A & \eta B & \eta C & \eta C \nonu \\
  & A & C &
C \nonu \\
   &  & A &
B \nonu \\
   &   &  & A \\
\end{array}
\right), u_2 = \left(
\begin{array}{ccccc}
   A & -\eta B & -\eta C & \eta C \nonu \\
  & A & C &
-C \nonu \\
   &  & A &
-B \nonu \\
   &   &  & A \\
\end{array}
\right), \nonu \\
&& u_3 = \left(
\begin{array}{ccccc}
   A & \eta B & -\eta C & -\eta C \nonu \\
 & A & -C &
-C \nonu \\
   &  & A &
B \nonu \\
   &  &  & A \\
\end{array}
\right), u_4 = \left(
\begin{array}{ccccc}
   D & \eta E & -\eta E & -\eta E \nonu \\
  & D & -E &
-E \nonu \\
   &  & D &
E \nonu \\
   &  &  & D \\
\end{array}
\right), \nonu \\
&& u_5 = \left(
\begin{array}{ccccc}
   D & -\eta E & \eta E & -\eta E \nonu \\
  & D & -E &
E \nonu \\
   &  & D &
-E \nonu \\
   &  &  & D \\
\end{array}
\right), u_6 = \left(
\begin{array}{ccccc}
   D & \eta E & \eta E & \eta E \nonu \\
  & D & E &
E \nonu \\
   &  & D &
E \nonu \\
  &  &  & D \\
\end{array}
\right), \nonu \\
&& u_7 = \left(
\begin{array}{ccccc}
   D & -\eta E & -\eta E & \eta E \nonu \\
  & D & E &
-E \nonu \\
   &  & D &
-E \nonu \\
   &  &  & D \\
\end{array}
\right),
 v_1 = -\varepsilon_{\pm}\left(
\begin{array}{ccccc}
  F & \eta G & -\eta H& -\eta H\nonu \\
  & F & -H&
-H\nonu \\
   &  & F &
G \nonu \\
  &  &  & F \\
\end{array}
\right), \nonu \\
& & v_2 = -\varepsilon_{\pm}\left(
\begin{array}{ccccc}
  F & -\eta G & \eta H& -\eta H\nonu \\
  & F & -H&
H\nonu \\
   &  & F &
-G \nonu \\
   &  &  & F \\
\end{array}
\right),  v_3 = -\varepsilon_{\pm}\left(
\begin{array}{ccccc}
 F & \eta G & \eta H& \eta H\nonu \\
  & F & H&
H\nonu \\
   &  & F &
G \nonu \\
   & &  & F \\
\end{array}
\right), \nonu \\
& & v_4 =\varepsilon_{\pm} \left(
\begin{array}{ccccc}
  I & \eta J & -\eta J & -\eta J \nonu \\
  & I & -J &
-J \nonu \\
   &  & I &
J \nonu \\
   &  &  & I \\
\end{array}
\right), v_5 = \varepsilon_{\pm}\left(
\begin{array}{ccccc}
  I & -\eta J & \eta J & - \eta J \nonu \\
  & I & -J &
J \nonu \\
   &  & I &
-J \nonu \\
   &  &  & I \\
\end{array}
\right), \nonu \\
& & v_6 = \varepsilon_{\pm}\left(
\begin{array}{ccccc}
  I & \eta J & \eta J & \eta J \nonu \\
 & I & J &
J \nonu \\
  &  & I &
J \nonu \\
  &  &  & I \\
\end{array}
\right),  v_7 = \varepsilon_{\pm}\left(
\begin{array}{ccccc}
  I & -\eta J & -\eta J &\eta J \nonu \\
  & I & J &
-J \nonu \\
   &  & I &
-J \nonu \\
   &  &  & I \\
\end{array}
\right), \\
\nonu
\eea
where
\begin{eqnarray}
A & = & \left(-1+2\cosh \left(\frac{2s}{3}\right)\right)
\cosh^{3}\left(\frac{s}{3}\right),\qquad B  = \cosh
\left(s\right)\sinh^{2}\left(\frac{s}{3}\right),
\nonu \\
C & = &
\left(2\cosh\left(\frac{s}{3}\right)+\cosh\left(s\right)\right)
\sinh^{2}\left(\frac{s}{3}\right), \qquad
D  =  \cosh^{3}\left(\frac{s}{3}\right),\nonu \\
E & = &
\cosh\left(\frac{s}{3}\right)\sinh^{2}\left(\frac{s}{3}\right),
\qquad F
 =  \left(1+2\cosh\left(\frac{2s}{3}\right)\right)\sinh^{3}
\left(\frac{s}{3}\right),\nonu \\ G & = &
\cosh^{2}\left(\frac{s}{3}\right)\sinh\left(s\right),
\qquad H =
\frac{1}{4}\left(\sinh\left(\frac{s}{3}\right)-\sinh
\left(\frac{5s}{3}\right)\right),\nonu
\\
 I & = & \sinh^{3}\left(\frac{s}{3}\right), \qquad
J  =
\frac{1}{4}\left(\sinh\left(\frac{s}{3}\right)+\sinh\left(s\right)\right).\nonu
\end{eqnarray}

$\bullet$
$SO(4)^{\pm} \times SO(4)^{\pm}$ Invariant Sectors:


\bea
&& u_1 = A  {\bf 1}_{4 \times 4} =
u_2 =u_3, \qquad u_4 = u_5 =u_6=u_7 ={\bf 1}_{4 \times 4}, \nonu \\
&& v_1 = -\varepsilon_{\pm}\left(
\begin{array}{ccccc}
  0 & \eta B & 0 &  0 \nonu \\
  & 0 & 0 &
0 \nonu \\
   &  & 0 &
B \nonu \\
   &  &  & 0 \\
\end{array}
\right) =-v_2 = v_3, \qquad v_4 = v_5 =v_6= v_7 =0,
\nonu
\eea
where
\begin{eqnarray}
A = \cosh s, \qquad B=\sinh s. \nonu
\end{eqnarray}

$\bullet$
$SO(3)^{\pm} \times SO(5)^{\pm}$ Invariant Sectors:


\bea && u_1 = \left(
\begin{array}{ccccc}
  A & -\eta B & \eta C & -\eta C \nonu \\
 & A & -C &
C \nonu \\
   &  & A &
-B \nonu \\
   &   &  & A \\
\end{array}
\right), u_2 = \left(
\begin{array}{ccccc}
   A & \eta B & -\eta C & -\eta C \nonu \\
  & A & -C &
-C \nonu \\
   &  & A &
B \nonu \\
  &   &  & A \\
\end{array}
\right), \nonu \\
&& u_3 = \left(
\begin{array}{ccccc}
   A & -\eta B & -\eta C & \eta C \nonu \\
  & A & C &
-C \nonu \\
   &  & A &
-B \nonu \\
  &  &  & A \\
\end{array}
\right), u_4 = \left(
\begin{array}{ccccc}
   D & \eta E & -\eta E & -\eta E \nonu \\
  & D & -E &
-E \nonu \\
  &  & D &
E \nonu \\
   &  &  & D \\
\end{array}
\right), \nonu \\
&& u_5 = \left(
\begin{array}{ccccc}
   D & -\eta E & -\eta E & \eta E \nonu \\
  & D & E &
-E \nonu \\
   &  & D &
-E \nonu \\
   &  &  & D \\
\end{array}
\right), u_6 = \left(
\begin{array}{ccccc}
   D & -\eta E & \eta E & -\eta E \nonu \\
  & D & -E &
E \nonu \\
   &  & D &
-E \nonu \\
   &  &  & D \\
\end{array}
\right), \nonu \\
&& u_7 = \left(
\begin{array}{ccccc}
   D & \eta E & \eta E & \eta E \nonu \\
  & D & E &
E \nonu \\
   &  & D &
E \nonu \\
   &  &  & D \\
\end{array}
\right),
 v_1 = \varepsilon_{\pm}\left(
\begin{array}{ccccc}
  F & -\eta G & \eta H & -\eta H \nonu \\
  & F & -H &
H \nonu \\
   &  & F &
-G \nonu \\
  &  &  & F \\
\end{array}
\right), \nonu \\
& & v_2 = \varepsilon_{\pm}\left(
\begin{array}{ccccc}
  F & \eta G & -\eta H & -\eta H \nonu \\
  & F & -H &
-H \nonu \\
   &  & F &
G \nonu \\
   &  &  & F \\
\end{array}
\right),  v_3 = \varepsilon_{\pm}\left(
\begin{array}{ccccc}
  F & -\eta G & -\eta H & \eta H \nonu \\
  & F & H &
-H \nonu \\
   &  & F &
-G \nonu \\
   &  &  & F \\
\end{array}
\right), \nonu \\
& & v_4 = -\varepsilon_{\pm}\left(
\begin{array}{ccccc}
  I & \eta J & - \eta J & -\eta J \nonu \\
  & I & -J &
-J \nonu \\
   &  & I &
J \nonu \\
   &  &  & I \\
\end{array}
\right), v_5 = -\varepsilon_{\pm}\left(
\begin{array}{ccccc}
  I & -\eta J & -\eta J & \eta J \nonu \\
  & I & J &
-J \nonu \\
   &  & I &
-J \nonu \\
  &  &  & I \\
\end{array}
\right), \nonu \\
& & v_6 = -\varepsilon_{\pm}\left(
\begin{array}{ccccc}
  I & -\eta J & \eta J & -\eta J \nonu \\
  & I & -J &
J \nonu \\
   &  & I &
-J \nonu \\
   &  &  & I \\
\end{array}
\right),  v_7 = -\varepsilon_{\pm}\left(
\begin{array}{ccccc}
  I & \eta J & \eta J & \eta J \nonu \\
  & I & J &
J \nonu \\
   &  & I &
J \nonu \\
   &  &  & I \\
\end{array}
\right), \\
\nonu
\eea
where
\begin{eqnarray}
A & = & \left(-1+2\cosh\left(\frac{2s}{5}\right)\right)
\cosh^{3}\left(\frac{s}{5}\right),\qquad B  =
\cosh\left(\frac{3s}{5}\right)\sinh^{2}\left(\frac{s}{5}\right),
\nonu \\
C & = &
\frac{1}{4}\left(\cosh\left(s\right)-\cosh\left(\frac{s}{5}\right)\right),
\qquad D  =  \cosh^{3}\left(\frac{s}{5}\right),\nonu \\
E  & =
 & \cosh\left(\frac{s}{5}\right)\sinh^{2}\left(\frac{s}{5}\right),
\qquad F  = \left(1+2\cosh\left(\frac{2s}{5}\right)\right)
\sinh^{3}\left(\frac{s}{5}\right),\nonu
\\ G & =
& \cosh^{2}\left(\frac{s}{5}\right)\sinh\left(\frac{3s}{5}\right),
\qquad H  =
\frac{1}{4}\left(\sinh\left(s\right)-\sinh\left(\frac{s}{5}\right)\right),
\nonu
\\
 I & = & \sinh^{3}\left(\frac{s}{5}\right),
\qquad J  =
\cosh^{2}\left(\frac{s}{5}\right)\sinh\left(\frac{s}{5}\right).
\nonu
\end{eqnarray}
All these functions of $s$ can be obtained from those in
$SO(5)^{\pm} \times SO(3)^{\pm}$ by replacing $s$ with $3s/5$ and using
the properties of hyperbolic functions. For example, each $C$ that seems to
look different is the same by a simple change of variable.

$\bullet$
$SO(2)^{\pm} \times SO(6)^{\pm}$ Invariant Sectors:


\bea && u_1 = \left(
\begin{array}{ccccc}
   A & -\eta B & \eta B & -\eta B \nonu \\
  & A & -B &
B \nonu \\
   &  & A &
-B \nonu \\
   &   &  & A \\
\end{array}
\right), u_2 = C {\bf 1}_{4 \times 4}=u_3=u_4=u_5=
u_6=u_7, \nonu \\
&& v_1 = \varepsilon_{\pm}\left(
\begin{array}{ccccc}
  F & -\eta G & \eta G &  -\eta G \nonu \\
  & F & -G &
G \nonu \\
  &  & F &
-G \nonu \\
   &   &  & F \\
\end{array}
\right), \nonu \\
&& v_2 =
\varepsilon_{\pm}\left(
\begin{array}{ccccc}
  0 & \eta H & 0 &  0 \nonu \\
  & 0 & 0 &
0 \nonu \\
  &  & 0 &
H \nonu \\
   &   &  & 0 \\
\end{array}
\right)
=-v_3 = -v_4 = v_5 = v_6 =-v_7,
\nonu
\eea
where
\begin{eqnarray}
A & = & \cosh^{3}\left(\frac{s}{3}\right),
\qquad B=\cosh\left(\frac{s}{3}\right)\sinh^{2}\left(\frac{s}{3}\right),
\qquad C=\cosh\left(\frac{s}{3}\right),
\nonu \\
 F  & = & \sinh^{3}\left(\frac{s}{3}\right),
\qquad G=\sinh\left(\frac{s}{3}\right)\cosh^{2}\left(\frac{s}{3}\right),
\qquad
 H   =  \sinh\left(\frac{s}{3}\right).
\nonu
\end{eqnarray}
All these functions of $s$ can be obtained from those in
$SO(6)^{\pm} \times SO(2)^{\pm}$ by replacing $s$ with $s/3$.

$\bullet$
$SO(1)^{\pm} \times SO(7)^{\pm}$ Invariant Sectors:


\bea && u_1 = \left(
\begin{array}{ccccc}
  A & -\eta B & \eta B & -\eta B \nonu \\
  & A & -B &
B \nonu \\
   &  & A &
-B \nonu \\
   &  &  & A \\
\end{array}
\right) = u_3 =u_4,
u_2 = \left(
\begin{array}{ccccc}
   A & \eta B & -\eta B & -\eta B \nonu \\
  & A & -B &
-B \nonu \\
   &  & A &
B \nonu \\
   &  &  & A \\
\end{array}
\right)=u_6, \nonu \\
&& u_5 = \left(
\begin{array}{ccccc}
   A & \eta B & \eta B & \eta B \nonu \\
  & A & B &
B \nonu \\
   &  & A &
B \nonu \\
   &  &  & A \\
\end{array}
\right),
u_7 = \left(
\begin{array}{ccccc}
   A & -\eta B & -\eta B & \eta  B \nonu \\
  & A & B &
-B \nonu \\
   &  & A &
-B \nonu \\
   &  &  & A \\
\end{array}
\right), \nonu \\
&& v_1 = \varepsilon_{\pm}\left(
\begin{array}{ccccc}
  F & -\eta G & \eta G &  -\eta G \nonu \\
  & F & -G &
G \nonu \\
   &  & F &
-G \nonu \\
   &  &  & F \\
\end{array}
\right)=v_3=v_4,
v_2 = \varepsilon_{\pm}\left(
\begin{array}{ccccc}
  F & \eta G & -\eta G & -\eta G \nonu \\
  & F & -G &
-G \nonu \\
   &  & F &
G \nonu \\
  &  &  & F \\
\end{array}
\right)=v_6,  \nonu \\
&&
v_5 = \varepsilon_{\pm}\left(
\begin{array}{ccccc}
  F & \eta G & \eta G &  \eta G \nonu \\
  & F & G &
G \nonu \\
   &  & F &
G \nonu \\
   &  &  & F \\
\end{array}
\right),
v_7 = \varepsilon_{\pm}\left(
\begin{array}{ccccc}
  F & -\eta G & -\eta G &  \eta G \nonu \\
  & F & G &
-G \nonu \\
   &  & F &
-G \nonu \\
   &  &  & F \\
\end{array}
\right), \\
\nonu
\eea
where
\begin{eqnarray}
A & = & \cosh^{3}\left(\frac{s}{7}\right),
\qquad B=\cosh\left(\frac{s}{7}\right)\sinh^{2}\left(\frac{s}{7}\right),
\nonu \\
 F & = & \sinh^{3}\left(\frac{s}{7}\right),
\qquad G=\sinh\left(\frac{s}{7}\right)\cosh^{2}\left(\frac{s}{7}\right).
\nonu
\end{eqnarray}
All these functions of $s$ can be obtained from those in
$SO(7)^{\pm} \times SO(1)^{\pm}$ by replacing $s$ with $s/7$.

\section{Appendix F: Projectors of $SO(p)^{+} \times SO(q)^{+}$ Sectors
in $28 \times 28$ Matrices }

The projectors $ P_{\si, p, q}^{IJKL}(\si=\al, \be, \ga)$
of $SO(p)^{+} \times SO(q)^{+}$-invariant sectors
can be obtained explicitly.
We list $P_{\al, p, q}^{IJKL}$ and $P_{\be, p, q}^{IJKL}$ only because
$P_{\ga, p, q}^{IJKL}$
can be obtained from those:$P_{\ga, p, q}^{IJKL}=
1-P_{\al,p, q}^{IJKL}-P_{\be, p, q}^{IJKL}$.
\bea
P_{\al, 7, 1}^{IJKL} & = & \mbox{diag} \left(
F_1,F_2,F_1,F_3,F_4,F_2,F_1 \right), \nonu \\
 P_{\be, 7, 1}^{IJKL}  & = &  0,\nonu \\
P_{\al, 6, 2}^{IJKL} &  = &  \mbox{diag} \left(
F_1,F_9,F_{10},F_9,F_{10},F_9,F_{10} \right), \nonu \\
 P_{\be, 6, 2}^{IJKL} & = &
\mbox{diag} \left(
F_5,0,0,0,0,0,0 \right),
\nonu \\
P_{\al, 5, 3}^{IJKL} & = &  \mbox{diag} \left(
F_{10},F_9,F_{10},F_8,F_7,F_5,F_6 \right), \nonu \\
 P_{\be, 5, 3}^{IJKL} & = &
\mbox{diag} \left(
F_5,F_6,F_8,0,0,0,0 \right),
\nonu \\
P_{\al, 4, 4}^{IJKL} & = &  \mbox{diag} \left(
F_{10},F_9,F_{10},0,0,0,0 \right), \nonu \\
P_{\be, 4, 4}^{IJKL} &  = &
\mbox{diag} \left(
F_9,F_{10},F_9,0,0,0,0 \right),
\nonu \\
P_{\al, 3, 5}^{IJKL} & = & \mbox{diag} \left(
F_7,F_8,F_6,0,0,0,0 \right), \nonu \\
 P_{\be, 3, 5}^{IJKL} & = &
\mbox{diag} \left(
F_9,F_{10},F_9,F_8,F_6,F_7,F_5 \right),
\nonu \\
P_{\al, 2, 6}^{IJKL} & = &  \mbox{diag} \left(
F_7,0,0,0,0,0,0 \right), \nonu \\
 P_{\be, 2, 6}^{IJKL}&  = &
\mbox{diag} \left(
F_3,F_{10},F_9,F_9,F_{10},F_{10},F_9 \right),
\nonu \\
P_{\al, 1, 7}^{IJKL} & = &  0, \nonu \\
 P_{\be, 1, 7}^{IJKL} &  = &
\mbox{diag} \left(
F_3,F_4,F_3,F_3,F_1,F_4,F_2 \right),
\nonu
\eea
where the $4 \times 4$ block diagonal matrices $F_i$'s are
\bea
&& F_1 = \frac{1}{8} \left(
\begin{array}{ccccc}
 3  & -1 & -1 & -1 \nonu \\
 -1 & 3 & -1 & -1
 \nonu \\
 -1  &-1  & 3 &-1
 \nonu \\
 -1  & -1 & -1 & 3  \\
\end{array}
\right), \qquad
F_2 = \frac{1}{8} \left(
\begin{array}{ccccc}
  3  & 1 & 1 &-1  \nonu \\
 1 &3  &-1  &1
 \nonu \\
 1  & -1 & 3 &1
 \nonu \\
 -1  & 1 & 1 & 3 \\
\end{array}
\right), \nonu \\
&& F_3 = \frac{1}{8} \left(
\begin{array}{ccccc}
   3 & 1 & -1 & 1 \nonu \\
 1 &3  &1  &-1
 \nonu \\
 -1 & 1 & 3 &1
 \nonu \\
 1  & -1 & 1 & 3 \\
\end{array}
\right), \qquad
  F_4 = \frac{1}{8} \left(
\begin{array}{ccccc}
   3 & -1 & 1 & 1 \nonu \\
 -1 &3  &1  &1
 \nonu \\
 1 & 1 & 3 &-1
 \nonu \\
 1  & 1 & -1 & 3 \\
\end{array}
\right), \nonu \\
 && F_5 = \frac{1}{8} \left(
\begin{array}{ccccc}
   1 & 1 & 1 & 1 \nonu \\
 1 &1  &1  &1
 \nonu \\
 1 & 1 & 1 &1
 \nonu \\
 1  & 1 & 1 & 1 \\
\end{array}
\right), \qquad
  F_6 = \frac{1}{8} \left(
\begin{array}{ccccc}
   1 & -1 & -1 & 1 \nonu \\
 -1 &1  &1  &-1
 \nonu \\
 -1 & 1 & 1 &-1
 \nonu \\
 1  & -1 & -1 & 1 \\
\end{array}
\right), \nonu \\
&& F_7 = \frac{1}{8} \left(
\begin{array}{ccccc}
   1 & -1 & 1 & -1 \nonu \\
 -1 &1  &-1  &1
 \nonu \\
 1 & -1 & 1 &-1
 \nonu \\
 -1  & 1 & -1 & 1 \\
\end{array}
\right), \qquad
  F_8 = \frac{1}{8} \left(
\begin{array}{ccccc}
   1 & 1 & -1 & -1 \nonu \\
 1 &1  &-1  &-1
 \nonu \\
 -1 & -1 & 1 &1
 \nonu \\
 -1  & -1 & 1 & 1 \\
\end{array}
\right), \nonu \\
&& F_9 = \frac{1}{4} \left(
\begin{array}{ccccc}
   1 & 1 & 0 & 0 \nonu \\
 1 &1  &0  &0
 \nonu \\
 0 & 0 & 1 &1
 \nonu \\
 0  & 0 & 1 & 1 \\
\end{array}
\right), \qquad
  F_{10} = \frac{1}{4} \left(
\begin{array}{ccccc}
   1 & -1 & 0 & 0 \nonu \\
 -1 &1  &0  &0
 \nonu \\
 0 & 0 & 1 &-1
 \nonu \\
 0  & 0 & -1 & 1 \\
\end{array}
\right). \nonu
\eea

\section{Appendix G: Kinetic Terms, Superpotential and Potential
in $SO(p)^{+} \times SO(q)^{+} \times SO(r)^{+}$ Sectors  }

We list here 1) the kinetic terms in terms of original variables, $m$ and
$n$, 2) new variables, $\widetilde{m}$ and $\widetilde{n}$ in order to
have usual canonical expression of kinetic terms, 3) superpotential
in terms of new fields, and 4) scalar potential in
$SO(p)^{+} \times SO(q)^{+} \times SO(r)^{+}$ sectors.
In all cases, the scalar potential can be expressed in terms of
superpotential   as (\ref{pqrsuperpotential}).
In this revised version, we list only six cases due to space
limitations and refer to the original version in the hep-th archive
for remaining cases.

$\bullet$ $SO(1,7)^{+}  \rightarrow SO(2,6)^{+} \rightarrow CSO(1,1,6)^{+}$:
\bea
K_{1,1,6}(m,n) & = & -\frac{1}{3}
\pa^{\mu} m \pa_{\mu} m - \frac{2}{7} \pa^{\mu} m \pa_{\mu} n-
\frac{1}{7} \pa^{\mu} n \pa_{\mu} n, \nonu \\
m & = &
 -\frac{3\sqrt{2}}{4} \widetilde{m} -\frac{\sqrt{6}}{2} \widetilde{n},
\nonu \\
n & = & \frac{7\sqrt{2}}{4} \widetilde{m},
\nonu \\
W_{1,1,6}(\xi, \zeta;\widetilde{m}, \widetilde{n}) & = &
\frac{1}{8} e^{-\frac{2\sqrt{2} \widetilde{m} + \sqrt{6}
\widetilde{n}}{2}} \left( e^{2\sqrt{2} \widetilde{m}} + \xi +
6 e^{\sqrt{2} \widetilde{m} +\frac{2\sqrt{6}}{3} \widetilde{n}}
\xi \zeta \right),
\nonu \\
V_{1,1,6}(\xi, \zeta;\widetilde{m}, \widetilde{n}) & =
& \frac{1}{8} e^{-2\sqrt{2} \widetilde{m} -\sqrt{6} \widetilde{n}}
\left( e^{4 \sqrt{2} \widetilde{m}} - 2 e^{2 \sqrt{2}
\widetilde{n}} \xi -12 e^{3\sqrt{2} \widetilde{m} + \frac{2\sqrt{6}}{3}
\widetilde{n}} \xi \zeta + \xi^2 \right. \nonu \\
&& \left.  -12 e^{\sqrt{2} \widetilde{m} +
\frac{2\sqrt{6}}{3} \widetilde{n}} \xi^2 \zeta -24 e^{2\sqrt{2}
\widetilde{m} +\frac{4\sqrt{6}}{3} \widetilde{n}}
\zeta^2 \xi^2   \right).
\nonu
\eea
There exists a $SO(7)^{+}$-invariant
critical point of $SO(8)$ theory  for $\xi=1$ and $\zeta=1$
and  a $SO(2)^{+} \times SO(6)^{+}$-invariant
critical point for $\xi=1$ and $\zeta=0$.

$\bullet$ $SO(1,7)^{+}  \rightarrow SO(3,5)^{+} \rightarrow CSO(1,2,5)^{+}$:
\bea
K_{1,2,5}(m,n) & = & -\frac{3}{5} \pa^{\mu} m \pa_{\mu} m -
\frac{2}{7} \pa^{\mu} m \pa_{\mu} n-
\frac{1}{7} \pa^{\mu} n \pa_{\mu} n, \nonu \\
m & = &
 -\frac{5\sqrt{6}}{24} \widetilde{m} -\frac{5\sqrt{6}}{6} \widetilde{n},
\nonu \\
n & = & \frac{7\sqrt{6}}{8} \widetilde{m},
\nonu \\
W_{1,2,5}(\xi, \zeta;\widetilde{m}, \widetilde{n}) & = &
\frac{1}{8} e^{-\frac{2\sqrt{6} \widetilde{m} + \sqrt{30}
\widetilde{n}}{6}} \left( e^{\sqrt{6} \widetilde{m}} + 2\xi +
5 e^{\frac{\sqrt{6}}{3} \widetilde{m} +\frac{4\sqrt{30}}{15} \widetilde{n}}
\xi \zeta \right),
\nonu \\
V_{1,2,5}(\xi, \zeta;\widetilde{m}, \widetilde{n}) & =
& \frac{1}{8} e^{-\frac{\sqrt{6}}{3}
\widetilde{m} -\frac{\sqrt{30}}{3} \widetilde{n}}
\left( e^{\frac{5\sqrt{6}}{3} \widetilde{m}} - 4 e^{\frac{2\sqrt{6}}{3}
\widetilde{n}} \xi -10 e^{\sqrt{6} \widetilde{m} + \frac{4\sqrt{30}}{15}
\widetilde{n}} \xi \zeta
 -20 e^{
\frac{4\sqrt{30}}{15} \widetilde{n}} \xi^2 \zeta \right. \nonu \\
&& \left. -15 e^{\frac{\sqrt{6}}{3}
\widetilde{m} +\frac{8\sqrt{30}}{15} \widetilde{n}}
 \xi^2 \zeta^2   \right).
\nonu
\eea
There exists an $SO(3)^{+} \times SO(5)^{+} $-invariant
critical point for $\xi=1$ and $\zeta=-1$.

$\bullet$ $SO(1,7)^{+}  \rightarrow SO(4,4)^{+} \rightarrow CSO(1,3,4)^{+}$:
\bea
K_{1,3,4}(m,n) & = & - \pa^{\mu} m \pa_{\mu} m - \frac{2}{7}
\pa^{\mu} m \pa_{\mu} n-
\frac{1}{7} \pa^{\mu} n \pa_{\mu} n, \nonu \\
m & = &
 -\frac{\sqrt{3}}{6} \widetilde{m} -\frac{\sqrt{2}}{2} \widetilde{n},
\nonu \\
n & = & \frac{7\sqrt{3}}{6} \widetilde{m},
\nonu \\
W_{1,3,4}(\xi, \zeta;\widetilde{m}, \widetilde{n}) & = &
\frac{1}{8} e^{-\frac{2\sqrt{3} \widetilde{m} + 3\sqrt{2}
\widetilde{n}}{6}} \left( e^{\frac{4\sqrt{3}}{3} \widetilde{m}} + 3\xi +
4 e^{\frac{\sqrt{3}}{3} \widetilde{m} +\sqrt{2} \widetilde{n}}
\xi \zeta \right),
\nonu \\
V_{1,3,4}(\xi, \zeta;\widetilde{m}, \widetilde{n}) & = & \frac{1}{8}
\left( e^{2 \sqrt{3} \widetilde{m}-\sqrt{2} \widetilde{n}} - 6 e^{
\frac{2\sqrt{3}}{3} \widetilde{m}-  \sqrt{2}
\widetilde{n}} \xi -8 e^{\sqrt{3} \widetilde{m}} \xi \zeta -
e^{-\frac{2\sqrt{3}}{3} \widetilde{m}-\sqrt{2} \widetilde{n}}
\xi^2 \right. \nonu \\
&& \left. -24 e^{
-\frac{\sqrt{3}}{3} \widetilde{m}} \xi^2 \zeta -8 e^{\sqrt{2}
\widetilde{n}}
 \xi^2 \zeta^2   \right).
\nonu
\eea
There exists
an $SO(4)^{+} \times SO(4)^{+}$-invariant
critical point for $\xi=1$ and $\zeta=-1$, and
an $SO(5)^{+} \times SO(3)^{+}$-invariant
critical point for $\xi=-1$ and $\zeta=-1$.

$\bullet$ $SO(1,7)^{+}  \rightarrow SO(5,3)^{+} \rightarrow CSO(1,4,3)^{+}$:
\bea
K_{1,4,3}(m,n) & = & -\frac{5}{3} \pa^{\mu} m \pa_{\mu} m -
\frac{2}{7} \pa^{\mu} m \pa_{\mu} n-
\frac{1}{7} \pa^{\mu} n \pa_{\mu} n, \nonu \\
m & = &
 -\frac{3\sqrt{5}}{40} \widetilde{m} -\frac{\sqrt{30}}{10} \widetilde{n},
\nonu \\
n & = & \frac{7\sqrt{5}}{8} \widetilde{m},
\nonu \\
W_{1,4,3}(\xi, \zeta;\widetilde{m}, \widetilde{n}) & = &
\frac{1}{8} e^{-\frac{2\sqrt{5} \widetilde{m} + \sqrt{30}
\widetilde{n}}{10}} \left( e^{\sqrt{5} \widetilde{m}} + 4\xi +
3 e^{\frac{\sqrt{5}}{5} \widetilde{m} +\frac{4\sqrt{30}}{15} \widetilde{n}}
\xi \zeta \right),
\nonu \\
V_{1,4,3}(\xi, \zeta;\widetilde{m}, \widetilde{n}) & =
& \frac{1}{8} e^{-\frac{2\sqrt{5} \widetilde{m} + \sqrt{30}
\widetilde{n}}{5}}
\left( e^{2 \sqrt{5} \widetilde{m}} - 8 e^{
\sqrt{5} \widetilde{m}} \xi -6 e^{\frac{6\sqrt{5}}{5} \widetilde{m}+
\frac{4\sqrt{30}}{15} \widetilde{n} } \xi \zeta -8
\xi^2 \right. \nonu \\
& &  \left. -24 e^{
-\frac{\sqrt{5}}{5} \widetilde{m}+\frac{4\sqrt{30}}{15}
\widetilde{n}} \xi^2 \zeta -3 e^{\frac{2\sqrt{5}}{5} \widetilde{m}+
\frac{8\sqrt{30}}{15} \widetilde{n}}
 \xi^2 \zeta^2   \right).
\nonu
\eea
There exists
an $SO(5)^{+} \times SO(3)^{+}$-invariant
critical point for $\xi=1$ and $\zeta=-1$, and
a $SO(4)^{+} \times SO(4)^{+}$-invariant
critical point for $\xi=-1$ and $\zeta=-1$.

$\bullet$ $SO(1,7)^{+}  \rightarrow SO(6,2)^{+} \rightarrow CSO(1,5,2)^{+}$:
\bea
K_{1,5,2}(m,n) & = & -3 \pa^{\mu} m \pa_{\mu} m -
\frac{2}{7} \pa^{\mu} m \pa_{\mu} n-
\frac{1}{7} \pa^{\mu} n \pa_{\mu} n, \nonu \\
m & = &
 -\frac{\sqrt{30}}{60} \widetilde{m} -\frac{\sqrt{6}}{6} \widetilde{n},
\nonu \\
n & = & \frac{7\sqrt{30}}{20} \widetilde{m},
\nonu \\
W_{1,5,2}(\xi, \zeta;\widetilde{m}, \widetilde{n}) & = &
\frac{1}{8} e^{- \frac{\sqrt{30}}{15} \widetilde{m} - \frac{\sqrt{6}}{6}
\widetilde{n}} \left( e^{\frac{2\sqrt{30}}{5} \widetilde{m}} + 5\xi +
2 e^{\frac{\sqrt{30}}{15} \widetilde{m} +\frac{2\sqrt{6}}{3} \widetilde{n}}
\xi \zeta \right),
\nonu \\
V_{1,5,2}(\xi, \zeta;\widetilde{m}, \widetilde{n}) & =
& -\frac{1}{8} e^{-\frac{2\sqrt{30}}{15} \widetilde{m} -
\frac{\sqrt{6}}{3} \widetilde{n}}
\left( - e^{4 \frac{\sqrt{30}}{5} \widetilde{m}} + 10 e^{
\frac{2\sqrt{30}}{5} \widetilde{m}} \xi +
4 e^{\frac{7\sqrt{30}}{15} \widetilde{m}+
\frac{2\sqrt{6}}{3} \widetilde{n} } \xi \zeta + 15
\xi^2 \right. \nonu \\
&& \left. + 20 e^{
\frac{\sqrt{30}}{15} \widetilde{m}+\frac{2\sqrt{6}}{3}
\widetilde{n}} \xi^2 \zeta  \right).
\nonu
\eea
There exists
an $SO(3)^{+} \times SO(5)^{+}$-invariant
critical point for $\xi=-1$ and $\zeta=-1$.

$\bullet$ $SO(1,7)^{+}  \rightarrow SO(7,1)^{+} \rightarrow CSO(1,6,1)^{+}$:
\bea
K_{1,6,1}(m,n) & = & -7 \pa^{\mu} m \pa_{\mu} m - \frac{2}{7}
\pa^{\mu} m \pa_{\mu} n-
\frac{1}{7} \pa^{\mu} n \pa_{\mu} n, \nonu \\
m & = &
 -\frac{\sqrt{42}}{168} \widetilde{m} -\frac{\sqrt{14}}{14} \widetilde{n},
\nonu \\
n & = & \frac{7\sqrt{42}}{24} \widetilde{m},
\nonu \\
W_{1,6,1}(\xi, \zeta;\widetilde{m}, \widetilde{n}) & = &
\frac{1}{8} e^{- \frac{\sqrt{42}}{21} \widetilde{m} - \frac{\sqrt{14}}{14}
\widetilde{n}} \left( e^{\frac{\sqrt{42}}{3} \widetilde{m}} + 6\xi +
e^{\frac{\sqrt{42}}{21} ( \widetilde{m} + 4\sqrt{3} \widetilde{n})}
\xi \zeta \right),
\nonu \\
V_{1,6,1}(\xi, \zeta;\widetilde{m}, \widetilde{n}) & =
& \frac{1}{8} e^{-\frac{2\sqrt{42}}{21} \widetilde{m} -
\frac{\sqrt{14}}{7} \widetilde{n}}
\left( e^{2 \frac{\sqrt{42}}{3} \widetilde{m}} - 12 e^{
\frac{\sqrt{42}}{3} \widetilde{m}} \xi -
2 e^{\frac{4\sqrt{42}}{21} ( 2 \widetilde{m}+
\sqrt{3} \widetilde{n} )} \xi \zeta - 24
\xi^2 \right. \nonu \\
&& \left. -12 e^{
\frac{\sqrt{42}}{21} ( \widetilde{m}+ 4\sqrt{3}
\widetilde{n})} \xi^2 \zeta +e^{\frac{2\sqrt{42}}{21}
(\widetilde{m} + 4 \sqrt{3} \widetilde{n})} \xi^2 \zeta^2   \right).
\nonu
\eea
There are no critical points, in this case.

\vspace{2cm}
\centerline{\bf Acknowledgments}

This research was supported by Kyungpook National University
Research Fund, 2000 and
grant 2000-1-11200-001-3 from the Basic Research Program of the Korea
Science $\&$ Engineering
Foundation.
CA thanks Max-Planck-Institut f\"ur Gravitationsphysik,
Albert-Einstein-Institut where part of this work was undertaken and
thanks B. de Wit, C.M. Hull and H. Nicolai for discussions.

\end{document}